\documentclass[fleqn,usenatbib]{mnras}

\usepackage{newtxtext,newtxmath}

\usepackage[T1]{fontenc}

\DeclareRobustCommand{\VAN}[3]{#2}
\let\VANthebibliography\thebibliography
\def\thebibliography{\DeclareRobustCommand{\VAN}[3]{##3}\VANthebibliography}

\usepackage{graphicx}	
\usepackage{amsmath}	
\usepackage{subcaption} 
\usepackage{multirow}

\title[RGZ: FR classification]{Radio Galaxy Zoo: Morphological classification by Fanaroff-Riley designation using self-supervised pre-training}

\author[Buatthaisong et al.]{
Nutthawara Buatthaisong,$^{1}$\thanks{E-mail: nutthawara.buatthaisong@postgrad.manchester.ac.uk (NB)} 
Inigo Val Slijepcevic,$^{1}$
Anna M.~M.~Scaife,$^{1,2}$
Micah Bowles,$^{1,3}$ \and
Andrew Hopkins,$^{4}$ 
Devina Mohan,$^{1}$
Stanislav S Shabala,$^{5}$
O. Ivy Wong$^{6,7}$
\\
$^{1}$Jodrell Bank Centre for Astrophysics, University of Manchester, Oxford Road, Manchester M13 9PL, UK\\
$^{2}$Alan Turing Institute, Euston Road, London, UK\\
$^{3}$Department of Physics, University of Oxford, Oxford, UK\\
$^{4}$School of Mathematical and Physical Sciences, 12 Wally’s Walk, Macquarie University, NSW 2109, Australia\\
$^{5}$School of Natural Sciences, Private Bag 37, University of Tasmania, Hobart, TAS 7001, Australia\\
$^6$ CSIRO Space \& Astronomy, PO Box 1130, Bentley, Western Australia 6102, Australia \\ 
$^7$ ICRAR, University of Western Australia M468, 35 Stirling Highway, Crawley, WA 6009, Australia\\ 
}

\date{Accepted XXX. Received YYY; in original form ZZZ}

\pubyear{2025}

\begin{document}
\label{firstpage}
\pagerange{\pageref{firstpage}--\pageref{lastpage}}
\maketitle

\begin{abstract}
%
In this study, we examine over 14,000 radio galaxies finely selected from Radio Galaxy Zoo (RGZ) project and provide classifications for approximately 5,900 FRIs and 8,100 FRIIs. We present an analysis of these predicted radio galaxy morphologies for the RGZ catalogue, classified using a pre-trained radio galaxy foundation model that has been fine-tuned to predict Fanaroff-Riley (FR) morphology.
As seen in previous studies, our results show overlap between morphologically classified FRI and FRII luminosity-size distributions and we find that the model's confidence in its predictions is lowest in this overlap region, suggesting that source morphologies are more ambiguous. 
We identify the presence of low-luminosity FRII sources, the proportion of which, with respect to the total number of FRIIs, is consistent with previous studies. 
However, a comparison of the low-luminosity FRII sources found in this work with those identified by previous studies reveals differences that may indicate their selection is influenced by the choice of classification methodology. We investigate the impacts of both pre-training and fine-tuning data selection on model performance for the downstream classification task, and show that while different pre-training data choices affect model confidence they do not appear to cause systematic generalisation biases for the range of physical and observational characteristics considered in this work; however, we note that the same is not necessarily true for fine-tuning.  As automated approaches to astronomical source identification and classification become increasingly  prevalent, we highlight training data choices that can affect the model outputs and propagate into downstream analyses.

\end{abstract}




\section{Introduction}

\subsection{Radio galaxy morphology}

Radio galaxies, also known as radio-loud active galactic nuclei (AGN) or radio-loud quasars, are a type of active galaxy that generate jets of ions and magnetic fields, due to a combination of mechanisms involving the central supermassive black hole and its accretion processes \citep[see e.g.][for a review]{Hardcastle2020RadioJets}.
Charged particles in the jets produce synchrotron emission, which is detectable at radio frequencies. The morphology of these jets is thought to be linked to a number of different factors, both intrinsic and environmental, and hence separating populations based on their morphology has been used as a method for characterising relevant physical processes.  One widely utilised scheme used to classify radio galaxies is the Fanaroff-Riley dichotomy \citep{Fanaroff1974}. It divides radio sources into two main types: 
Fanaroff-Riley class I galaxies (FRIs) with luminous regions closer to the core than a distance equal to half their sizes, and Fanaroff-Riley class II galaxies (FRIIs) with high-luminosity regions further from the core than a distance equal to half their sizes. 
\cite{Fanaroff1974} also noted a connection between luminosity and morphology. 
It was reported that there is a sharp separation in luminosity between these FRIs and FRIIs that was evident at $\sim2 \times 10^{25}$ h$_{50}$ W Hz$^{-1}$ sr$^{-1}$ at 178 MHz. The difference in luminosity between FRIs and FRIIs was also reported in \cite{Ledlow1996}. 
However, the observed division of FRI/II luminosities discussed in earlier studies has now largely been discounted. \citet{Best2009} first suggested that the clear FRI/II separation found in \citet{Ledlow1996} might be influenced by selection effects, and the overlap between FRI and FRII sources has also been reported by many subsequent studies, such as \cite{Capetti2017A, Capetti2017B, Miraghaei2017, Macconi2020, Vardoulaki2021, Mingo2019, Mingo2022}.
The study conducted by \cite{Mingo2019} investigated the relation between the morphology of radio galaxies and their luminosity. Over half of total FRIIs were found to have a luminosity three orders of magnitude lower than the break line at 150 MHz, which is $1\times10^{26}$ W Hz$^{-1}$. Such results have confirmed that the FR dichotomy, i.e. the relationship between radio morphology and radio luminosity, is more complex than originally proposed.

\subsection{Morphological classification using deep learning}
There is a large amount of data in the field of astronomy that has been obtained from multi-wavelength surveys, and this volume of data is expected to increase over the years. 
For example, the Vera C. Rubin Observatory\footnote{\url{https://rubinobservatory.org/}}, also known as the Large Synoptic Survey Telescope (LSST), is expected to generate about 15 TB of raw data per night and will have collected approximately 60 PB at the end of the mission \citep{Ivezi2019}. The Square Kilometre Array Observatory (SKAO)\footnote{\url{https://www.skao.int/en}} is expected to generate around 600\,PB per year \citep{Scaife2020}. For comparison, the seventh data release of Sloan Sky Survey \citep[SDSS DR7;][]{Abazajian2009} contained 15\,TB of data, which were collected over ten years.
Accordingly, astronomers are increasingly adopting automated algorithms, including deep learning, to facilitate the processing of those data. 
Deep learning is a versatile tool for various tasks in astronomy. It has been applied in a number of contexts, including, but not limited to,  source and anomaly detection \citep[e.g.][]{Lochner2025} and object classification \citep[e.g.][]{Walmsley2022PracticalLearning, Wu2020}, telescope calibration and imaging \citep[e.g.][]{Lai2024}. 
In recent years, many studies have developed deep learning approaches to classify radio galaxies based on their morphology, in particular using the Fanaroff-Riley scheme, with different sample sizes, sources and various techniques \citep[e.g.][]{Aniyan2017, Alhassan2018, Wu2019, Tang2019, Wang2021ResearchYolov5, Bowles2021Attention-gatingClassification, Mohan2022QuantifyingClassification, Slijepcevic2021CanClassification, Mohale2024, BaronPerez2025, Lao2023, Lao2025,Barnes2025QuantifyingMorphology}. \citet{Galvin2020} used self-organizing maps, a form of unsupervised machine learning that reduces high-dimensional data to low-dimensional representations while maintaining the topological structure of the input data, to identify the radio components and their infrared host galaxies and successfully found objects with rare and unique morphologies, including giant radio galaxies.

\subsection{Foundation model approaches}
\label{sec:foundationmodels}

A recent paradigm shift in deep learning research has resulted in the field moving away from building ``from scratch'' models that solve a specific task without any prior information, towards an approach that uses a large-scale multi-purpose pre-trained (``up-stream'') \emph{foundational} representation as the basis for addressing multiple downstream tasks, including those that require an additional \emph{fine-tuning} of the pre-trained representation using a smaller labelled dataset \citep[see e.g.][]{Bommasani2021OnModels}.

The initial pre-training stage usually involves a relatively large compute budget and often takes advantage of large-scale unlabelled datasets by using a self-supervised training loss. The fine-tuning stage then requires significantly fewer labelled data to achieve equivalent or improved performance to a ``from scratch'' approach for a specific task using a much larger dataset of labelled data.

Following these advancements in the deep-learning literature, practitioners in astronomy have closely followed by making use of large volumes of unlabelled archival data by following the pre-train and fine-tune recipe. 
So far, most successful applications in astronomy have used a variation of contrastive, or \emph{view-based}, self-supervised learning, an approach that learns meaningful representations by contrasting features based on unlabelled data, with convolutional neural networks \citep{Lecun1998, Goodfellow2016}, 
to train their model. \cite{Huertas-Company2023AAstrophysics} provide a comprehensive overview of contrastive learning methods applied to astrophysics. For example, \citet[][]{Stein2022} and \citet[][]{Hayat2021} successfully adapted the Momentum Contrast (MoCo) technique \citep{He2020MomentumLearning} to data mine for gravitational lenses and improve galaxy morphology classification, respectively. The participation of citizen-scientists enabled the Galaxy Zoo project \citep{Lintott2008GalaxySurvey} to perform model pre-training on a labelled dataset comparable in size to ImageNet-1k \citep{Deng2010ImageNet:Database} through a custom joint classification loss \citep{Walmsley2022PracticalLearning, Walmsley2024ScalingImages}, but this model was also shown to benefit from a hybrid contrastive learning approach \citep{Walmsley2022TowardsLearning}. AstroCLIP \citep{Lanusse2023AstroCLIP:Models} use paired optical spectra and images of galaxies to learn a joint embedding space, enabling high zero and few shot performance, i.e. a model's ability to perform without prior examples or with a minimal amount of training data, on relevant downstream tasks. \citet{Riggi2024} employed self-supervised contrastive learning methods to analyse radio image data for various downstream tasks, such as classifying radio source morphology, detecting radio sources, and identifying peculiar objects. A representation learned with self-supervised learning can also be used without further fine-tuning, for tasks such as similarity search \citep{Slijepcevic2023RadioLearning} and mining unusual sources from large unlabelled datasets \citep{Walmsley2023RareRepresentations}.

Although self-supervised learning (SSL) has gained significant traction as a substitute for traditional supervised learning, especially in contexts where labelled data is sparse or challenging to acquire, the SSL model has various challenges and constraints in practical applications. Corrupted, warped, or noisy data in pre-training may impair the model's efficacy \citep{budach2022effects} and may result in adverse consequences for fine-tuned models, including insufficient generalisation and diminished performance \citep{chen2024on}.
In addition, a common problem in predictive modelling is dataset shift, which is where the distribution of inputs and outputs changes between the training and testing phases \citep[e.g.][]{QuinoneroCandela2008Datasetshift, MORENOTORRES2012}. Dataset shift might reduce the model's accuracy \citep[e.g.][]{ovadia2019, Finlayson2021, Bissoto2024}, but also for SSL models, dataset shift can adversely affect the model's performance and efficiency as the model may learn representations that are not able to transfer to the out-of-distribution data in case the pre-training data is biased towards particular data distributions. This potentially results in poor generalisation \citep{sarkar2023uncovering}. Additionally, self-supervised models often involve fine-tuning on a specific downstream task. The accuracy of the fine-tuning process can be impacted negatively by the dataset shift in pre-training data, as the model might not be able to adapt well to other data distributions \citep{kumar2022finetuning}.

In this work we use a pre-trained radio galaxy foundation model \citep{Slijepcevic2023RadioLearning} to provide predictive Fanaroff Riley classifications for the Radio Galaxy Zoo catalogue. The Radio Galaxy Zoo \citep[RGZ;][]{Banfield2015} dataset itself does not include FR classification, instead it provides morphological classification labels that describe the number of components and the total number of surface brightness peaks within those components. Our work uses the SSL radio galaxy foundation model from \cite{Slijepcevic2023RadioLearning}, which was pre-trained using the RGZ data, and fine-tuned for Fanaroff-Riley classification using the \textit{confidently} classified sources from the MiraBest dataset, in order to provide predictive FR labels for the catalogue of radio galaxies from the RGZ project.
We analyse these predictions in an astrophysical context, while also considering the effects of known potential biases due to dataset shift, as described in the preceding paragraph, in order to establish their impact in a downstream astronomical context.
The structure of the paper is as follows: in Section~\ref{sec:data}, we describe the datasets used in this work, and in Section~\ref{sec:class} we describe the application of the fine-tuned model; in Sections~\ref{sec:ld}~\&~\ref{sec:colours} we analyse the resulting distributions for the classified objects with respect to both physical and other observational characteristics, and in Section~\ref{sec:vote_fraction} we consider these distributions in the context of the model confidence for individual classification; in Section~\ref{sec:lotss} we cross-match the RGZ classifications with the similarly labelled catalogue from the LOFAR LoTSS DR1 \citep{Shimwell2019} survey by \citet{Mingo2019} and draw comparisons. In Section~\ref{sec:biases} we investigate the impact of different generalisation biases inherent in the machine learning model on these results, and in Section~\ref{sec:conclusion} we draw our conclusions.

In this study, we have used the cosmological values; $H_0 =$ 70\,Mpc$^{-1}$ km$^{-1}$, $\Omega_m =$ 0.3, and $\Omega_{\Lambda} =$ 0.7.

\section{Data}
\label{sec:data}

\subsection{Radio Galaxy Zoo (RGZ)}
\label{sec:rgz}

The RGZ project is an online citizen science project that is aimed at the morphological classification of extended radio galaxies and the identification of host galaxies \citep{Banfield2015}. The online RGZ programme operated for $\sim$5.5 years between December 2013 and May 2019.  Within this operational period, over 2.2 million independent classifications were registered for over 140,000 subjects.  
The RGZ subjects inspected by citizen scientists consist of images derived from the the Faint Images of the Radio Sky at Twenty-Centimeters (FIRST) survey \citep{becker1995, White1997, Franzen2015} and the ATLAS survey \citep{norris2006}, overlaid onto WISE \citep{Wright2010wise} $W1$ (3.6~$\mu$m) infrared images. Within the catalogue, 99.4\% of classifications used radio data from the FIRST survey, and the remainder used data from the ATLAS survey. 
The RGZ dataset used in this paper is from the RGZ Data Release 1 catalogue \citep[RGZ DR1;][]{Wong2025} and all image data are taken from the FIRST survey.

The RGZ DR1 catalogue contains approximately 100,000 source classifications with a user-weighted consensus fraction (consensus level) that is equal to or greater than 0.65.   
The largest angular size of each source in the RGZ DR1 is estimated by measuring the hypotenuse of a rectangle that encompasses the entire radio source at the lowest radio contour \citep{Wong2025}. This method is generally reliable if the radio components of a source are correctly-identified and the source components are distributed in a linear structure (in projection). 
The RGZ DR1 catalogue matches up all related radio components to a host galaxy observed in the $W1$ infrared image, but does not provide FR class labels for these radio components. In this study, we used the catalogue of cross-matched host galaxy positions\footnote{\url{https://doi.org/10.5281/zenodo.14195049}} as image centroids to define our dataset.

\subsection{MiraBest} \label{sec:MiraBest}

The MiraBest machine learning dataset \citep{PorterMiraBest2023} consists of 1,256 images of radio galaxies pre-processed for deep learning tasks. The dataset was constructed using the sample selection and classification of 1,329 extended radio sources as described in \cite{Miraghaei2017}, who made use of the parent galaxy sample from \cite{Best2012OnProperties}. Optical data from data release 7 of the Sloan Digital Sky Survey \citep[SDSS DR7;][]{sdssdr7} was cross-matched with NRAO VLA Sky Survey  \citep[NVSS;][]{condon1998} and FIRST radio surveys. Parent galaxies were selected such that their radio counterparts had an AGN host rather than emission dominated by star formation. To enable classification of sources based on radio morphology, sources with multiple components in either of the radio catalogues were considered. 

In \citet{Miraghaei2017}, the morphological classification was done by visual inspection at three levels: (i) The sources were first classified as FRI/FRII based on the original classification scheme of \citet{Fanaroff1974}. Additionally, 35 \emph{Hybrid} sources were identified as sources having FRI-like morphology on one side and FRII-like on the other \citep{Gopal-Krishna2000}. Of the 1329 extended sources inspected, 40 were determined to be unclassifiable. (ii) Each source was then flagged as `Confident' or `Uncertain' to represent the degree of belief in the human classification and, although this qualification was not extensively explained in the original paper, \citet{Mohan2021WeightClassification} have shown that the original catalogue qualification of a source label as being confident or uncertain is correlated with model posterior variance over the dataset. 
We note that these “uncertain” MiraBest sources are generally fainter in terms of their peak and total flux density, but are not significantly smaller compared to the confident sources.
(iii) Some of the sources which did not fit exactly into the standard FRI/FRII dichotomy were given additional tags to identify their sub-type. These sub-types include 53 Wide Angle Tail (WAT), 9 Head Tail (HT) and 5 Double-Double (DD) sources. To represent these three levels of classification, each source was given a three digit identifier as shown in Table~\ref{tab:digits_mirabest}.

\begin{table}
\centering
\caption[Short table caption.]{Three digit identifiers for sources in \citet{Miraghaei2017}}
	\begin{tabular}{llll}
    Digit 1 & Digit 2 & Digit 3 \\
    
    \hline
    
    0: FRI & 0: Confident  &  0: Standard \\
    1: FRII & 1: Uncertain & 1: Double Double \\
    2: Hybrid &         & 2: Wide Angle Tail \\ 
    3: Unclassifiable & & 3: Diffuse \\
    &  &                   4: Head Tail \\ 
	\end{tabular}
   \label{tab:digits_mirabest}
\end{table}

\begin{table}

\centering
\caption[MiraBest Class-wise Composition]{MiraBest Class-wise Composition}
	\begin{tabular}{cccc}
	\hline
    Class & Confidence & No. \\
    \hline
    \multirow{2}{2em}{FRI}  & Confident & 397\\
     & Uncertain & 194\\
    \hline
    \multirow{2}{2em}{FRII}  & Confident & 436\\
     & Uncertain & 195\\
    \hline
    \multirow{2}{2em}{Hybrid}  & Confident & 19\\
     & Uncertain & 15\\
    \hline
    
	\end{tabular}
   \label{tab:mb_classes}
\end{table}

To ensure the integrity of the ML dataset, the following 73 objects out of the 1329 sources identified in the catalogue were not included: (i) 40 unclassifiable objects; (ii) 28 objects with extent greater than the chosen image size of $150\times150$ pixels; (iii) 4 objects which were found in overlapping regions of the FIRST survey; (iv) 1 object in category 103 (FRII Confident Diffuse). Since this was the only instance of this category, it would not have been possible for the test set to be representative of the training set. The composition of the final dataset is shown in Table~\ref{tab:mb_classes}. We do not include the sub-types in this table as we do not consider their classification in this work.

\section{Classification}
\label{sec:class}

\subsection{RGZ FR classification model}
\label{sec:finetune}

Although FR labels are not available for the RGZ dataset, the model can be pre-trained on these data with self-supervised learning, which allows the model to see a larger number of images, resulting in better generalisation. We use the foundational model from \cite{Slijepcevic2023RadioLearning}, which has been pre-trained on the RGZ DR1 unlabelled catalogue. 

The pre-training of this model uses the Bootstrap Your Own Latent \citep[BYOL; ][]{Grill} algorithm, which is based on solving an instance differentiation task. During training, the distance in the representation space of the model between augmentations of the same image is minimised, resulting in a model which has learned a semantically structured representation of the training dataset, which can then be used for a specific task (e.g. classification). Details of how this model was trained specifically for radio galaxy images can be found in \cite{Slijepcevic2023RadioLearning}.

We use a fine-tuned version of this foundational radio galaxy model with the addition of a \emph{classification head}, i.e. an additional fully-connected layer responsible for generating predictions, to classify sources using the FR classification scheme. The model is fine-tuned using the confidently classified sources from the MiraBest dataset \citep{Miraghaei2017, PorterMiraBest2023}. The fine-tuning procedure trains all layers of the model with a layer wise learning rate decay (earlier layers have a lower learning rate). Further details of the fine-tuning recipe can be found in \cite{Slijepcevic2023RadioLearning}. This fine-tuned model achieved good performance with 1.92\% test error when the model was fine-tuned using the full 729 \textit{confidently} labelled sources from the MiraBest training data set.
In addition, the model outperformed baseline supervised methods in this scenario \citep{Slijepcevic2023RadioLearning}. 

A potential limitation in these results is that the model training (both pre-training and fine-tuning) only used data from the FIRST survey, unlike the original classifications of \citet{Miraghaei2017} that utilised both FIRST and NVSS radio images. Although FIRST has better resolution, NVSS is more sensitive to extended low-surface-brightness emission. Therefore the model performance will be limited in that regard, and this is potentially reflected in the reported  test classification error.

\subsection{RGZ FR catalogue}
\label{sec:frcat}



\textcolor{blue}{The RGZ\,DR1 catalogue contains} a total of 141,679 radio subjects \citep{Wong2025}. In this study, we combine data from various observations across different wavelengths, including the radio band from VLA FIRST, the wide-infrared wavelength from AllWISE, and the visible bands from SDSS. 



\begin{figure*}
\centerline{
\includegraphics[width=0.49\textwidth]{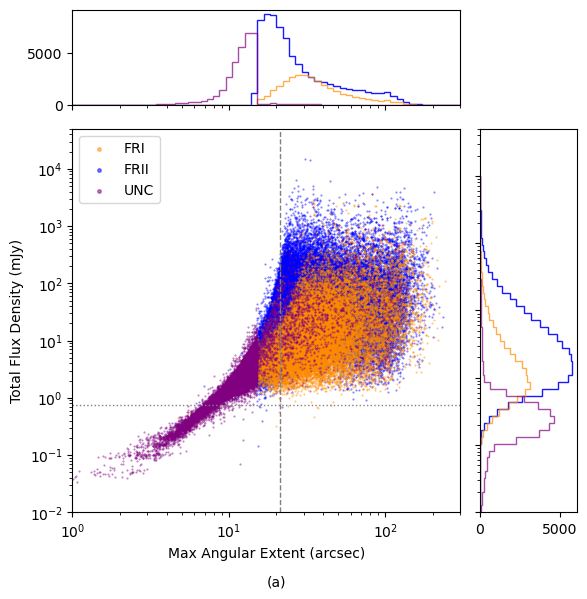}
\includegraphics[width=0.49\textwidth]{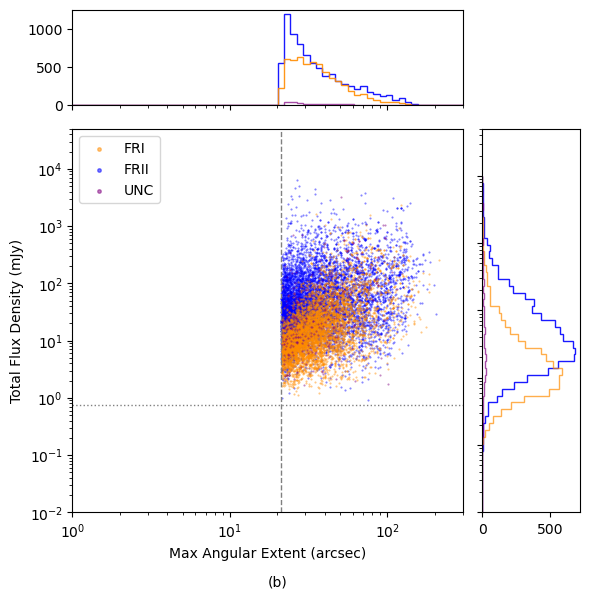}
}
\caption{The distribution as a function of total flux density and angular size with the angular size and flux density thresholds of 21.2\,arcsecond (dashed line) and 0.75\,mJy (dotted line), respectively. (a) all RGZ sources (141,678 sources); (b) the RGZ\,FR sources (14,375 sources) with redshift information and angular size and flux density above the thresholds, see Section~\ref{sec:frcat} for more details.}
\label{fig:flux_angsize}
\end{figure*}

Using the RGZ foundation model fine-tuned for Fanaroff-Riley classification as described in Section~\ref{sec:finetune} we identify 32,241 FRIs and 76,304 FRIIs from the full RGZ catalogue. The remaining 33,133 sources have been defined as \emph{unclassifiable} sources, referring to sources having an angular size of less than 15\,arcsecond (31,034 sources) or which are duplicated sources (2,099 sources), i.e. identical sources with multiple entries or IDs. The full distribution of RGZ sources is shown in Figure~\ref{fig:flux_angsize} (a).
We then select all sources with an angular size larger than $\Delta \theta > 21.2$\,arcsecond and a total flux density greater than $S_{\rm tot}  > 0.75$\,mJy, equivalent to three times the resolution and five times the rms noise in FIRST, respectively. 
%

Subsequently, the RGZ\,DR1 provides a cross-matching of the sources with either spectroscopic or photometric redshift from the SDSS database by \citet{Rowan-Robinson2013}. 
The redshift cross-matched RGZ\,DR1 dataset includes 7,486 sources (approximately 52\%) with spectroscopic redshifts; all remaining sources have photometric redshifts. 
We require that the photometric redshifts matched to the RGZ\,FR sources should have relatively low uncertainty, following the criteria outlined in \cite{Duncan2022allpurpose}:
\begin{equation}
\frac{\sigma_z}{1+z_{\rm phot}} < 0.2,
\end{equation}
where $\sigma_z$ refers to the uncertainty of the photometric redshift and $z_{\rm phot}$ denotes the photometric redshift value. This quality threshold excludes only one source from our final sample.
Following this selection, the final sample comprises 5,933 FRIs, 8,135 FRIIs, and 307 unclassifiable sources. The resulting distribution of redshift-matched sources as a function of angular size and flux density is shown in Figure~\ref{fig:flux_angsize} (b). 
We refer to the sample of 14,375 sources with redshift information as ``RGZ\,FR''.

Additionally, the FRII objects can be categorised into low- and high-luminosity subclasses, which will be denoted as FRII-Low and FRII-High hereinafter, based on the traditional luminosity break line at $\sim 2\times 10^{25}$ W\,/\,Hz at 1.4\,GHz, calculated by power-law slope of a non-thermal synchrotron emission with $\alpha$ = 0.7 from originally defined at $1\times 10^{26}$ W\,/\,Hz at 150\,MHz by \citep{Fanaroff1974}.
As a result, we obtain 4,719 FRII-Low and 3,416 FRII-High sources. The number of sources in each class is summarised in Table~\ref{tab:FR_class}.

\begin{table*}
\centering
\caption{The radio galaxy classes in the RGZ\,FR catalogue with the number of the total sources and the sources with spectroscopic redshift in each class.}
\label{tab:FR_class}
\begin{tabular}{cccc}
\hline
\multicolumn{1}{|c|}{Class Name} & \multicolumn{1}{c|}{Class code} & \multicolumn{1}{c|}{Number of Total Sources} & \multicolumn{1}{l|}{Number of Sources with $z_{\text{sp}}$} \\ \hline
FRI                              & 1                               & 5,933                                        & 3,264                                                \\
FRII-Low                         & 2                               & 4,719                                       & 2,498                                                \\
FRII-High                        & 2                               & 3,416                                        & 1,559                                                \\
Unclassifiable                   & -1                              & 307                                          & 165                                                  \\ \hline
Total                            &                                 & 14,375                                       & \multicolumn{1}{c|}{7,486}                          \\ \hline
\end{tabular}
\end{table*}

The full catalogue of RGZ\,FR-$z$ data containing 14,375 radio sources will be made publicly available at \url{https://doi.org/10.5281/zenodo.14031760}.
 



\section{Luminosity-distance analysis}
\label{sec:ld}

\begin{figure}
\centerline{
\includegraphics[width=0.5\textwidth]{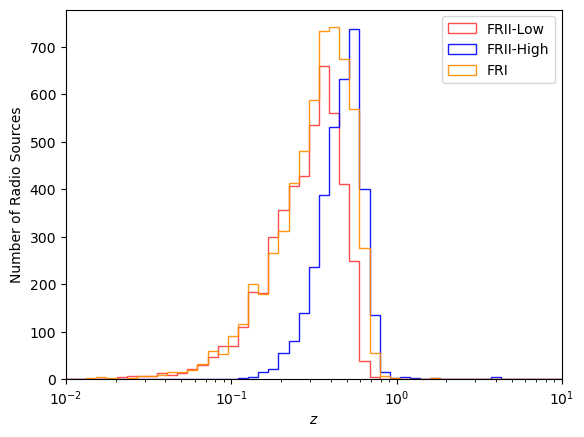}
}
\caption{Redshift distribution of  RGZ\,FR sources, including FRIs (orange), FRII-Low objects (red), and luminous FRIIs (blue).}
\label{fig:z}
\end{figure}

The luminosity of each source from the RGZ\,FR catalogue was calculated using the total flux density obtained from the VLA FIRST data at a frequency of 1.4\,GHz,
and the matched SDSS redshift, where the redshift distribution for the RGZ\,FR sources is shown in Figure~\ref{fig:z}.

\begin{figure*}
\centerline{
\includegraphics[width=0.6\textwidth]{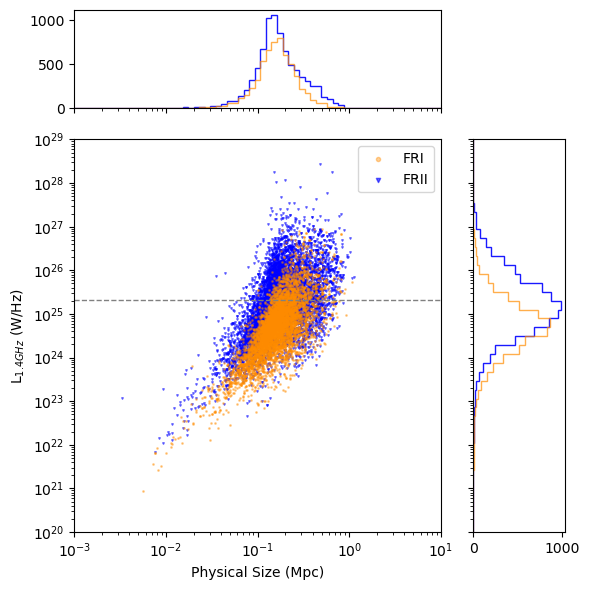}
}
\caption{Luminosity (W/Hz) at 1.4\,GHz versus physical size (Mpc) of 14,375 RGZ\,FR sources, the dashed line represents the historical FRI/II break line ($\sim2\times10^{25}$ W/Hz).}
\label{fig:lum_physize}
\end{figure*}

The physical size of each source was calculated based on the angular extent as listed in the RGZ\,DR1 catalogue. The distribution of sources with respect to luminosity and physical size for FRIs and FRIIs in the RGZ\,FR sources is depicted in Figure~\ref{fig:lum_physize}.

It can be seen from Figure~\ref{fig:lum_physize} that FRII sources cover a wide range from low to high luminosity and from compact to large physical size, while FRI sources tend to be clustered at slightly lower luminosities and intermediate sizes. 
For physical size, FRIIs have mean and median values of 0.20\,Mpc and 0.16\,Mpc, compared to FRIs with a mean of 0.18\,Mpc and a median of 0.16\,Mpc. 
On the other hand, FRIs have a smaller mean luminosity of $1.63\times10^{25}$\,W/Hz and a smaller median of $5.20\times10^{24}$\,W/Hz, compared to the FRII mean and median luminosity of $1.53\times10^{26}$\,W/Hz and $1.52\times10^{25}$\,W/Hz, respectively.
This indicates that  FRIIs tend to be slightly larger on average and are generally more luminous compared to FRIs.  
However, the size estimates for FRIs can be significantly affected by observational sensitivity, as jetted FRI sources are core-brightened compared to lobed FRI sources \citep{Leahy1993}. \citet{Heesen2017} showed that the sizes of jetted FRIs, e.g. 3C 31, are larger than previously thought. In addition, 
\citet{Turner2017} suggested that the sizes estimates from simulated models of lobed FRIs and FRIIs are not as severely affected by observational sensitivity, in contrast to those of jetted FRIs.

In this work, we study a combined population of lobed and jetted FRIs, which may affect the classification results due to the differences in observable properties for these subclasses. In future work, classifying these subclasses separately could enhance our understanding of the nature and evolutionary processes of these objects.

There is a significant overlap region between FRIs and FRIIs in this distribution, which contradicts the widely accepted understanding initially proposed in the research of \cite{Fanaroff1974}. 
The presence of this overlap also appeared in previous studies conducted, e.g. \cite{Capetti2017A, Miraghaei2017, Mingo2019}. 
From the data presented in Figure~\ref{fig:lum_physize}, approximately 15\% of the total FRIs (908 sources out of 5,933) have luminosities in excess of the break line. This finding aligns with the proportion reported by \cite{Mingo2019} of 11\%.
A significant proportion ($\sim 58\%$) of the FRIIs lie below the break-line (low-luminosity FRIIs or FRII-Low objects), while the remaining sources, roughly 42\%, are referred to as luminous FRIIs or FRII-High.
\cite{Mingo2019} suggested that the presence of low-luminosity FRIIs is a challenge to the understanding of the jet dynamic model of FRI/II galaxies. Therefore, it is important to conduct further investigations in order to fully understand this ambiguous separation between FRIs and FRIIs.

\section{Colour analysis}
\label{sec:colours}

\begin{figure*}
\centerline{
\includegraphics[width=\columnwidth]{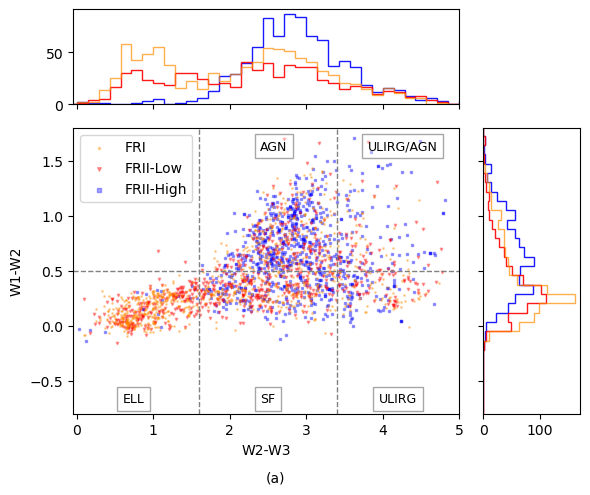}
\includegraphics[width=\columnwidth]{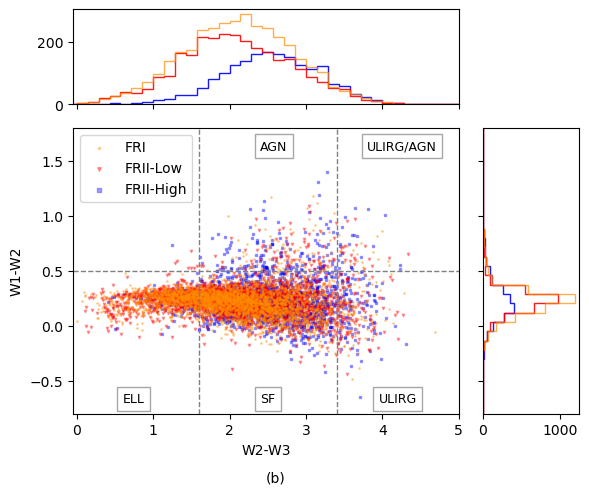}
}
\caption{WISE colour/colour plot comparing among selected RGZ FRI (yellow), FRII-Low (red), and FRII-High (blue) objects with the W1 and W2 SNR above 5 and the W3 SNR (a) above 3 (2,494 sources) and (b) lower than 3 (8,358 sources).}
\label{fig:wise_cc}
\end{figure*}

WISE colour analysis can be used to study the properties of the host galaxy of each radio source, as originally proposed by \cite{Wright2010wise} and investigated further by many studies \citep[e.g.][]{Lake2012WiseColour, Assef2013Midinfrared, Mingo2019, Mingo2022}.
The WISE colour/colour plots illustrate the distribution of galaxies based on their magnitude differences (colour indices) in the W1 ($\lambda_{\text{eff}}$ = 3.4 $\mu$m), W2 ($\lambda_{\text{eff}}$ = 4.6 $\mu$m), and W3 ($\lambda_{\text{eff}}$ = 12 $\mu$m) bands for each source.
This colour analysis can indicate the physical properties and characteristics of the galaxies hosting radio sources, allowing for a better understanding of their nature and evolution. 
By examining the WISE colour/colour plots, we can depict the host star formation rate and radiation of AGNs, thus we can roughly predict the types of host galaxies, including ultra-luminous infrared galaxies (ULIRGs), AGNs and quasi-stellar objects, elliptical galaxies (Ells), and spiral galaxies (star forming galaxies; SFGs), as shown in Figure~\ref{fig:wise_cc}. 
The sources are selected using the W1 and W2 thresholds set at a signal-to-noise ratio (SNR) of 5 and 3 for W3.
The lines dividing the population in terms of their W2$-$W3 colour at 1.6 and 3.4 mag distinguish different regimes in the star formation rate, and the division at 0.5 mag in the W1$-$W2 colour represents the degree of dominance of AGNs \citep{Mingo2016WiseColour}.

Approximately 60\% of the sources in the RGZ\,FR sources are found in the SFG area. Our study shows that 87\% of FRIs are distributed within the range of W1$-$W2 $< 0.5$ mag and W2$-$W3 $< 3.4$ mag, covering the Ell and SFG regions, while FRIIs are more dominant for W2$-$W3 $> 1.6$ mag, especially in the bright high-excitation radio galaxy (HERG) region (W1$-$W2 $> 0.5$ mag).
On the other hand, FRIIs with high luminosity and low luminosity tend to appear in different areas on the diagram. Low luminosity FRIIs tend to cluster in the Ell and SFG regions, similar to FRIs, while more luminous FRIIs are mainly found in the SF and AGN regions.
The overlap, particularly in the SF region, is readily apparent in Figure~\ref{fig:wise_cc}. Figures~\ref{fig:wise_cc} (a) and (b) show the colour/colour diagram for RGZ\,FR sources with W1 and W2 signal-to-noise ratio (SNR) greater than 5, and W3 SNR greater than or equal to 3, and below 3, respectively.
A W3 SNR below three indicates that the position in the horizontal direction may shift further to the left (towards the Ell area) than that the position indicated in the diagram. As represented in Figure~\ref{fig:wise_cc} (a), the overlap remains apparent even after considering the W3 upper limit.


The figure also shows the overlap of FRIs and FRIIs. Even though we applied the W3 SNR limit, Figure~\ref{fig:wise_cc} (a) still shows the overlap region among all FR classes.
However, it is readily apparent that luminous FRIIs are more dominant for W2$-$W3 $> 1.6$, especially in the W1$-$W2 $> 0.5$ area, while FRIs lie predominantly in the elliptical and star-forming source area (W1$-$W2 $< 0.5$; W2$-$W3 $< 3.4$). On the other hand, FRII-Low sources appear to follow a more similar colour distribution to FRIs than FRII-High sources.



The distribution of RGZ\,FR sources across the colour space shows that FRII-Low objects are found pre-dominantly in the SFG and ELL regions, while luminous FRIIs are concentrated in the AGN or SFG regions. The possible explanation is that the host galaxy properties of AGNs appear differently in various analyses depending on the proportion of AGN emission contributing to each diagnostic \citep{Prathap2024}. Radio sources lying in the SFG/ULIRG regions have stellar populations that are dominating the IR light, while those with redder W1$-$W2 colours have measurable AGN contributions to the IR light that dominate over the stellar contribution. It is possible that the sources in the SFG/ULIRG regions are low radio luminosity AGNs, and their host galaxy colours are dominated by the star-forming stellar population rather than AGNs. No rest frame correction is applied to the WISE magnitudes in Figure~\ref{fig:wise_cc}. Thus the distribution in the intermediate (SFG) W2$-$W3 region might be affected by redshift, whereby sources become redder at higher redshift causing their W1 and W2 magnitudes to appear brighter, moving them towards the upper left of the colour-colour diagram. 
\citet{Jarrett2023} suggested that sources having z $>$ 0.2, which would include 83\% of the RGZ\,FR sources, may require a rest frame correction. 


\section{RGZ FR vote fraction: luminosity distance and colour analysis} \label{sec:vote_fraction}


Training with finite data volumes can sometimes lead to unpredictable results during inference. For this reason, in this work the pre-trained model is fine-tuned 10 times with a different weight initialisation in the classification head, with prediction results being aggregated across all 10 fine-tuned models. These fine-tuning realisations provide a ``vote fraction'' (VF) metric: the proportion of the fine-tuned models that voted for the majority class in each case. This proportion can be interpreted as a simple measure of the model's confidence in a given prediction.

\begin{figure*}
\centerline{
\includegraphics[width=0.49\textwidth]{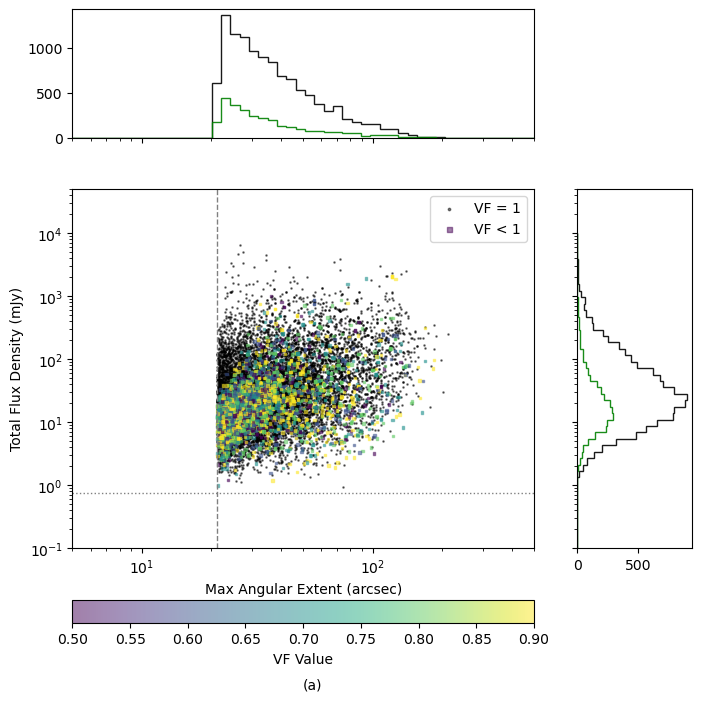}
\includegraphics[width=0.49\textwidth]{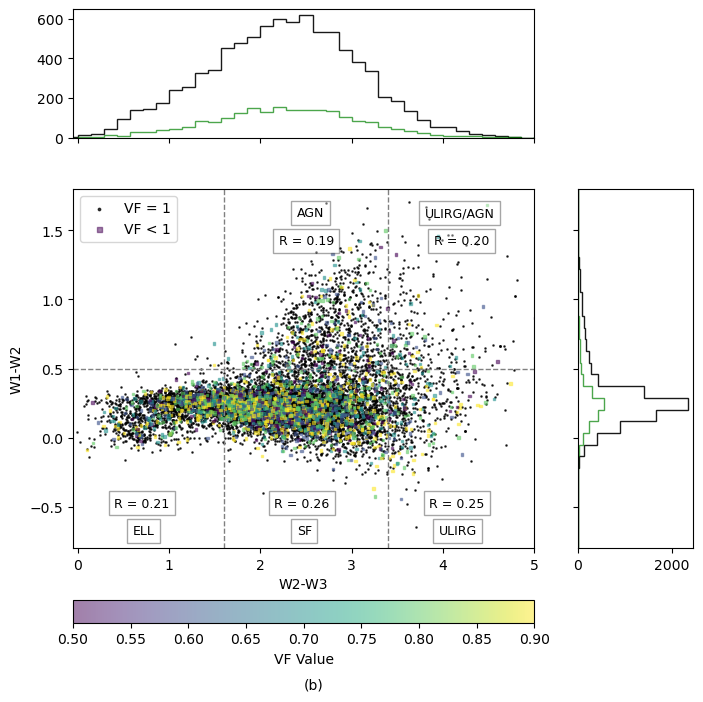} 
}
\caption{Comparison between RGZ\,FR sources with a vote fraction (VF) equal to 1 (black) and the sources with the VF value less than 1 (coloured). (a) Total flux density (mJy) at 1.4 GHz versus angular size (\,arcsecond) with with the angular size and flux density thresholds of 21.2\,arcsecond and 0.75 mJy, indicated by the dashed lines, respectively.
(b) WISE colour/colour plot comparing RGZ\,FR sources with VF equal to 1 and sources with VF value less than 1.}
\label{fig:votefrac}
\end{figure*}

Figure~\ref{fig:votefrac} illustrates the distribution of VF values. Black points represent sources with VF equal to 1 (agreement across all realisations), while gradient-coloured points (ranging from blue to yellow) denote sources with VF less than one. The colour bar on the bottom of each plot indicates the value of the vote fraction, which ranges from 0.50 (purple) to 0.90 (yellow). The coloured data points are densely clustered, particularly in the region where the angular size is between 21.2\,arcsecond and 55\,arcsecond and the total flux density ranges between 1 and 100\,mJy.  This area covers small and faint radio sources, as well as a boundary between FRIs and FRIIs, as depicted in Figure~\ref{fig:flux_angsize}. It indicates that the classification results of the models are more ambiguous in this region.

We also investigated the mid-infrared colour distribution of the data with a vote fraction equal to or less than one, with W1 and W2 SNR greater than 5 and no limit applied on W3. 
Figure~\ref{fig:votefrac} (b) depicts the distribution of sources with a VF < 1 on the WISE colour/colour plot, which can indicate the type of host galaxy. Sources with VF < 1 are present in all blocks; however, there is a greater density of these sources in the lower region of the plot (W1$-$W2 $< 0.5$), which includes the ELL, SF, and ULIRG regions. The ratio denoted by the parameter $R$ in Figure~\ref{fig:votefrac} (b) represents the fraction (ratio) of
the sources with VF < 1 (coloured) to the sources with VF = 1 in each region. 
We compared the number of the sources with VF < 1  and ratio $R$ and found that the SFG region with $R$ = 0.26 accounts for 64\% of the total VF < 1 sources followed by the sources in the ELL region with $R$ = 0.21, which accounts for approximately 19\%. 
Although the ratio $R$ seems relatively high in the ULIRG region, the proportion of sources with VF < 1 is significantly lower than in the SFG and ELL regions. More precisely, the number of VF < 1 sources in the ULIRG region only accounts for 7.6\% and 25\% of sources in the SFG and ELL regions. 
Since the ratio could be used as an indicator of the ambiguity in the classification results, our findings indicate that the SFG region, where we witnessed the overlap of FRIs and FRIIs, is the most difficult region for the algorithm to classify. Additionally, the high ratio in the ULIRG region could be attributed to the small size and lower luminosity of the sources. In contrast, the total number of sources in this region is lower compared to other regions.


\section{RGZ-LoTSS Cross-matching}
\label{sec:lotss}



We cross-match the RGZ\,FR sources with the FR classification catalogue produced by \cite{Mingo2019}. This catalogue consists of 5,805 radio galaxies that were obtained from the LoFAR Two-Metre Sky Survey DR1 value-added catalogue \citep[LoTSS DR1;][]{Williams2019}. The data were analysed using an automated algorithm called \texttt{LoMorph} \citep{Mingo2019}, which is implemented in \texttt{python}, following source finding with the Python Blob Detector and Source Finder (\texttt{PYBSDF}\footnote{\url{https://github.com/lofar-astron/PyBDSF}}).
The catalogue provides six subclass labels for FR classification: FRI, FRII, indeterminate, small, double-double, and bent-tailed sources. Indeterminate sources refer to sources that cannot be categorised into FRI or FRII. This class includes hybrids, fuzzy blobs, sources with unreliable identification, and unresolved sources. Small sources can be described as compact objects that have an angular size below 27\,arcsecond \citep{Mingo2019}. Bent-tailed sources are the sources classified as FRIs that possess either a wide-angle tail or a narrow-angle tail. Double-double (D-D) sources refer to double-lobed radio sources or restarted radio galaxies \citep{Schoenmakers2000, Mahatma2019LoTSS}.

In this study, we found 513 sources that are present in both the RGZ\,FR (14,375 sources) and LoTSS\,FR catalogues. The RGZ-LoTSS sample includes 225 FRIs, 158 low-luminosity FRIIs, and 130 high-luminosity FRIIs according to RGZ\,FR classification. By comparing RGZ\,FR labels to LoTSS FR labels, as shown in Table~\ref{tab:cross-match}, it was found that around 40\% of the catalogued objects have matching classifications in both datasets, and these will be referred to as \emph{class-matched} sources. 
Out of all RGZ-FRIs, 68\% were identified as LoTSS-FRIs, whereas 17\% of RGZ-FRIIs were classified as LoTSS-FRIIs. Furthermore, it was found that 22\% of high-luminosity RGZ-FRIIs were categorised as LoTSS-FRIIs, and  13\% of low-luminosity RGZ-FRIIs were classified as LoTSS-FRIIs. On the other hand, around 50\% of total FRIIs possess LoTSS-FRI labels. These sources with different FR classifications in the two catalogues will be referred to as \emph{class-changed} sources. 
It can be concluded that class-matched sources account for 40\% of the total RGZ-LoTSS sample, including 76\% of RGZ FRIs and 24\% RGZ FRIIs. On the other hand, class-changed sources are the remaining 60\%, with including 23\% FRIs and 77\% FRIIs in the RGZ\,FR catalogue.
We note that it is possible that a change in FR classification between LoTSS and RGZ datasets is not only due to differences in observational parameters such as sensitivity and resolution, but there are also factors linked to the synchrotron emission mechanism that may contribute. For example, the synchrotron energy loss timescales are different between the low frequency LOFAR data and the higher frequency VLA data, resulting in increased CRe life times, and hence larger path lengths, at lower frequencies \citep[e.g.][]{mulcahyM51}. This can cause sources to appear larger at lower frequencies. In particular this may account for the ``fried egg'' like structure exhibited in many of the LOFAR images for the class-changed sources shown in Figure~\ref{fig:FR2_FR1}.

\begin{figure*}
        \centering
        \setlength{\tabcolsep}{1pt}
        \setlength{\doublerulesep}{1pt}
        \begin{tabular}{p{2mm}ccccc}
             & RGZJ144219.2+504357 & RGZJ133336.2+542749 & RGZJ135142.1+555943 & RGZJ120458.6+455937 & RGZJ112942.1+542529\\
            \rotatebox{90}{LoTSS}
            & \includegraphics[width = 0.18\textwidth]{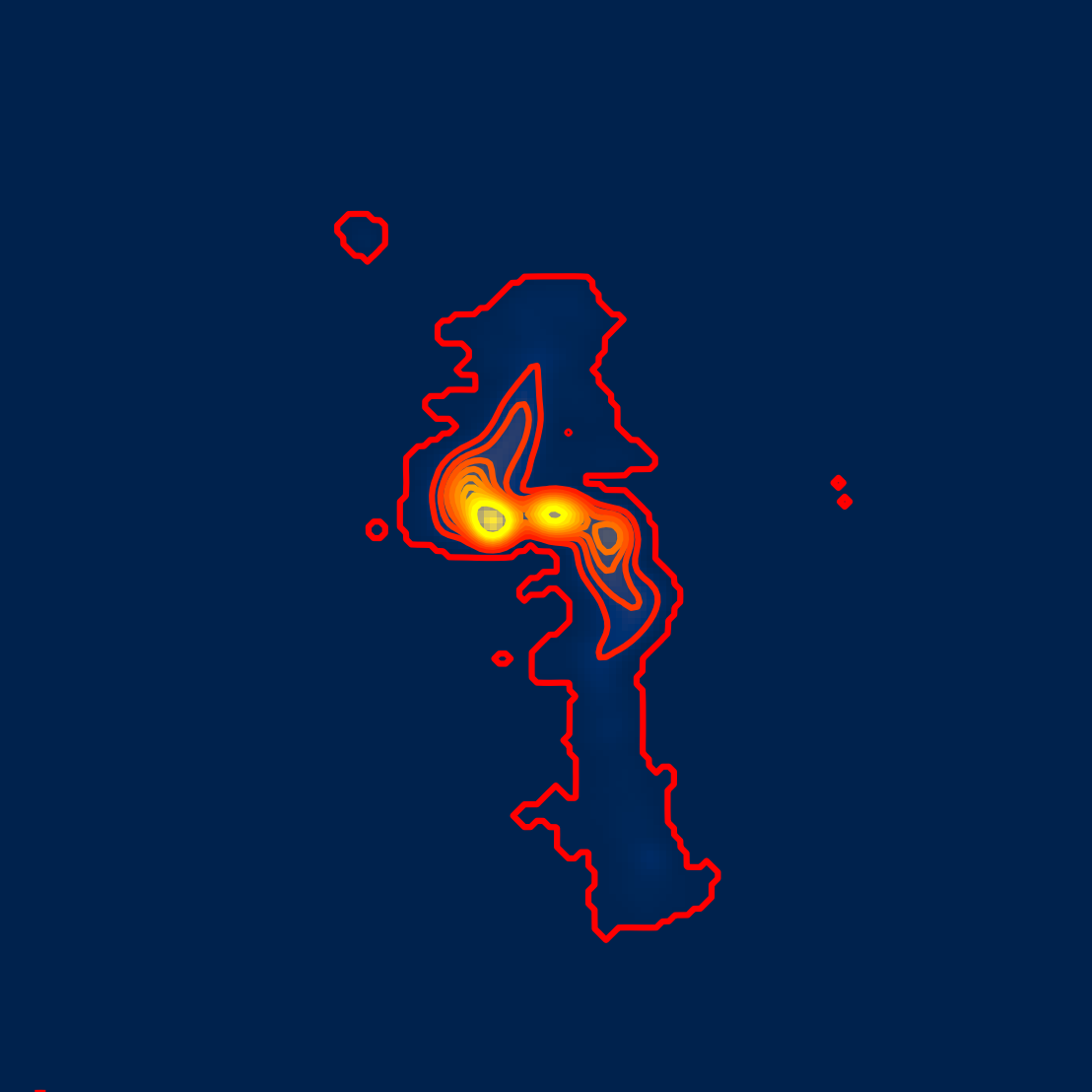}
            & \includegraphics[width = 0.18\textwidth]{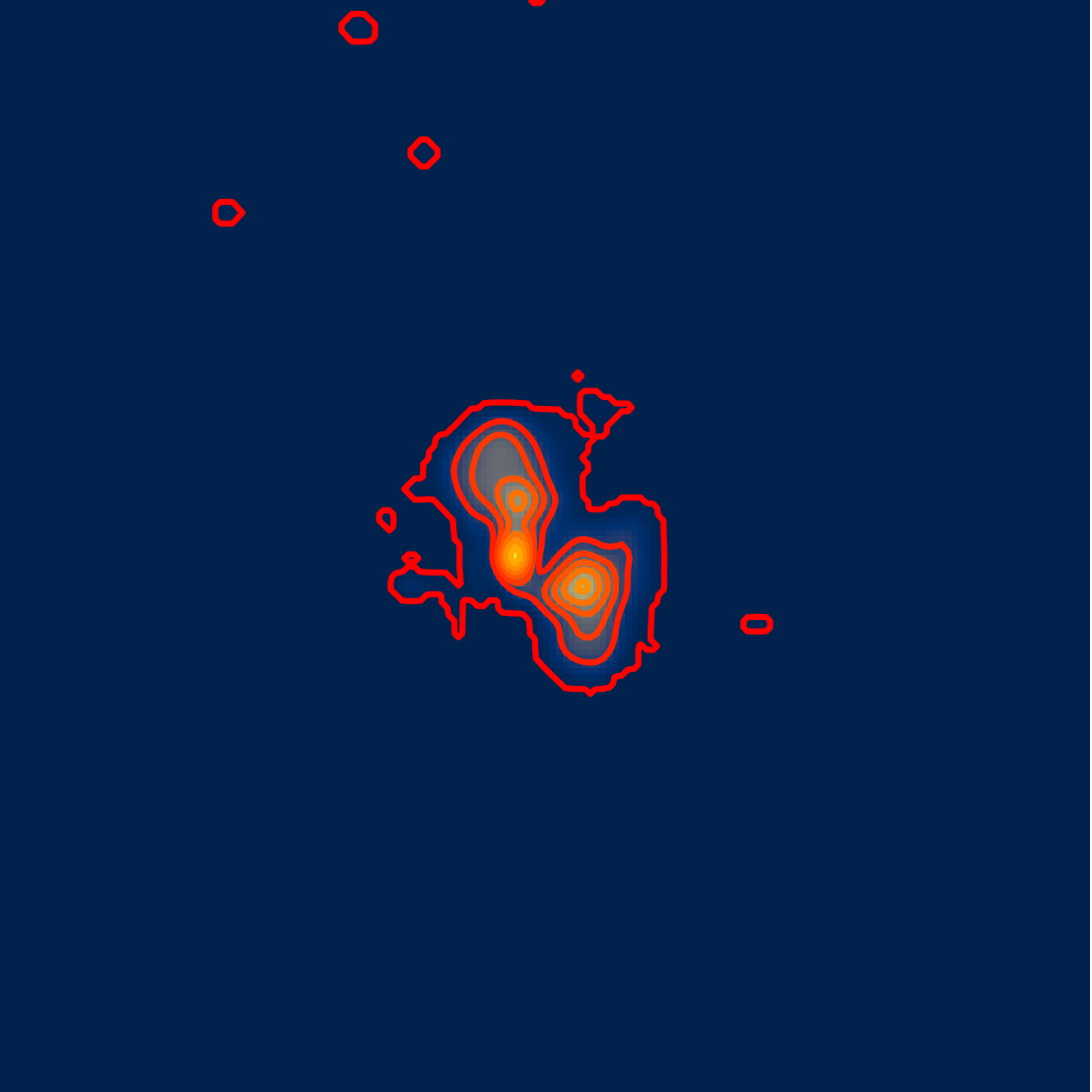}
            & \includegraphics[width = 0.18\textwidth]{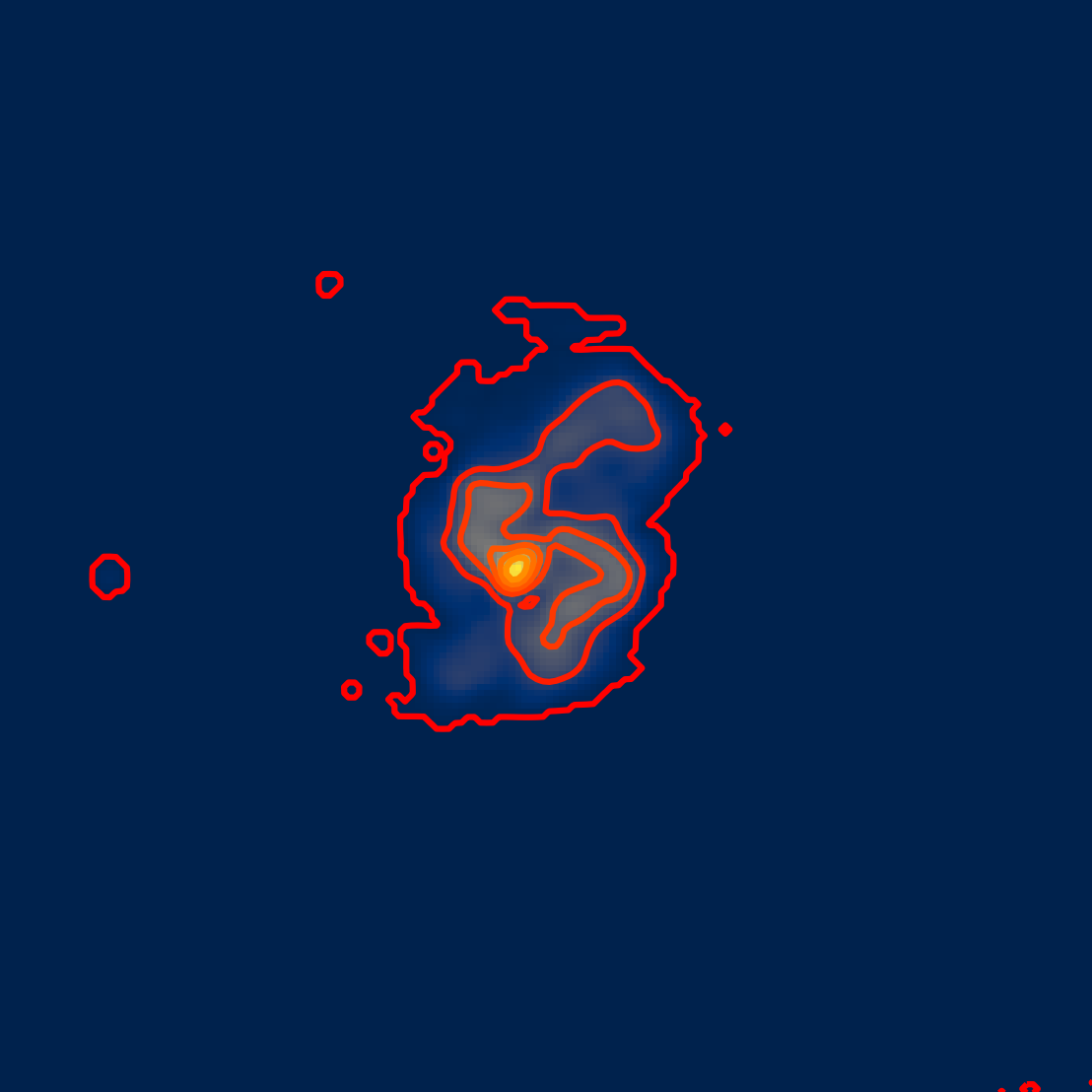}
            & \includegraphics[width = 0.18\textwidth]{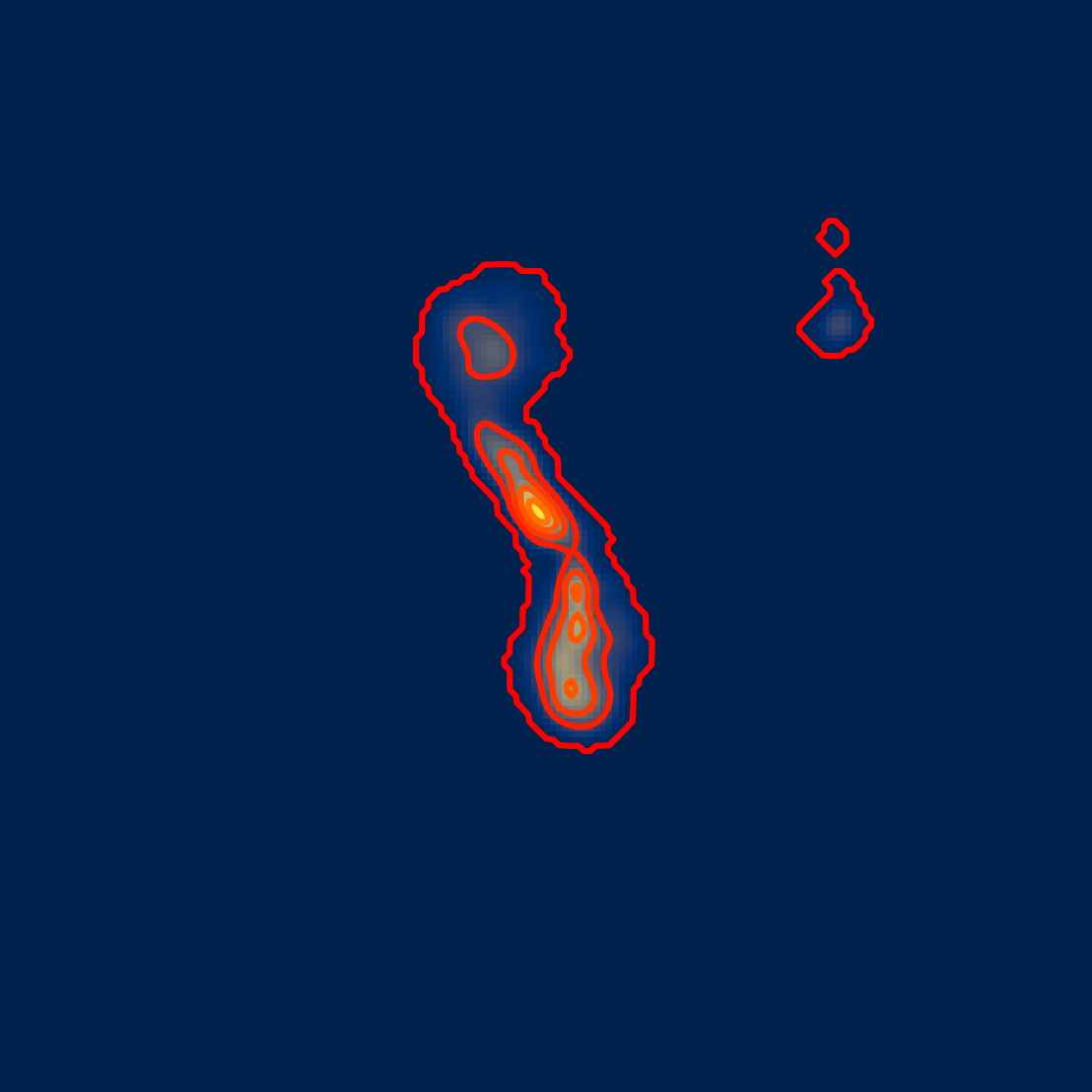}
            & \includegraphics[width = 0.18\textwidth]{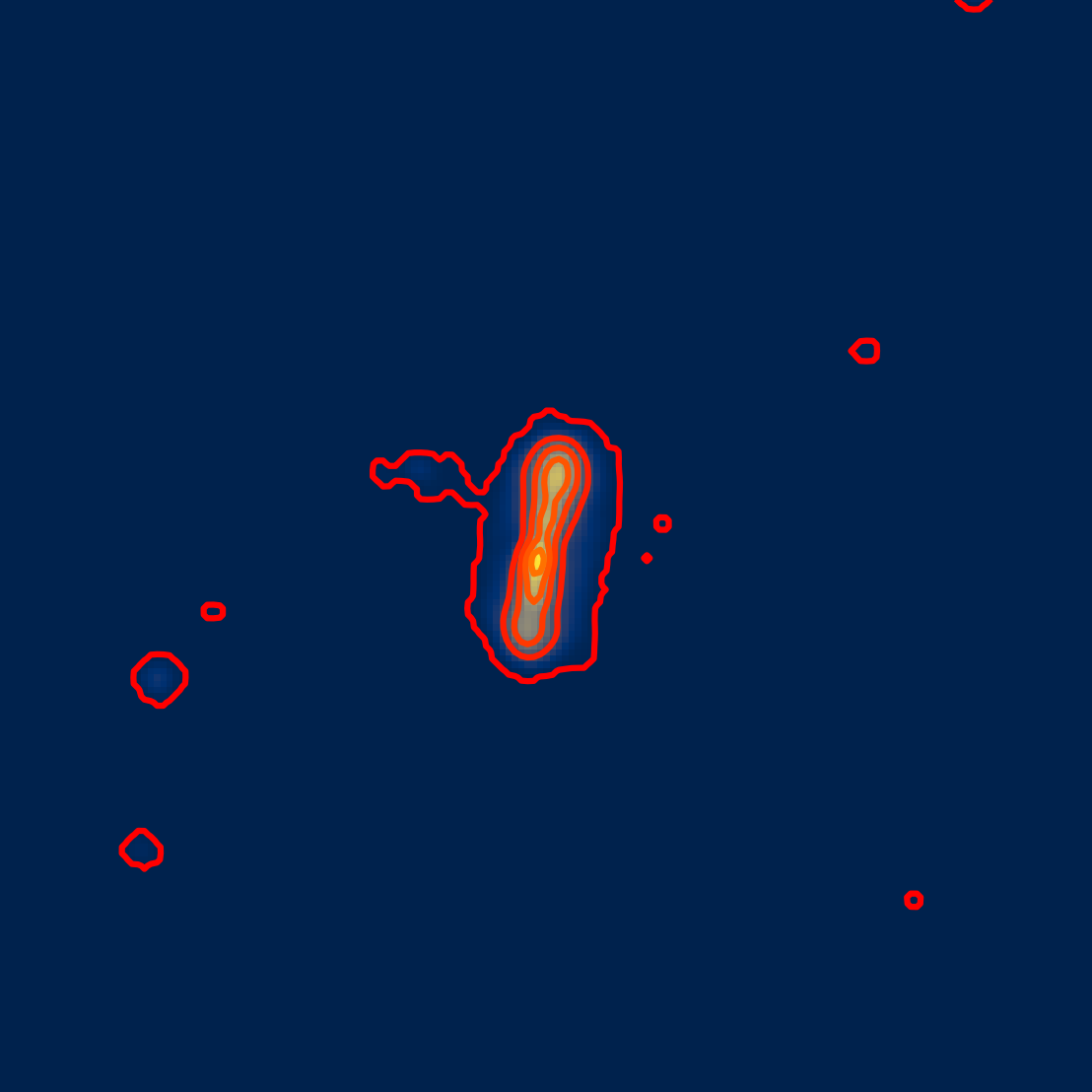}
            \\ 
            \rotatebox{90}{FIRST}
            & \includegraphics[width = 0.18\textwidth]{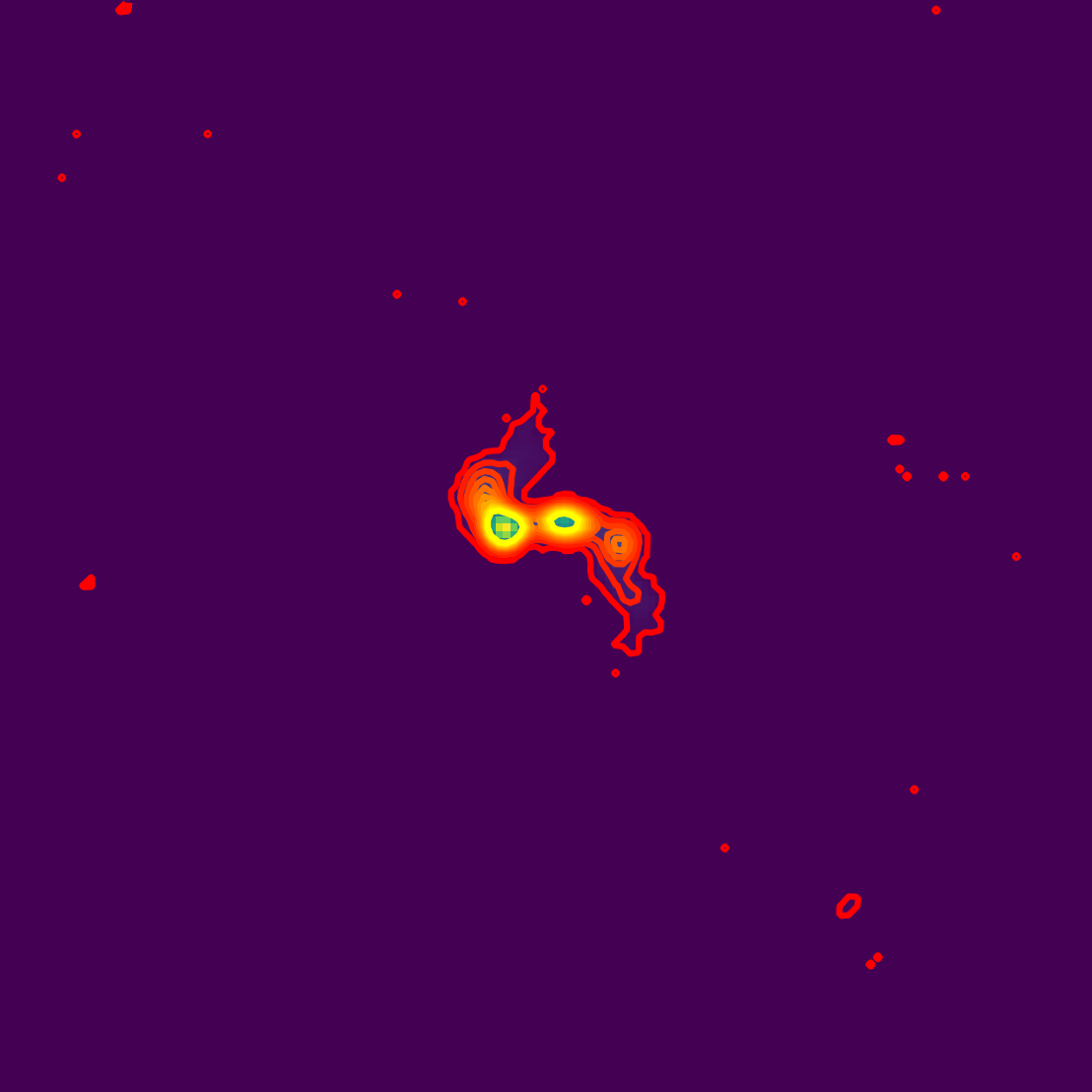}
            & \includegraphics[width = 0.18\textwidth]{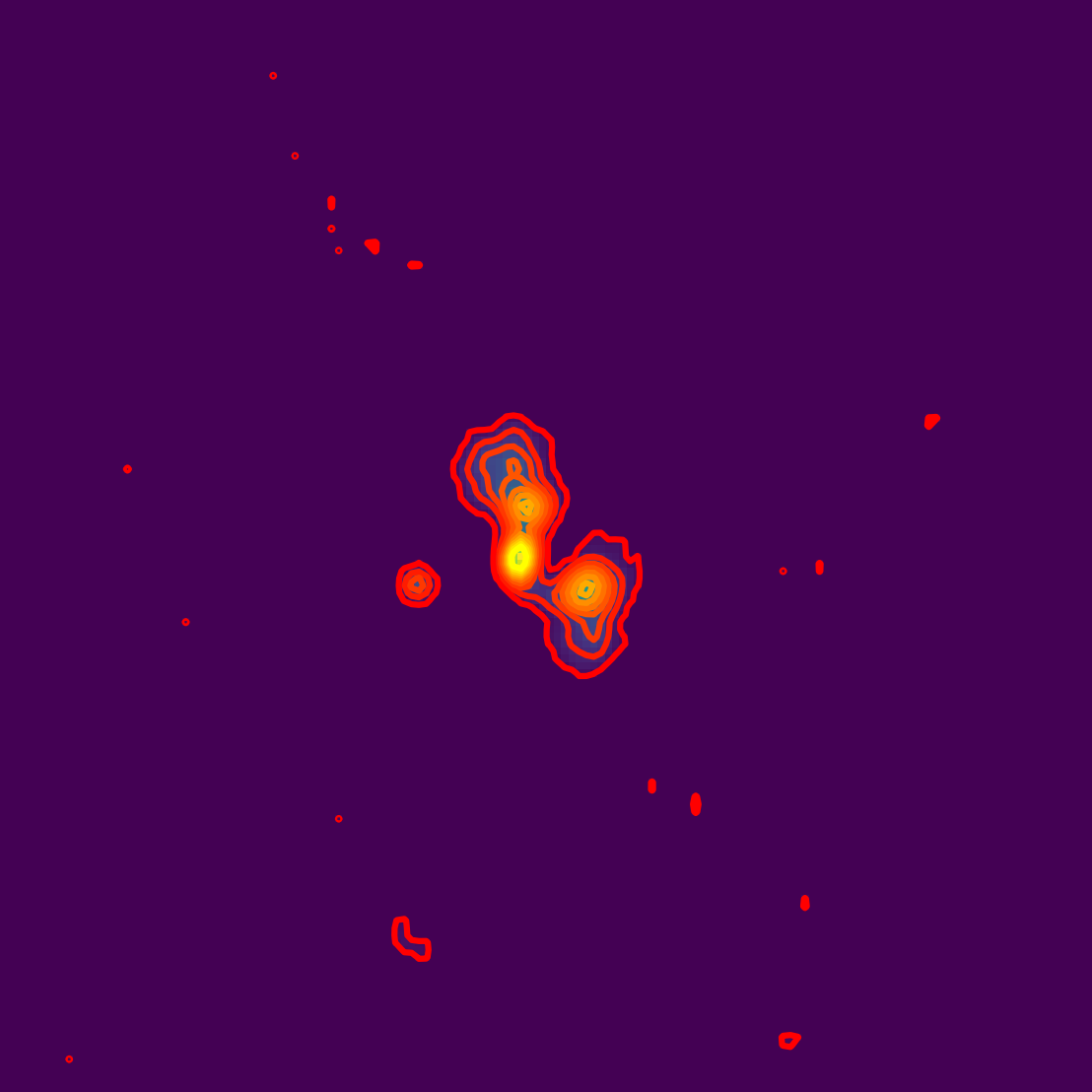}
            & \includegraphics[width = 0.18\textwidth]{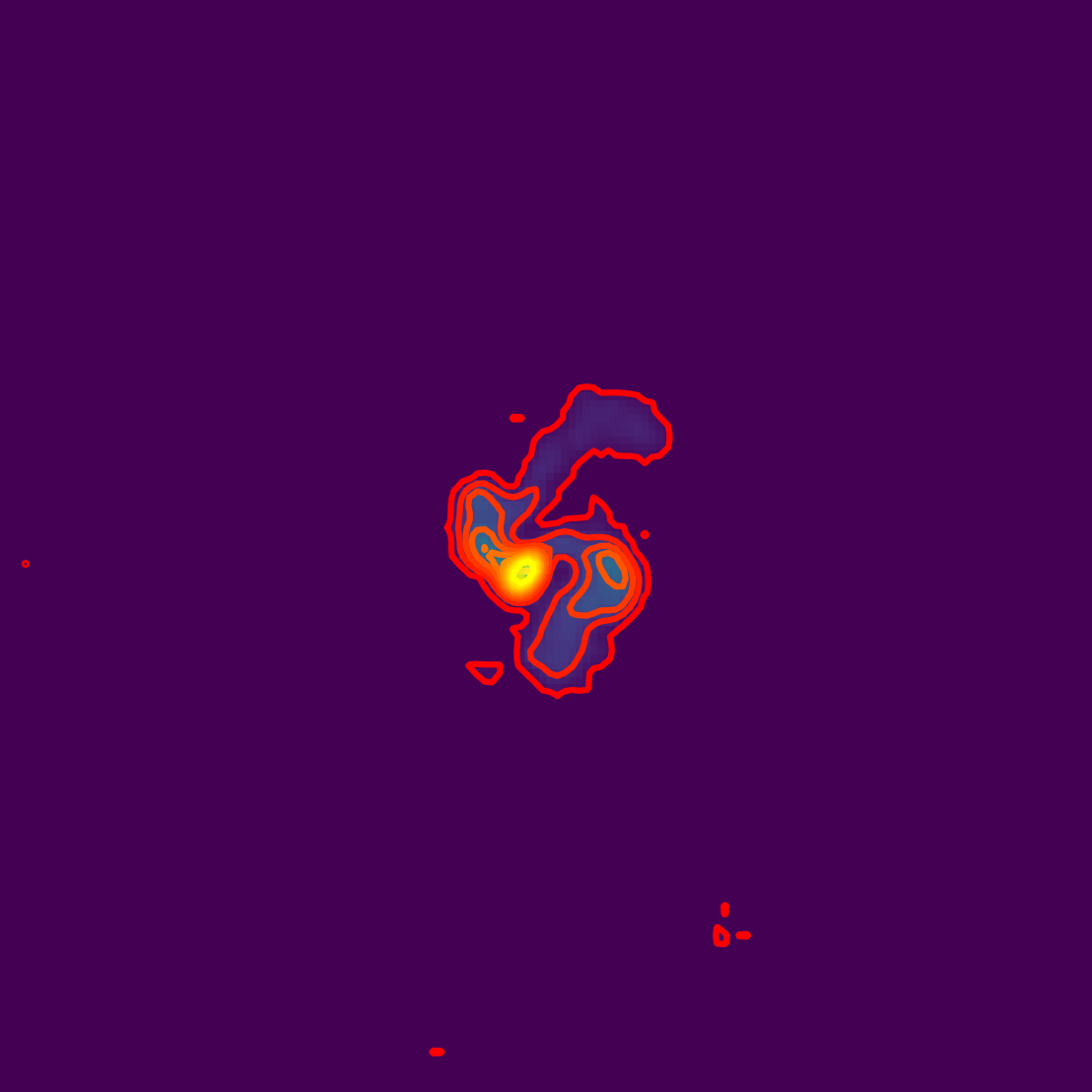}
            & \includegraphics[width = 0.18\textwidth]{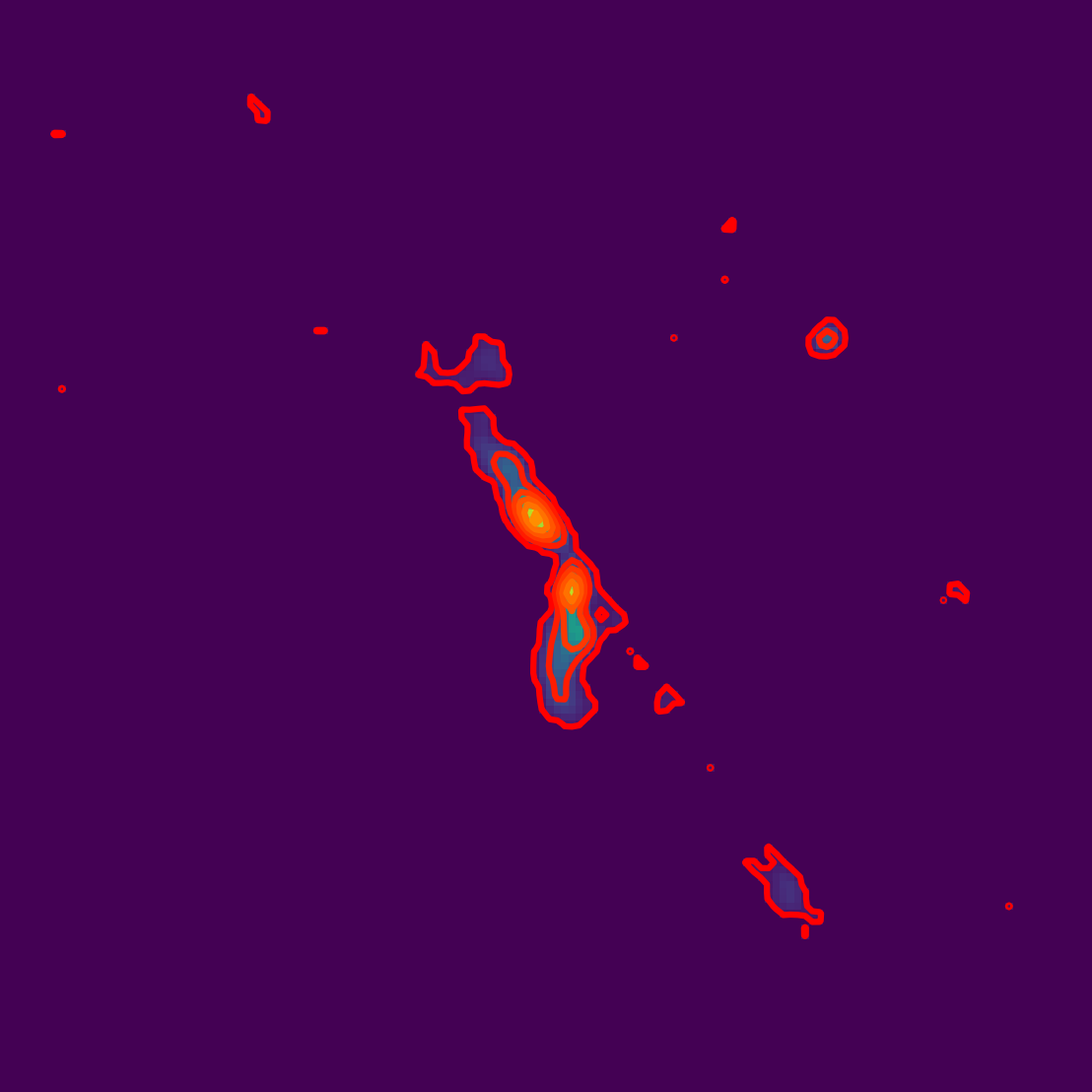}
            & \includegraphics[width = 0.18\textwidth]{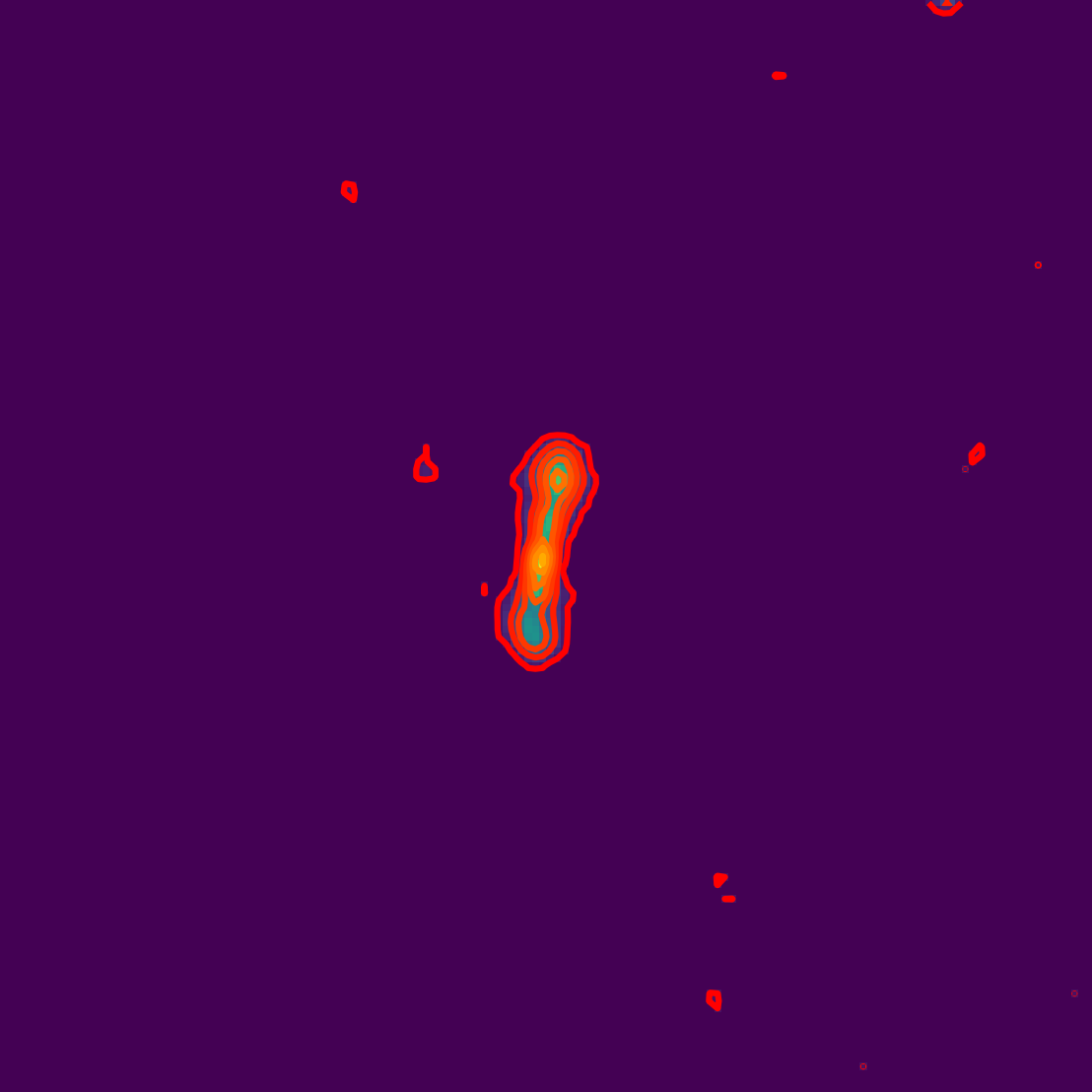}     
        \end{tabular}
        \caption{Examples of class-matched FRIs. The 4.5-arcminute radio images were obtained from LoTSS (upper) and VLA FIRST (lower) overlaid by 10 levels of contour maps spanning 3$\sigma$ to 100$\sigma$ for FIRST images and 3$\sigma$ to 1000$\sigma$ for LoTSS images.}
        \label{fig:FR1}
    \end{figure*}

\begin{figure*}
        \centering
        \setlength{\tabcolsep}{1pt}
        \setlength{\doublerulesep}{1pt}
        \begin{tabular}{p{2mm}ccccc}
             & RGZJ144219.2+504357 & RGZJ133336.2+542749 & RGZJ135142.1+555943 & RGZJ120458.6+455937 & RGZJ112942.1+542529\\
            \rotatebox{90}{LoTSS}
            & \includegraphics[width = 0.18\textwidth]{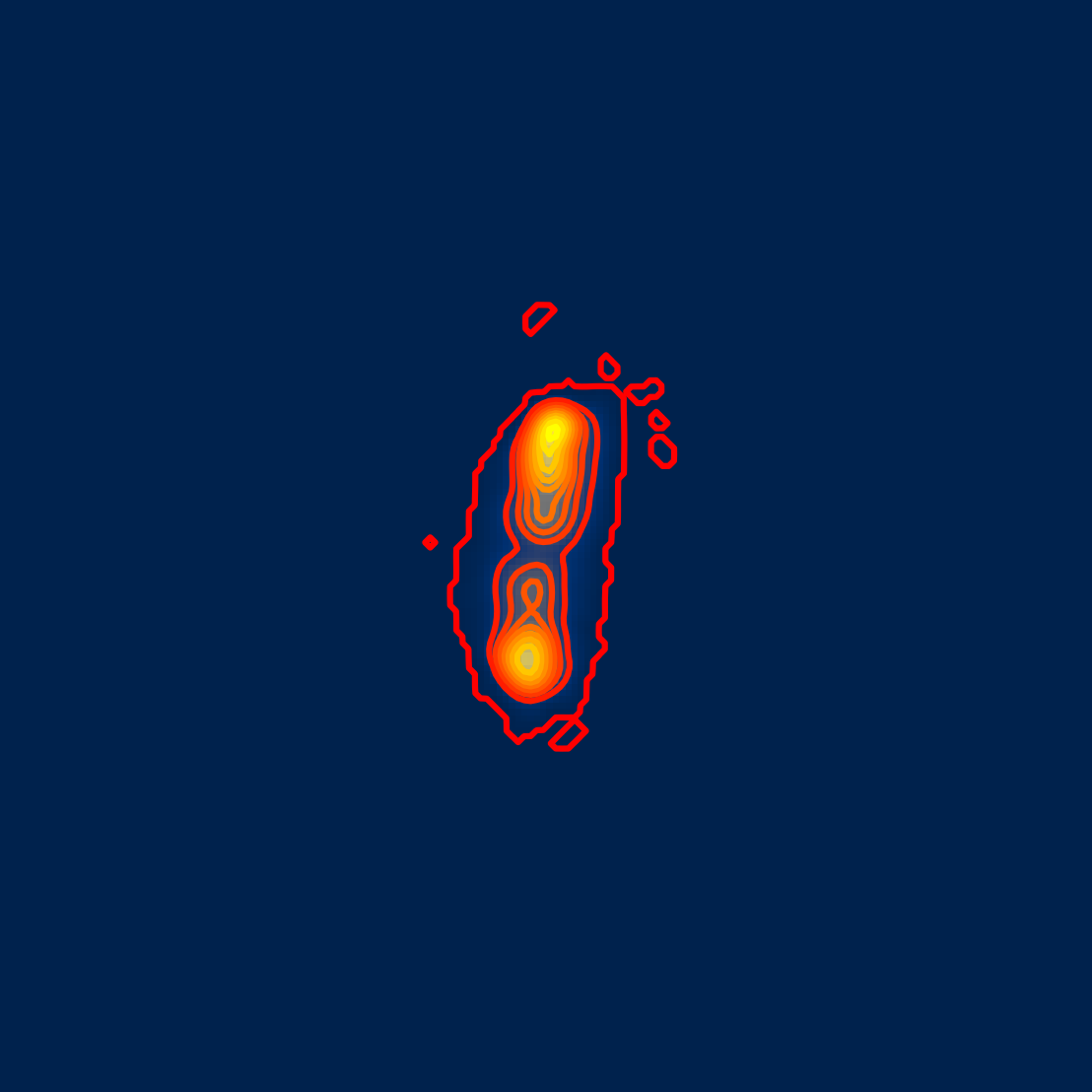}
            & \includegraphics[width = 0.18\textwidth]{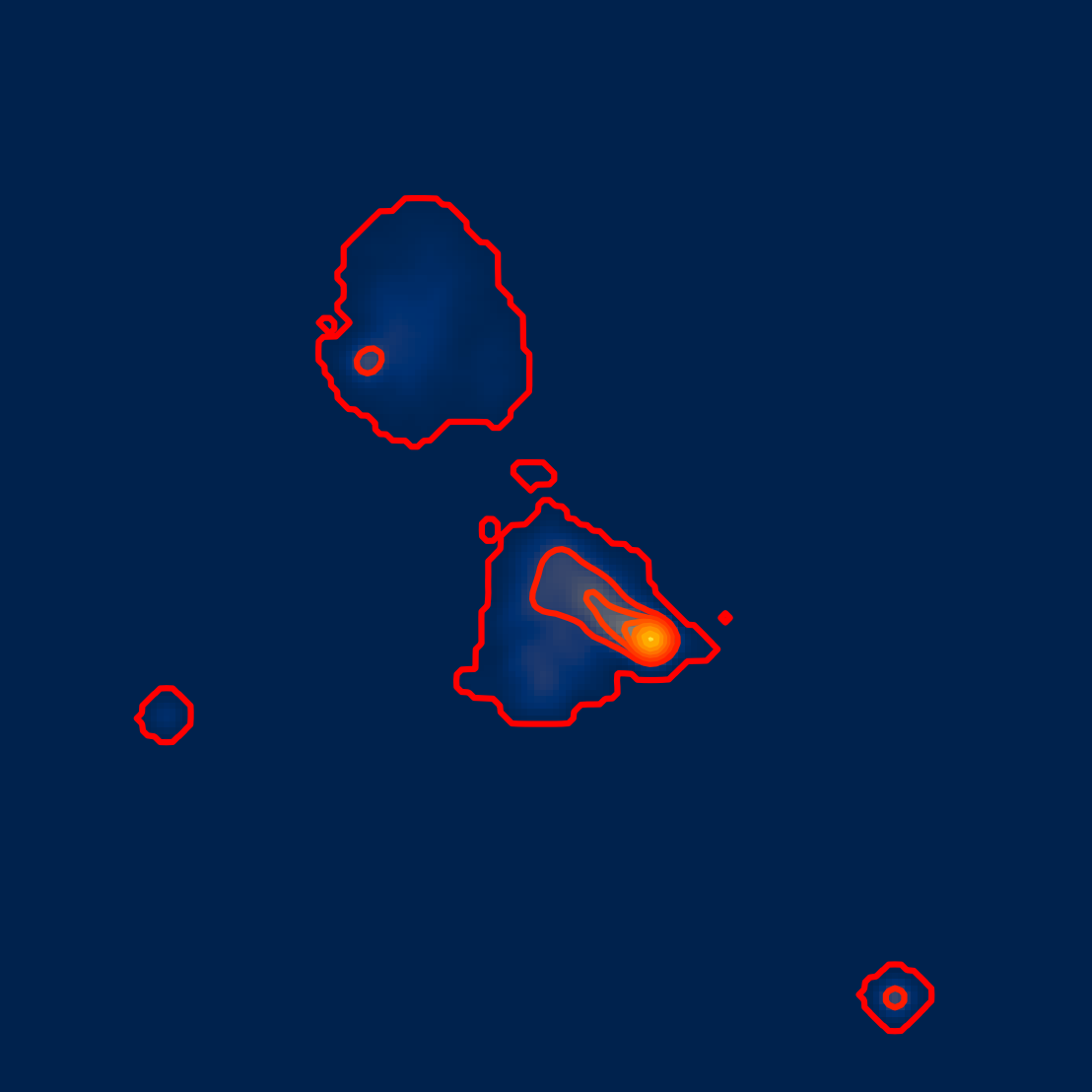}
            & \includegraphics[width = 0.18\textwidth]{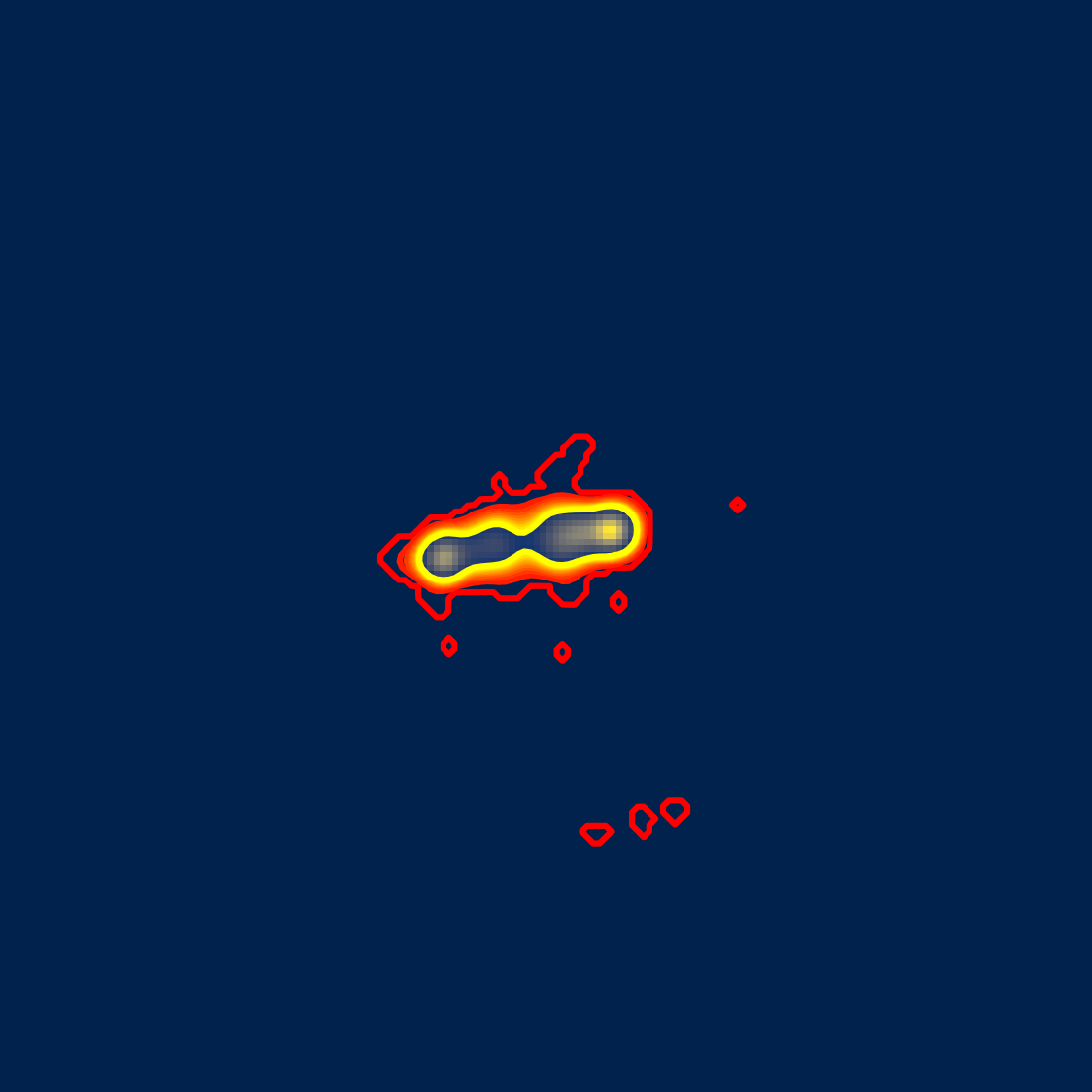}
            & \includegraphics[width = 0.18\textwidth]{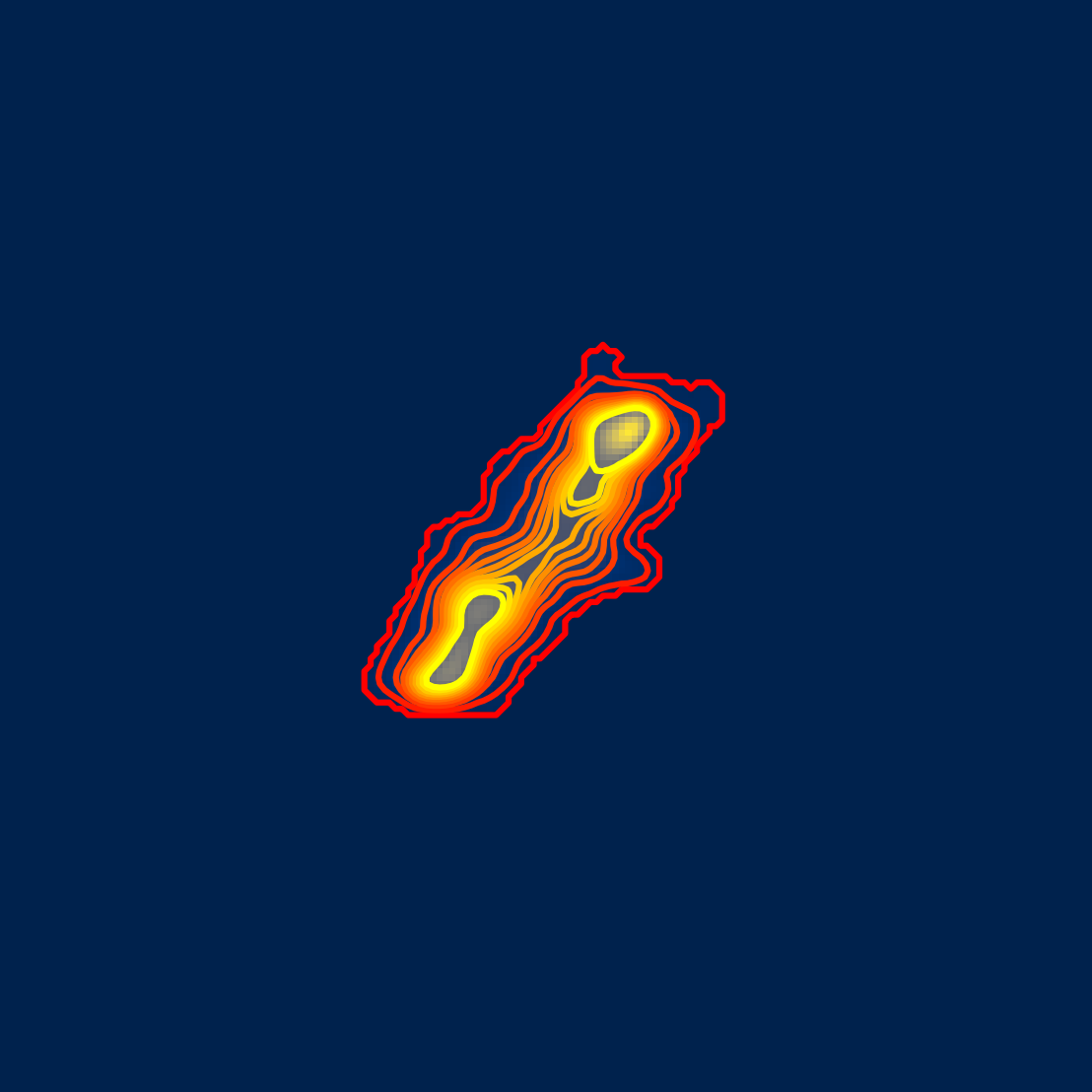}
            & \includegraphics[width = 0.18\textwidth]{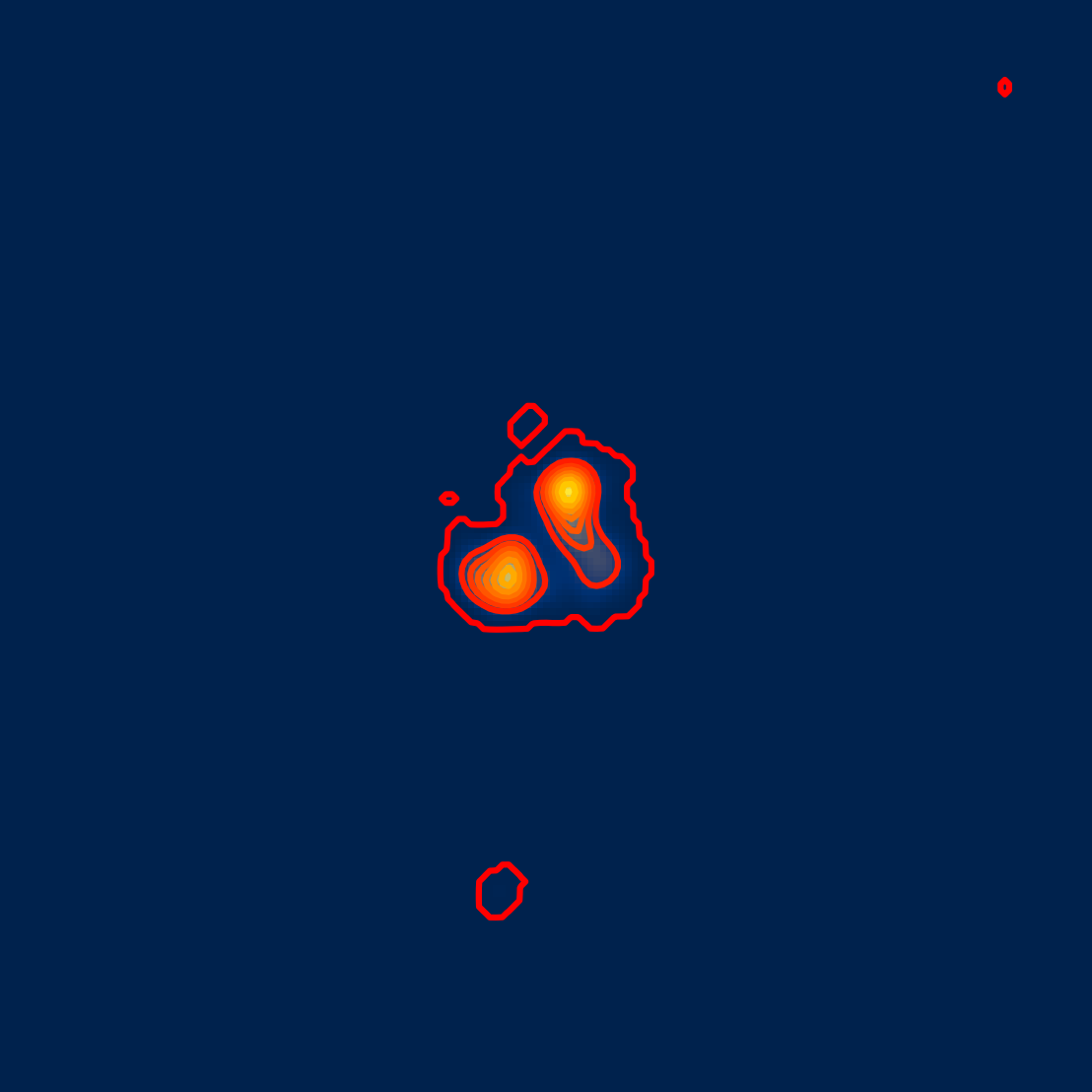}
            \\ 
            \rotatebox{90}{FIRST}
            & \includegraphics[width = 0.18\textwidth]{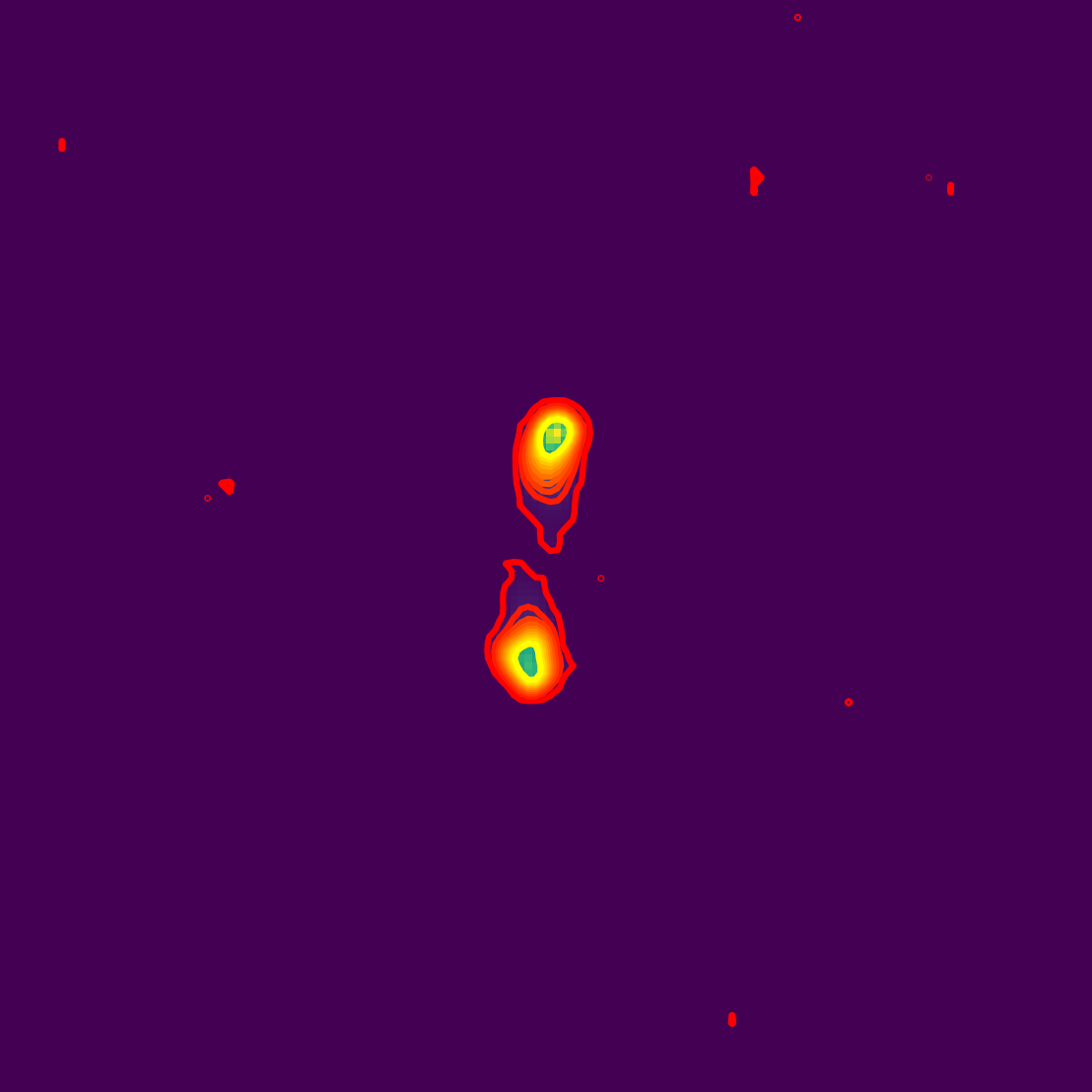}
            & \includegraphics[width = 0.18\textwidth]{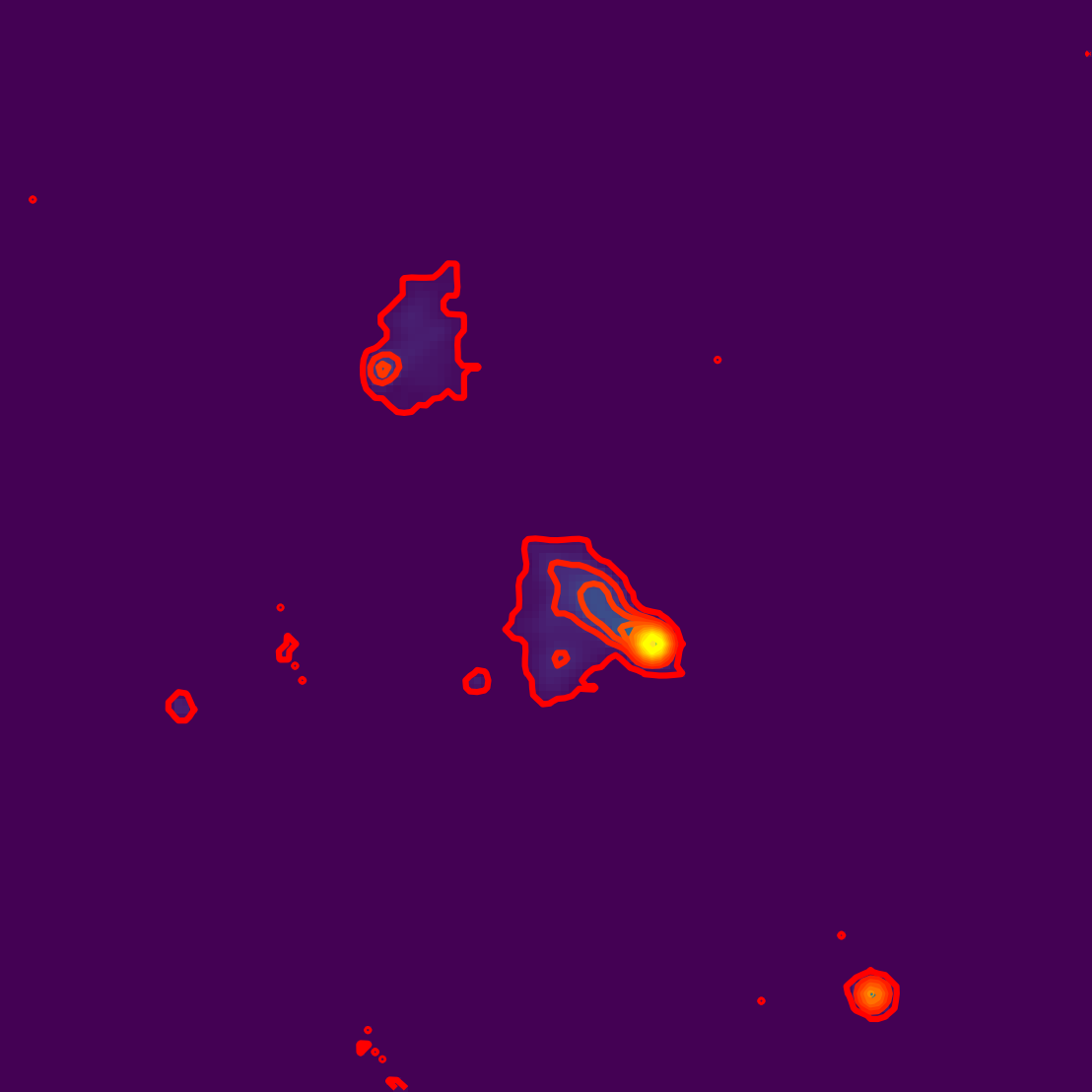}
            & \includegraphics[width = 0.18\textwidth]{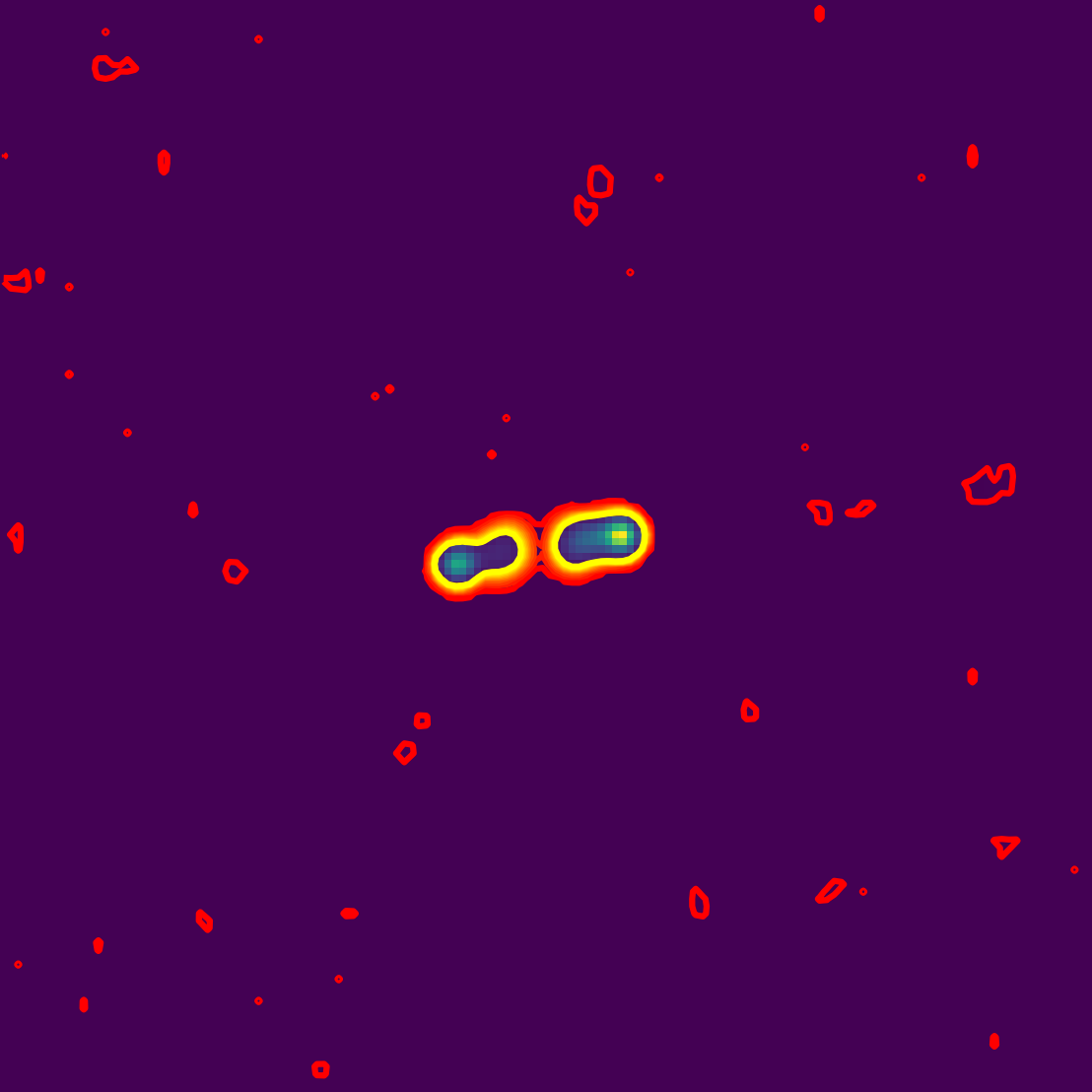}
            & \includegraphics[width = 0.18\textwidth]{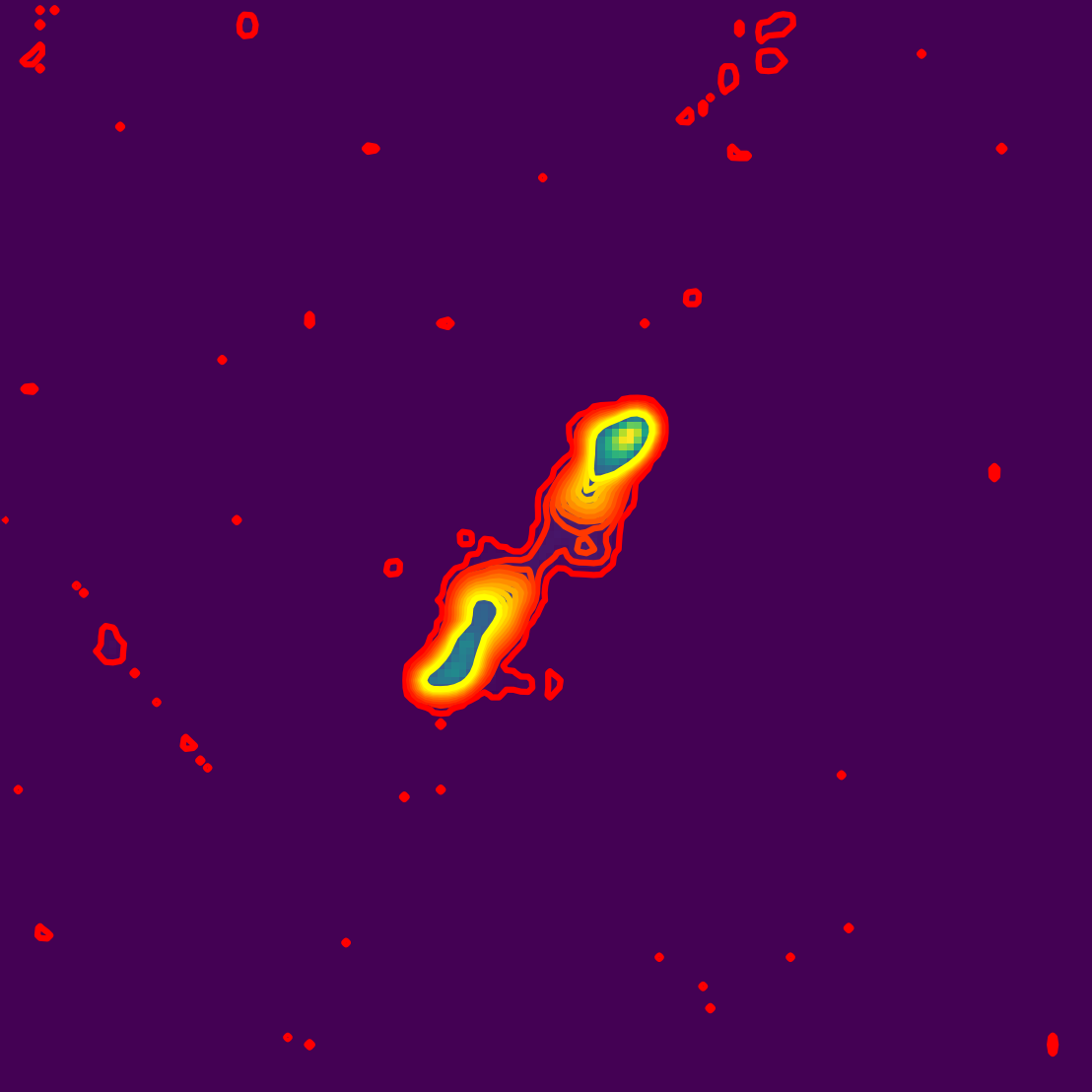}
            & \includegraphics[width = 0.18\textwidth]{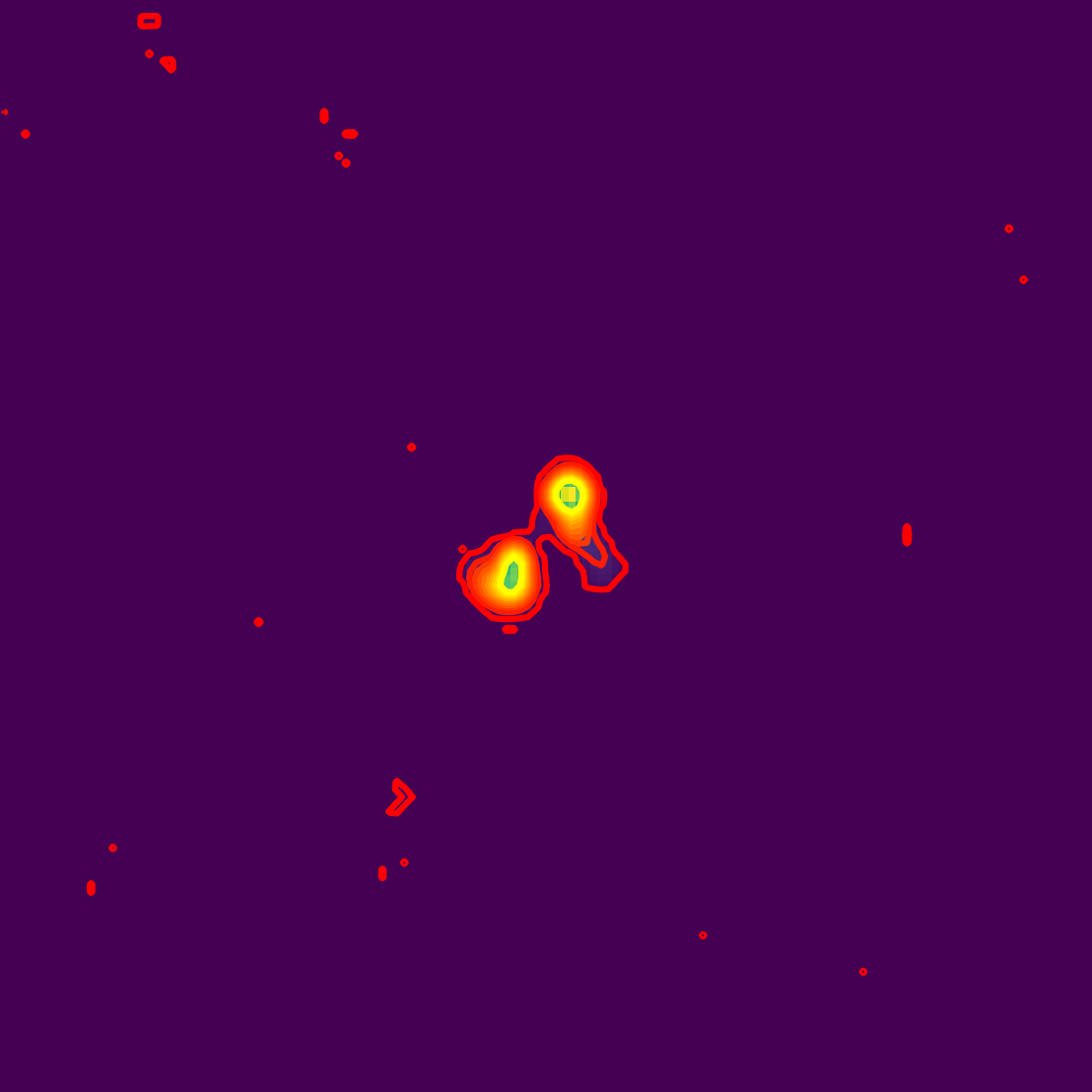}     
        \end{tabular}
        \caption{Examples of class-matched FRIIs. The 4.5-arcminute radio images were obtained from LoTSS and VLA FIRST with 10 levels of contour map spanning 3$\sigma$ to 100$\sigma$ for FIRST images and 3$\sigma$ to 2000$\sigma$ for LoTSS images.}
        \label{fig:FR2}
    \end{figure*}

Figures \ref{fig:FR1} and \ref{fig:FR2} depict images of class-matched RGZ-LoTSS FRIs and FRIIs, respectively. The 4.5-arcminute images were acquired from LoTSS at 150\,MHz (top) and VLA FIRST at 1.4\,GHz (bottom).
For LoTSS images, the contours were applied at a level of 3 to 1,000 times the rms noise level for FRI images and 3 to 2,000 times for FRII images.
For FIRST images, the contours were applied at 10 levels ranging from 3 to 100 times the typical rms noise level of FIRST.
The images show the structure of the radio sources in each FR class: FRIs typically have jets emanating from the core with the brightest areas close to the central regions, whereas FRIIs tend to have separate bright lobes and faint central regions.
It can be seen from Figures \ref{fig:FR1} and \ref{fig:FR2} that LoTSS images illustrated greater details in the surrounding area of the sources. 

Overall, the total RGZ-LoTSS sources show a strong correlation with the trends found in the RGZ\,FR data.
This includes the overlap of FRIs and FRIIs as well as the host galaxy types that were roughly predicted by WISE colour analysis. Additionally, we find a similarity in spectral indices, especially between FRI and FRII-Low objects. However, we also identify a significant number of class-changed sources. These objects are discussed further in this section.

\begin{table}
\centering
\caption{RGZ-LoTSS cross-matching sources}
\label{tab:cross-match}
\begin{tabular}{|cc|ccccc|}
\hline
\multicolumn{2}{|c|}{\multirow{2}{*}{\textbf{RGZ}}} & \multicolumn{5}{c|}{\textbf{LoTSS}}   \\ 
\multicolumn{2}{|c|}{}             & FRI & FRII & Ind & Small & D-D \\ \hline
\multicolumn{1}{|r|}{FRI}          & 225   & 154 & 6    & 55  & 7     & 3   \\
\multicolumn{1}{|r|}{FRII-Low}     & 158   & 83 & 20   & 44  & 11    & 0   \\
\multicolumn{1}{|r|}{FRII-High}    & 130   & 60  & 29   & 24  & 17    & 0   \\ \hline
\multicolumn{1}{|r|}{Total}        & 513   & 297 & 55   & 123 & 35    & 3   \\ \hline
\end{tabular}
\end{table}


\subsection{Luminosity-size distribution for RGZ-LoTSS sample}
\label{sec:match_lumdis}
\begin{figure*}
\centerline{
\includegraphics[width=0.48\textwidth]{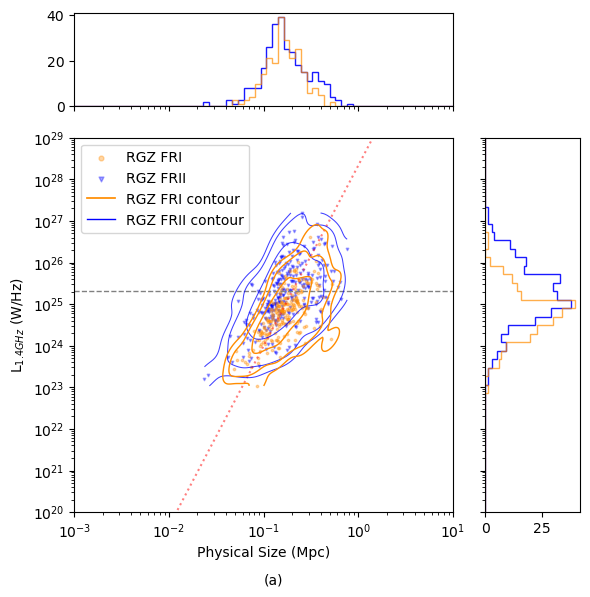}
\includegraphics[width=0.48\textwidth]{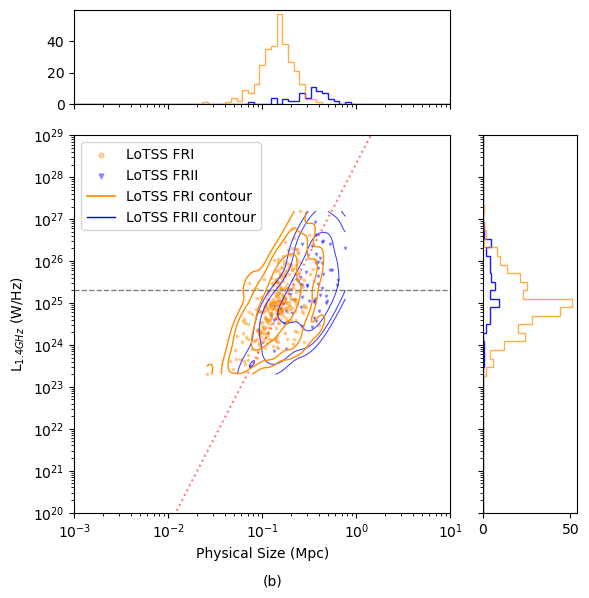}
}
\caption{Luminosity (W/Hz) at 1.4\,GHz versus physical size (Mpc) of 514 RGZ-LoTSS sources. (a) is labelled by the RGZ\,FR catalogue and (b) is labelled by the FR LoTSS catalogue \citep{Mingo2019}. The horizontal dashed grey line represents the historical FRI/II break line ($\sim2.0\times10^{25}$ W/Hz). The RGZ\,FR FRII distribution covers the physical size range from small to large, while LoTSS FRII objects mostly populate the right region of the distribution, roughly outlined by the dotted red line defined by Equation \ref{eq:log_class-changed}.}
\label{fig:lum_ps_rgz-lm}
\end{figure*}


Figure~\ref{fig:lum_ps_rgz-lm} illustrates the distribution of RGZ-LoTSS sources based on luminosity at 1.4\,GHz and physical size.  Considering the RGZ\,FR label, the distribution of the RGZ-LoTSS sample is consistent with the overall trends observed in the RGZ\,DR1 dataset, as shown in Figure~\ref{fig:lum_physize}. By comparing RGZ-LoTSS with all RGZ\,DR1 sources using the Kolmogorov-Smirnov test \citep{Massey1951}, we obtain p-values of 0.11 for luminosity and 0.91 for physical size, indicating that both values exceed the normal statistical significance threshold of 0.05 that the null hypothesis can be rejected. 
On the other hand, LoTSS FRIIs are mostly located in the right region of the distribution, which we segment empirically using the boundary indicated as a red dotted line in Figure~\ref{fig:lum_ps_rgz-lm},
%
%
\begin{equation}
\log_{10} [L_{1.4} / \text{Watts\,Hz$^{-1}$}] = 4.34 \log_{10} [d / {\rm Mpc}] + 28.30
\label{eq:log_class-changed}
\end{equation}
where $L_{1.4}$ represents luminosity at 1.4\,GHz and $d$ denotes the physical size in Mpc.
Furthermore, LoTSS FRIIs tend to populate above the breakline, accounting for 52\% of total LoTSS FRIIs, in contrast to LoTSS FRIs with 27\% of total LoTSS FRIs.
This indicates that LoTSS FRIIs tend to have higher luminosity and are larger in physical size compared to LoTSS FRIs, and LoTSS FRIIs are also distinct from RGZ\,FR FRIIs. Nonetheless,  Figure~\ref{fig:lum_ps_rgz-lm} shows the overlap between these two classes, similar to that shown in Figure~\ref{fig:lum_physize}.

    \begin{figure*}
        \centering
        \setlength{\tabcolsep}{1pt}
        \setlength{\doublerulesep}{1pt}
        \begin{tabular}{p{2mm}ccccc}
             & RGZJ144705.9+492502 & RGZJ142547.3+515451 & RGZJ105923.7+511839 & RGZJ143723.7+473614 & RGZJ120331.6+543009\\
            \rotatebox{90}{LoTSS}
            & \includegraphics[width = 0.18\textwidth]{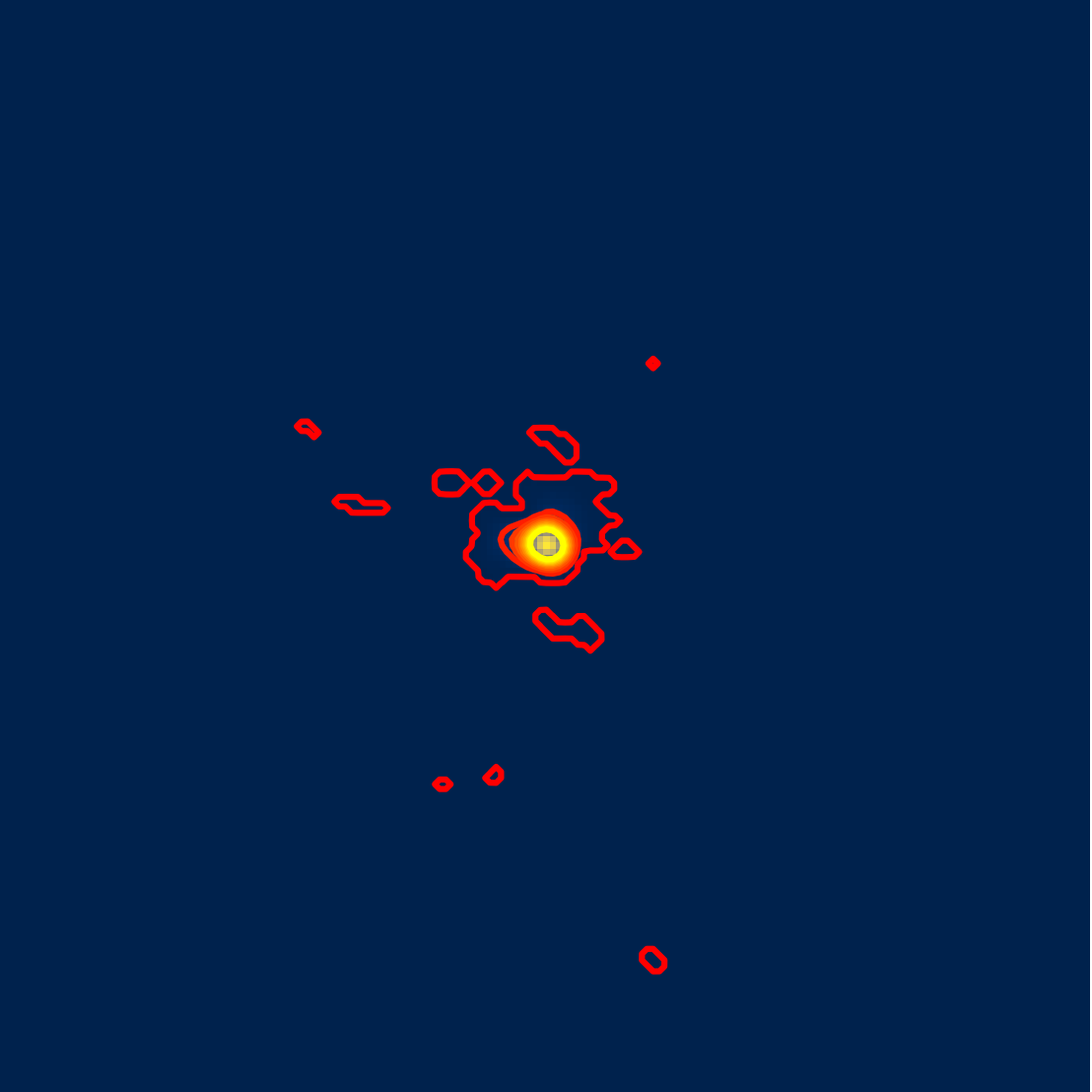}
            & \includegraphics[width = 0.18\textwidth]{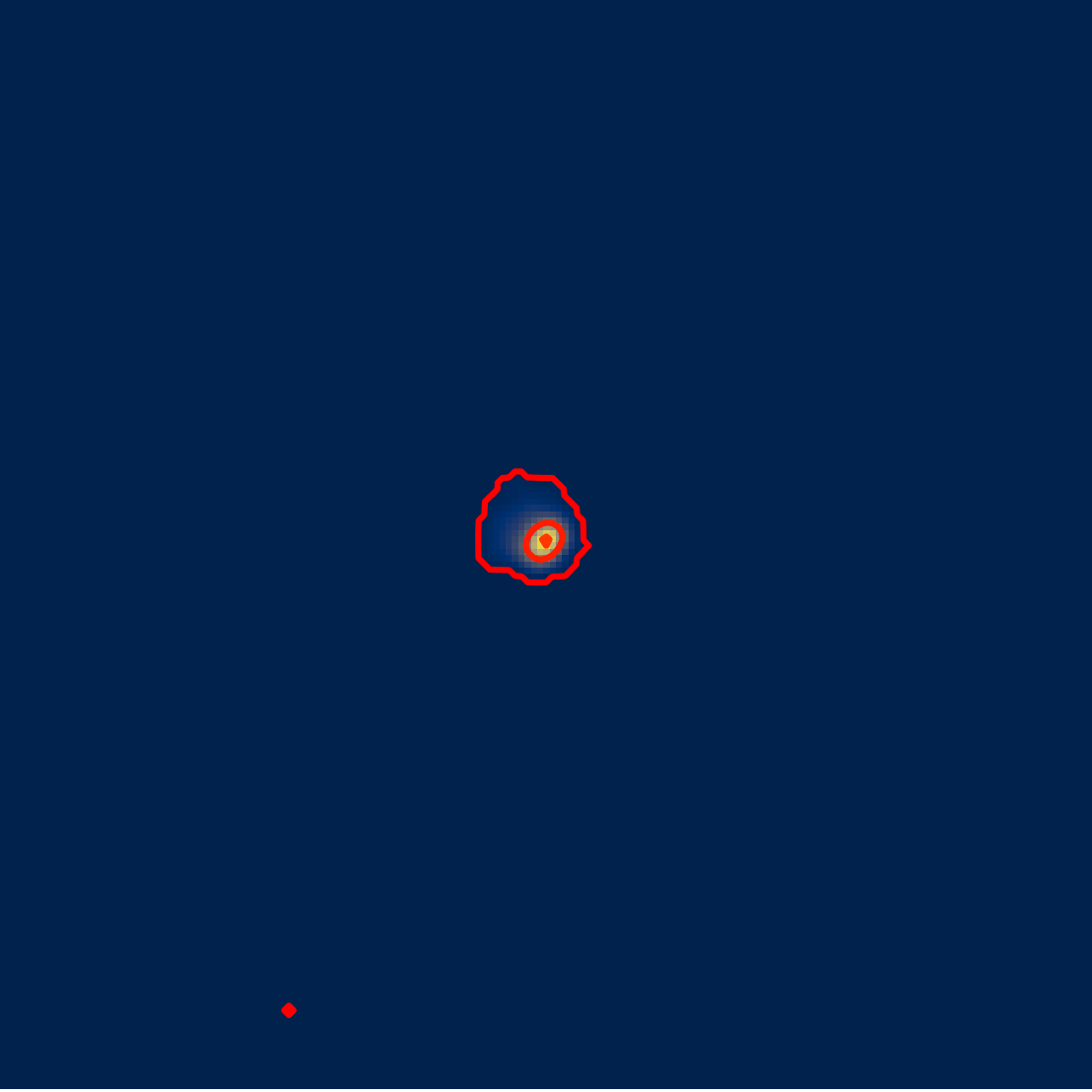}
            & \includegraphics[width = 0.18\textwidth]{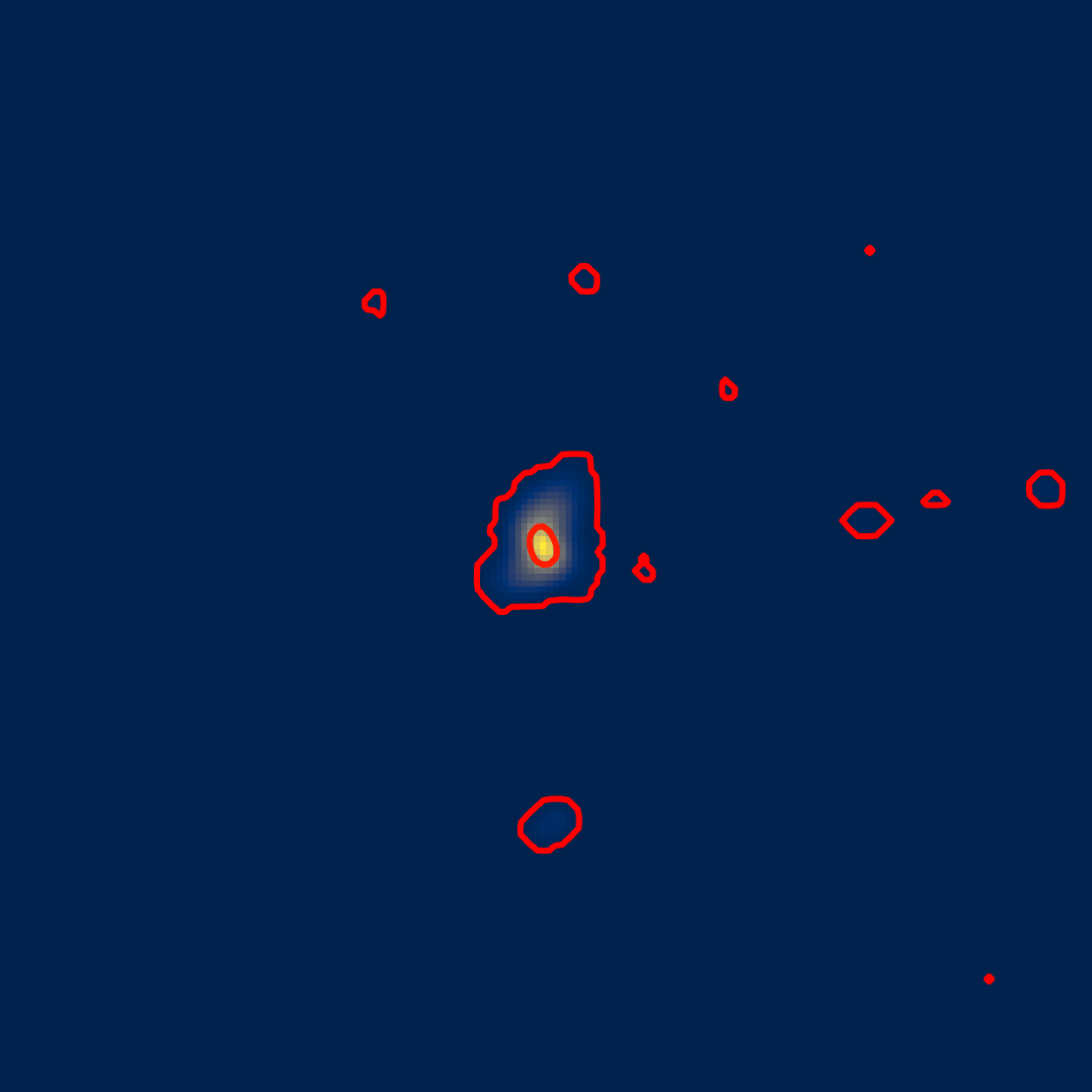}
            & \includegraphics[width = 0.18\textwidth]{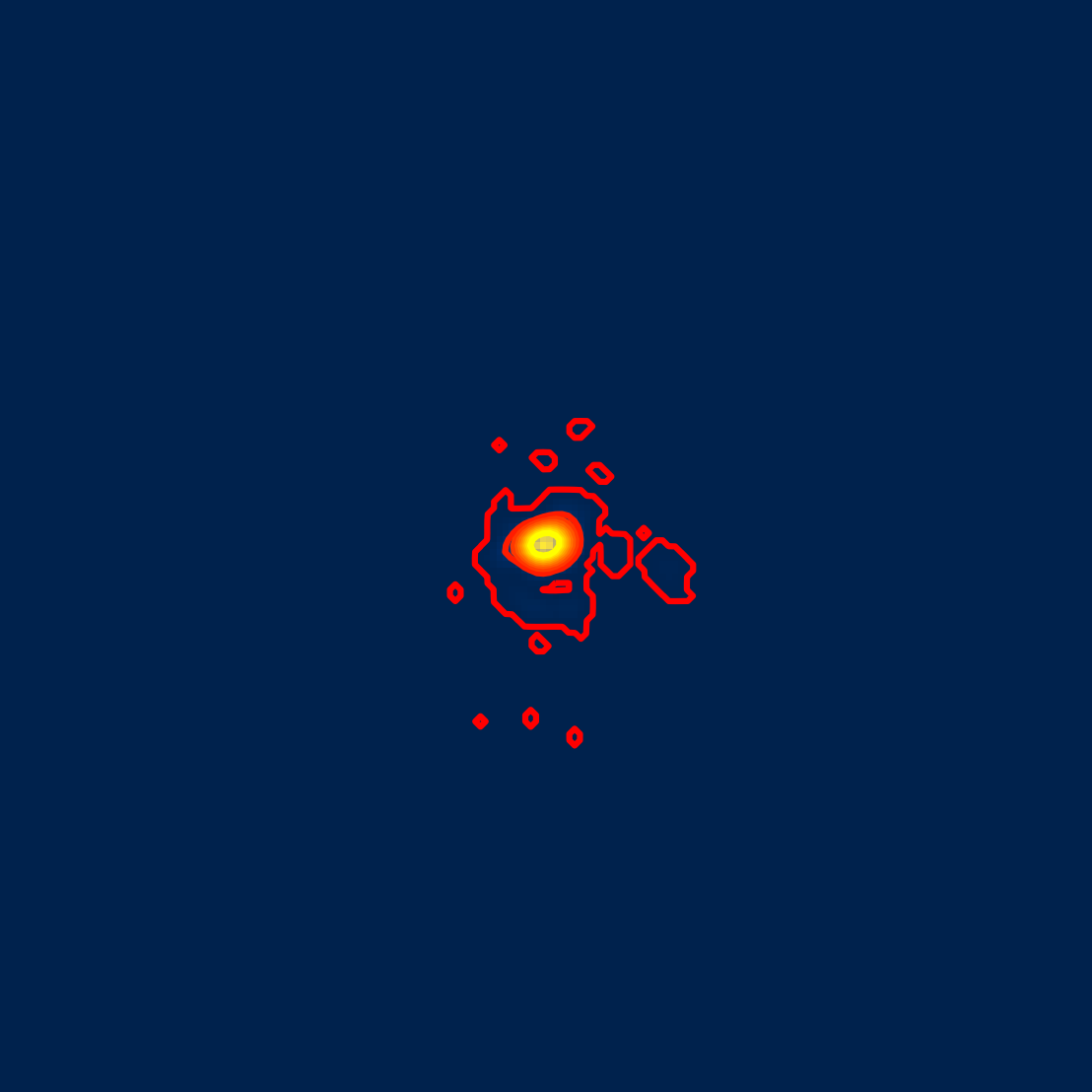}
            & \includegraphics[width = 0.18\textwidth]{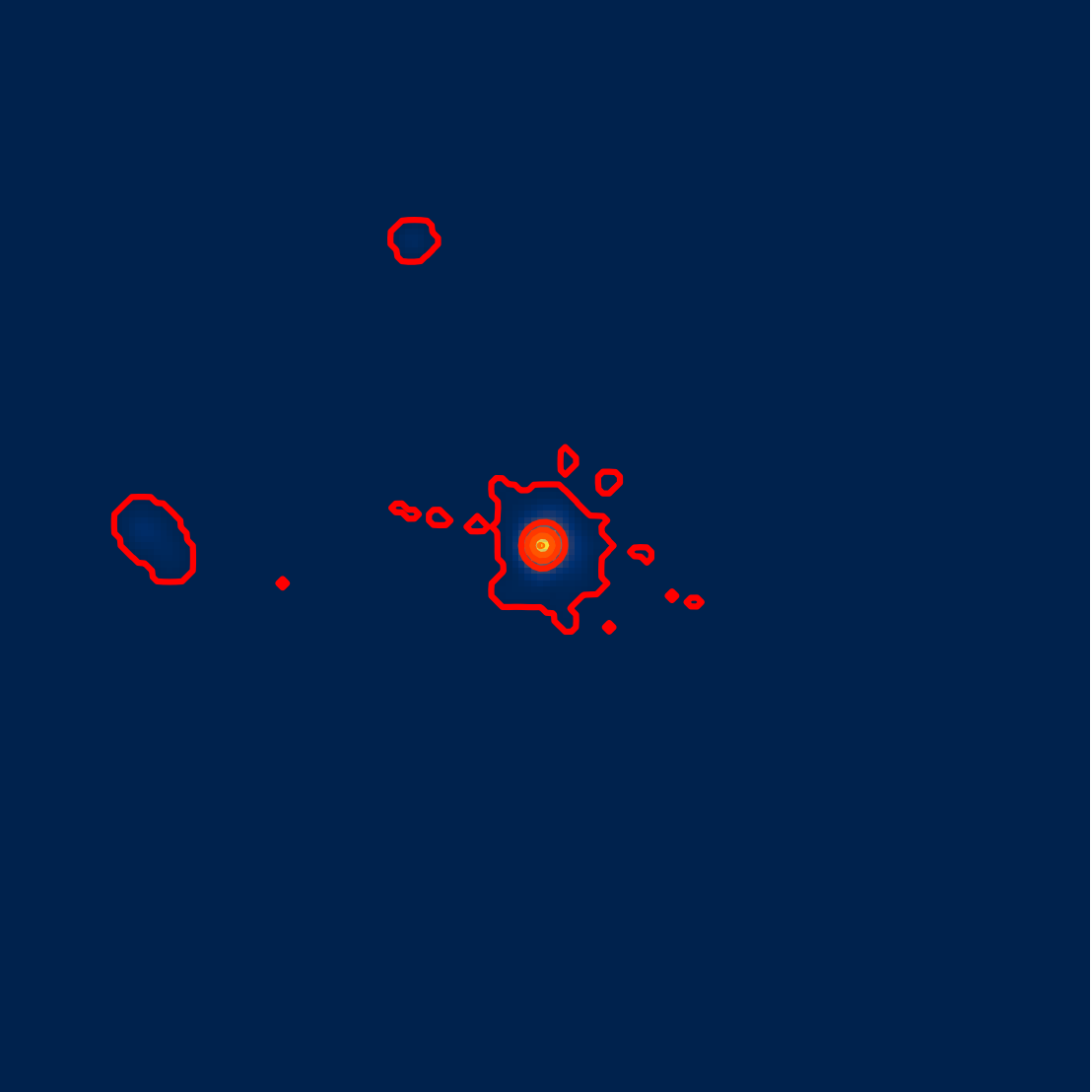}
            \\ 
            \rotatebox{90}{FIRST}
            & \includegraphics[width = 0.18\textwidth]{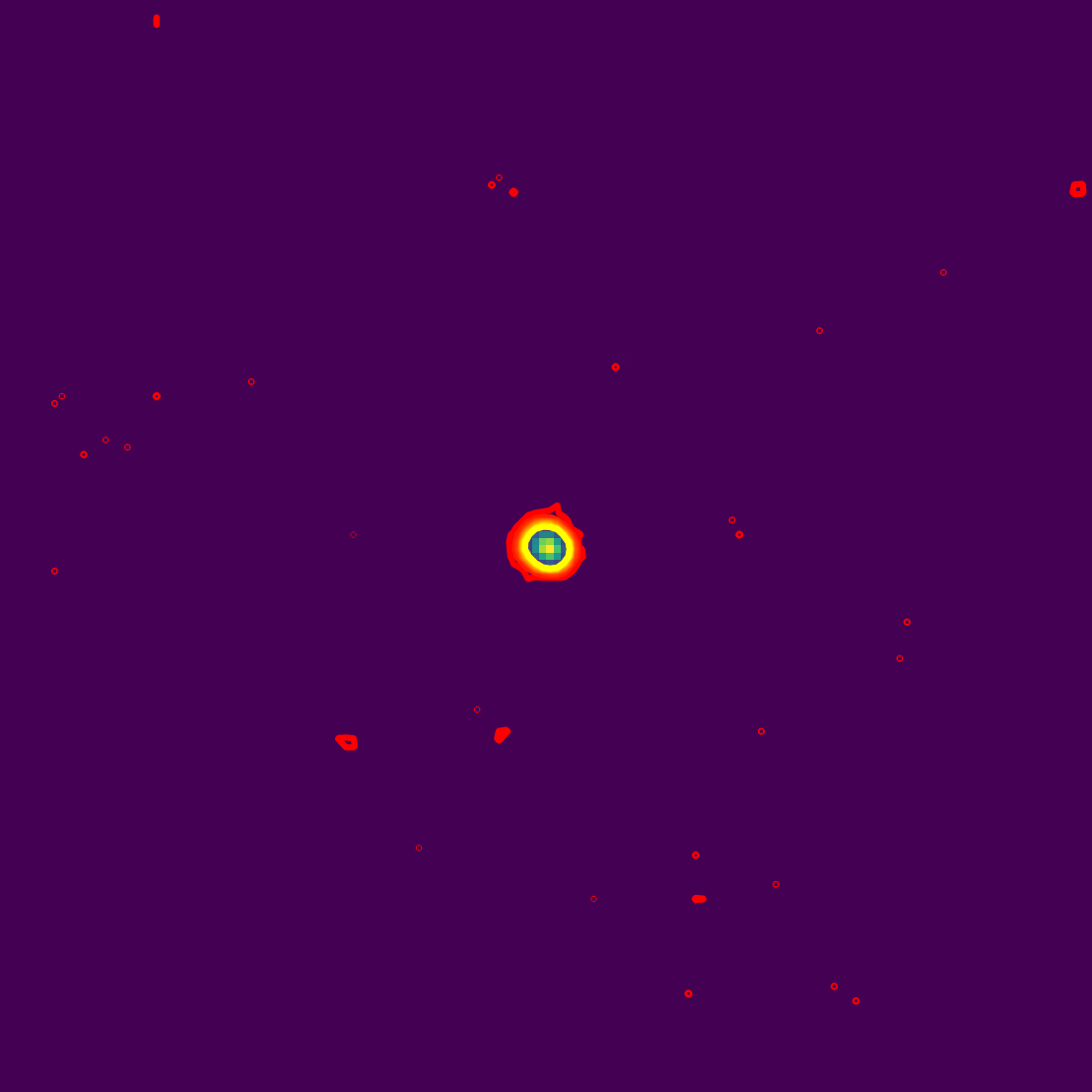}
            & \includegraphics[width = 0.18\textwidth]{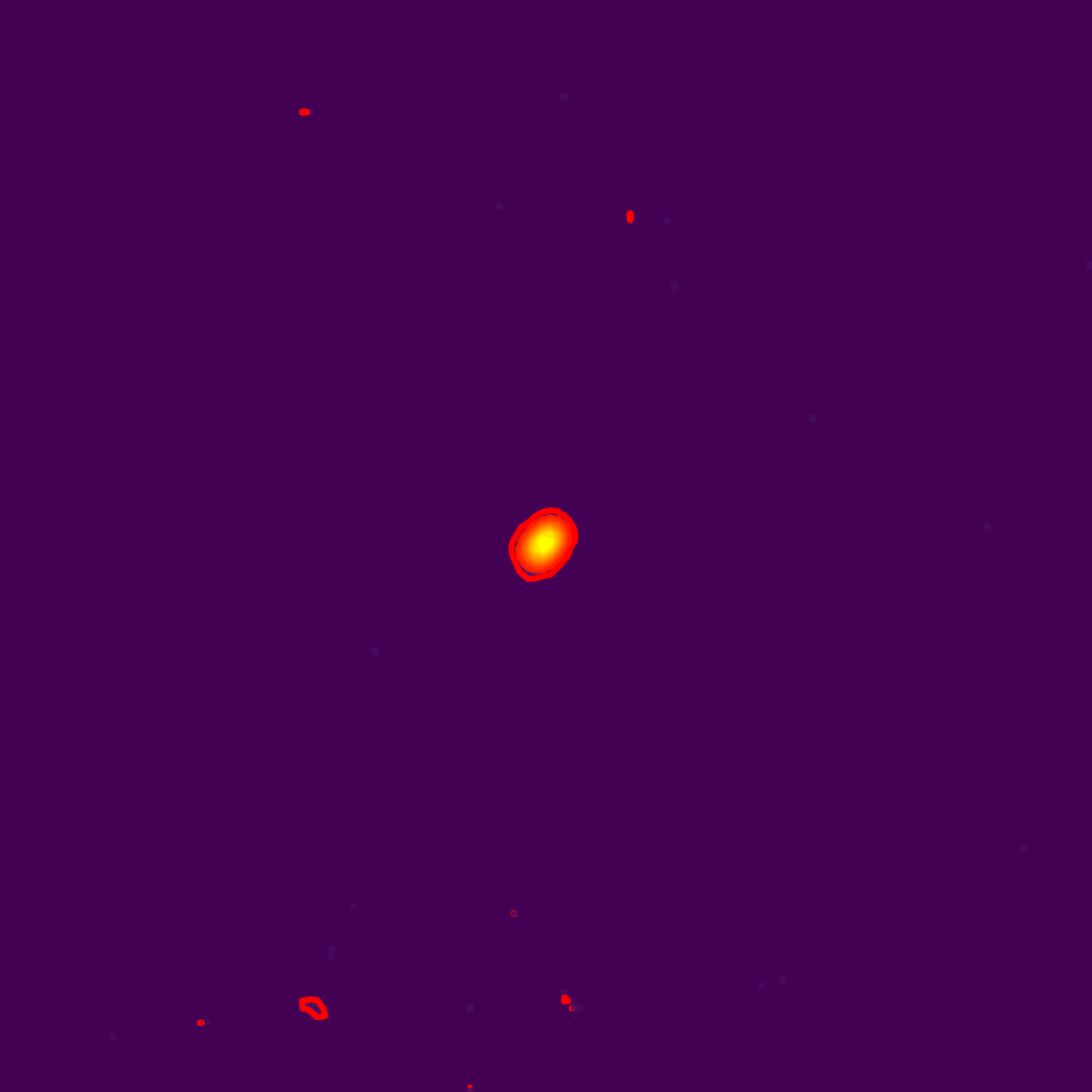}
            & \includegraphics[width = 0.18\textwidth]{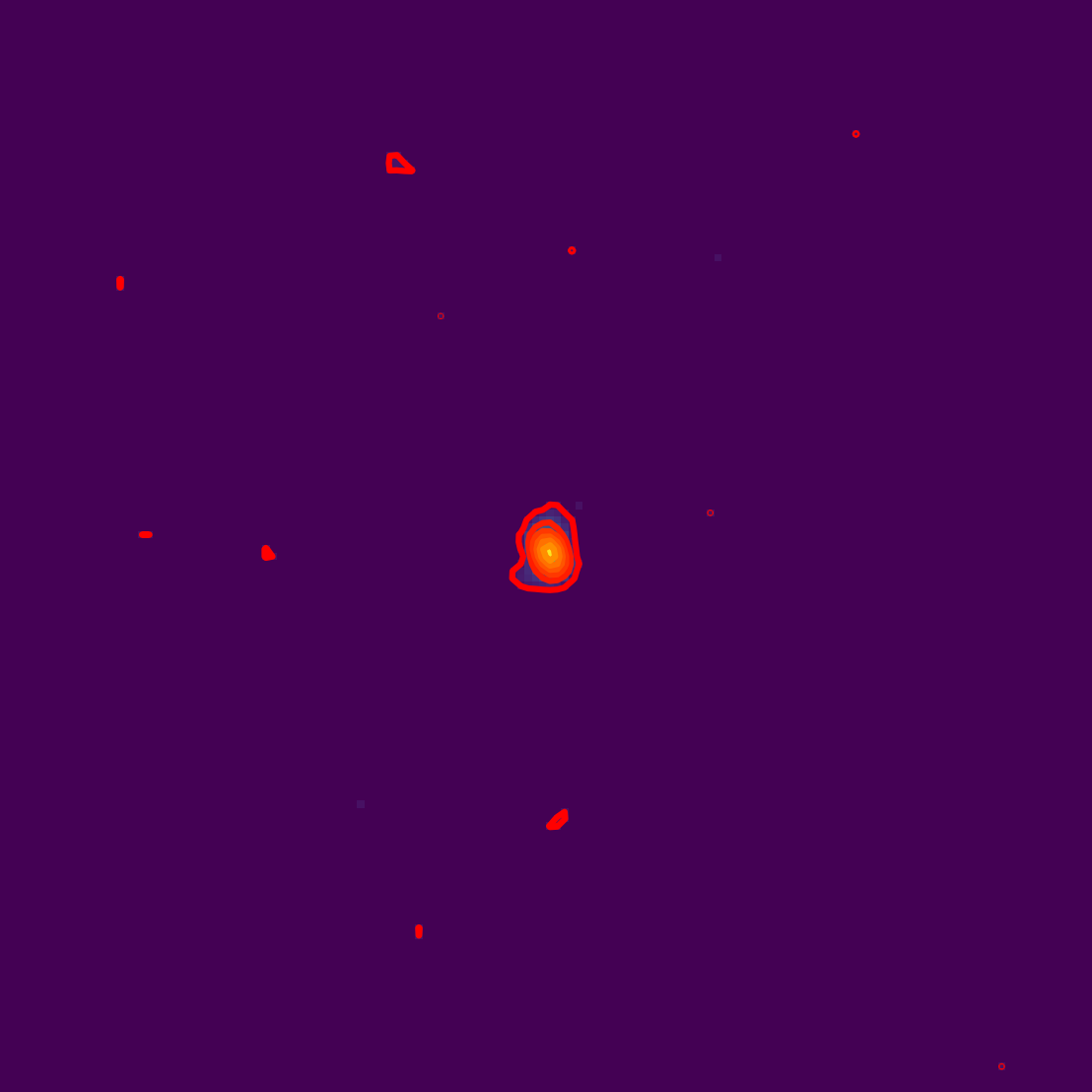}
            & \includegraphics[width = 0.18\textwidth]{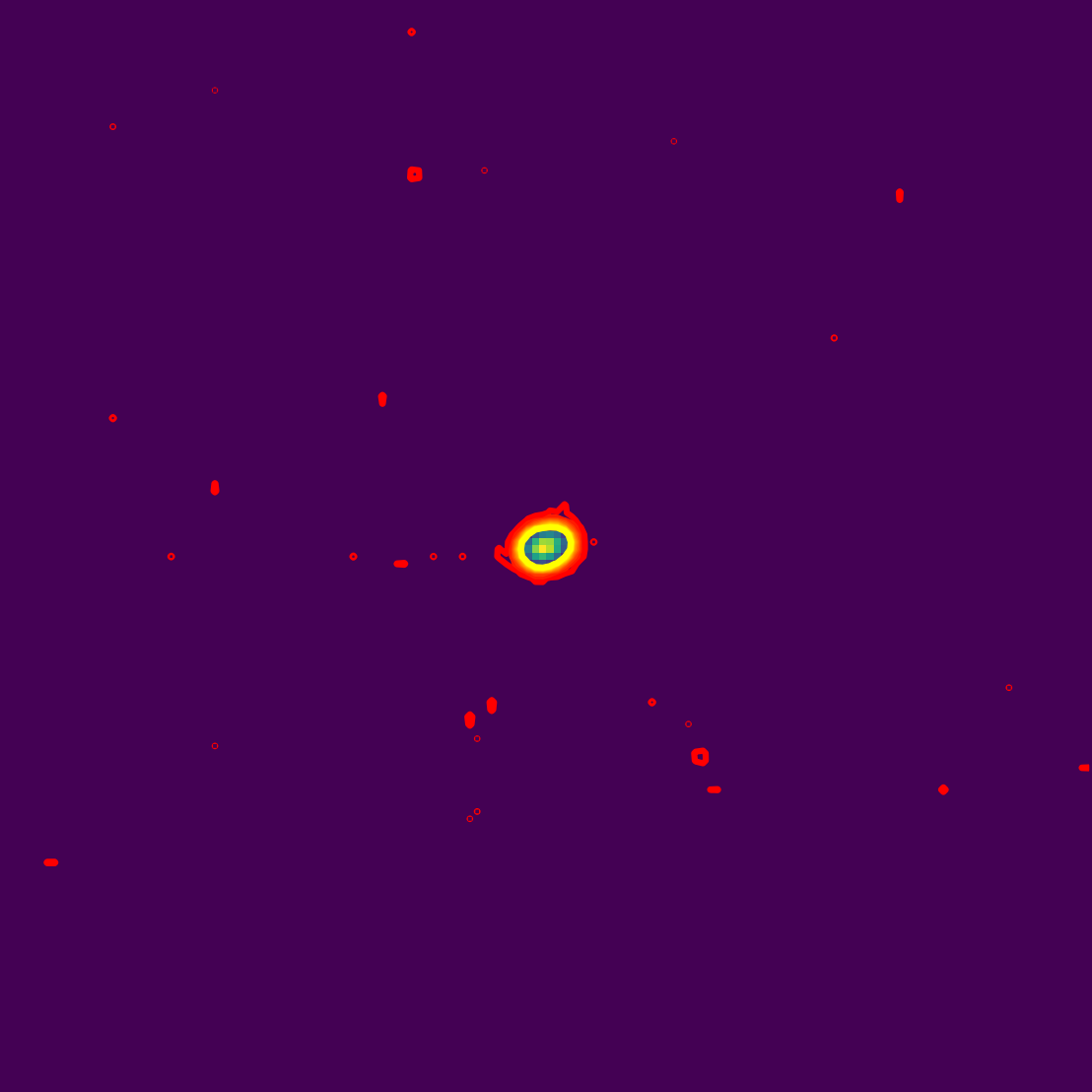}
            & \includegraphics[width = 0.18\textwidth]{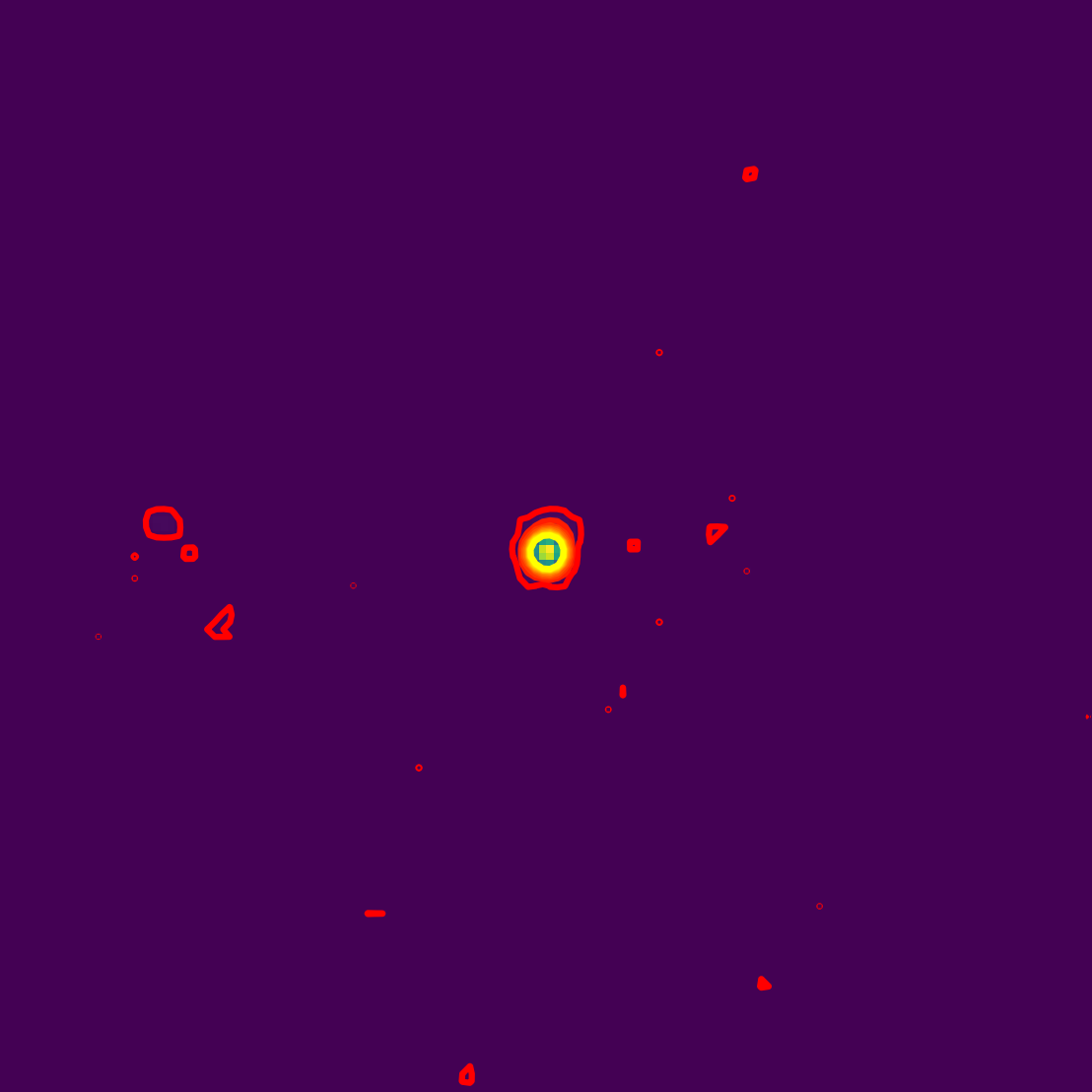}
            \\
            & RGZJ120050.1+494709 & RGZJ115427.1+524720 & RGZJ133230.8+524629	& RGZJ112313.8+515509 & RGZJ124459.2+490452 \\
            \rotatebox{90}{LoTSS}
            & \includegraphics[width = 0.18\textwidth]{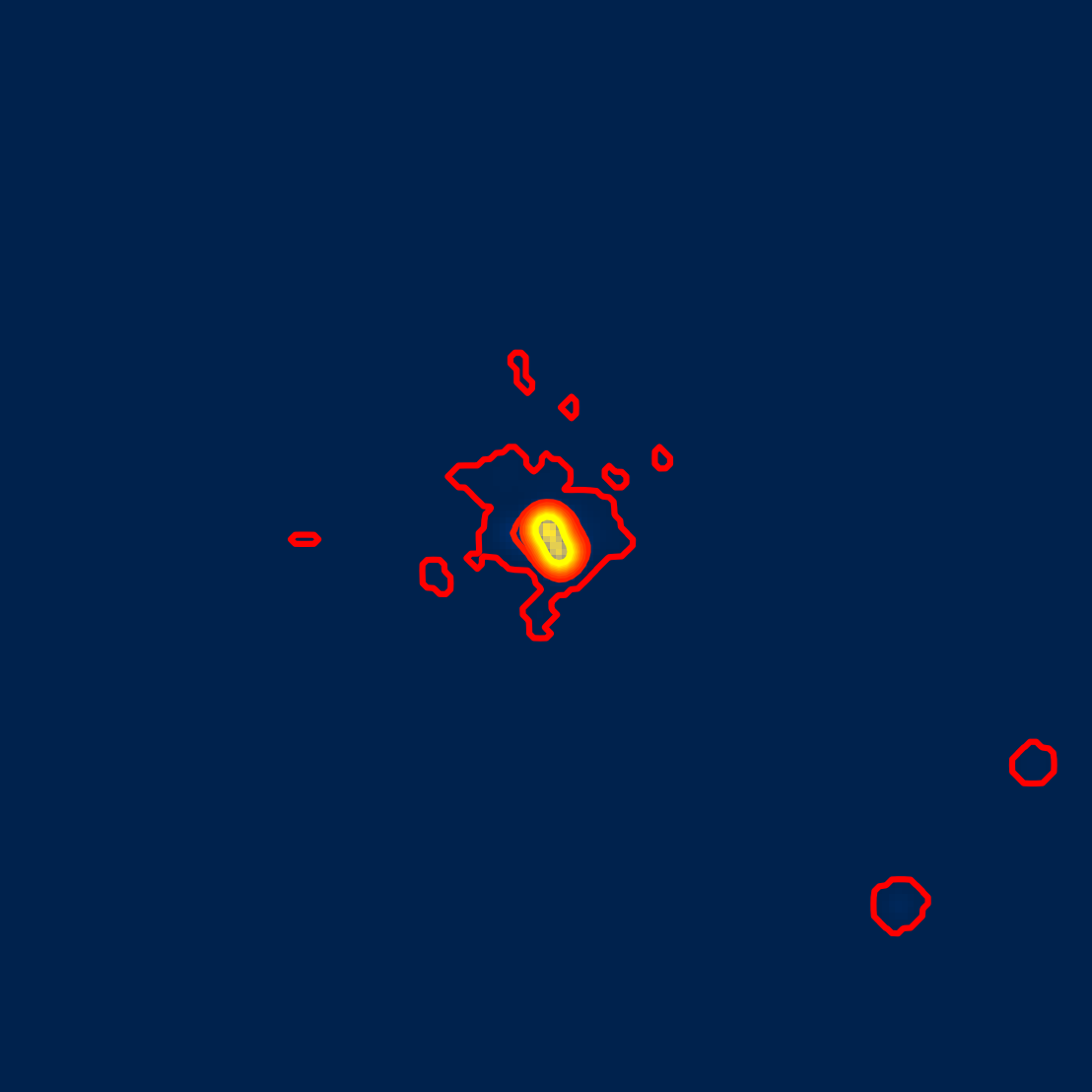}
            & \includegraphics[width = 0.18\textwidth]{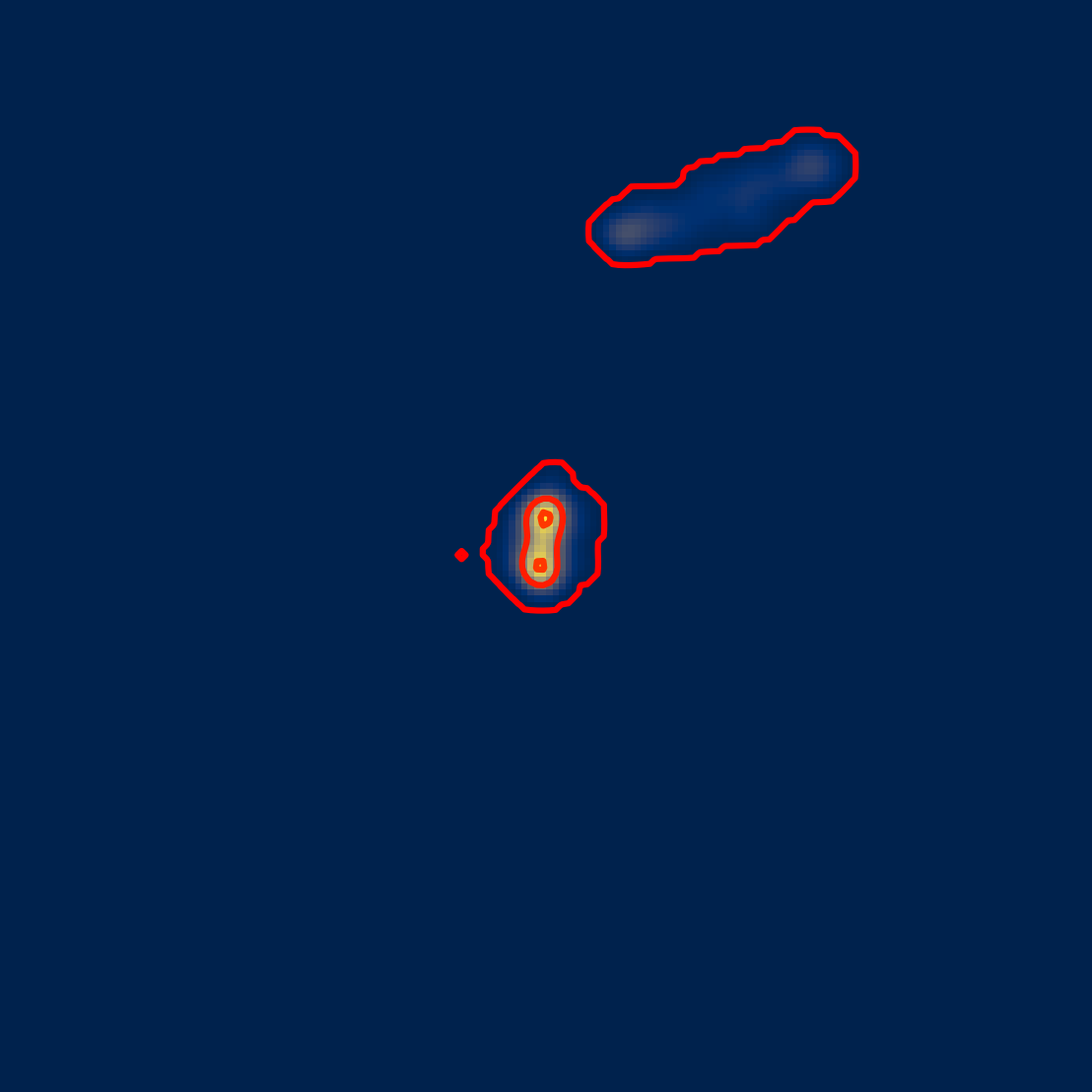}
            & \includegraphics[width = 0.18\textwidth]{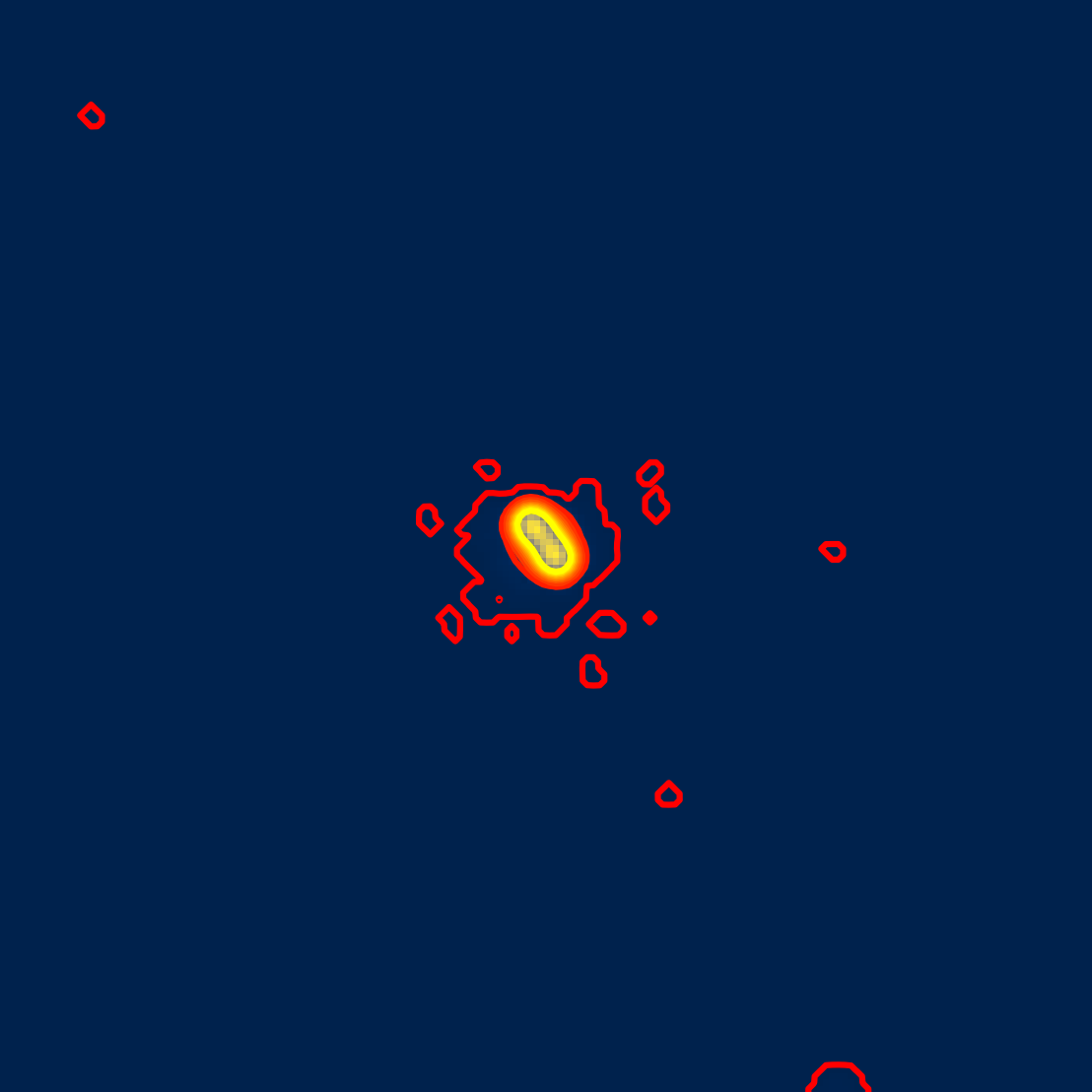}
            & \includegraphics[width = 0.18\textwidth]{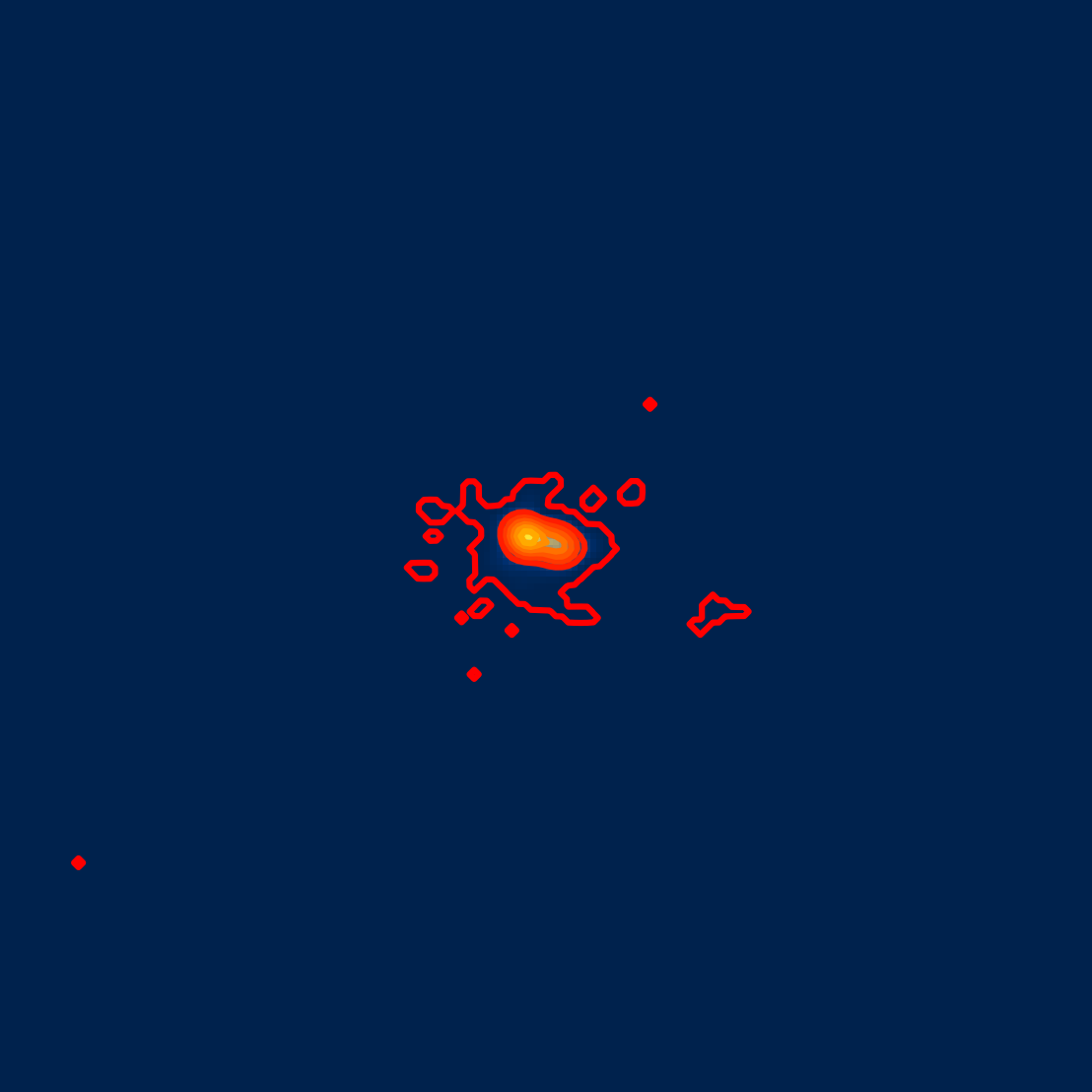}
            & \includegraphics[width = 0.18\textwidth]{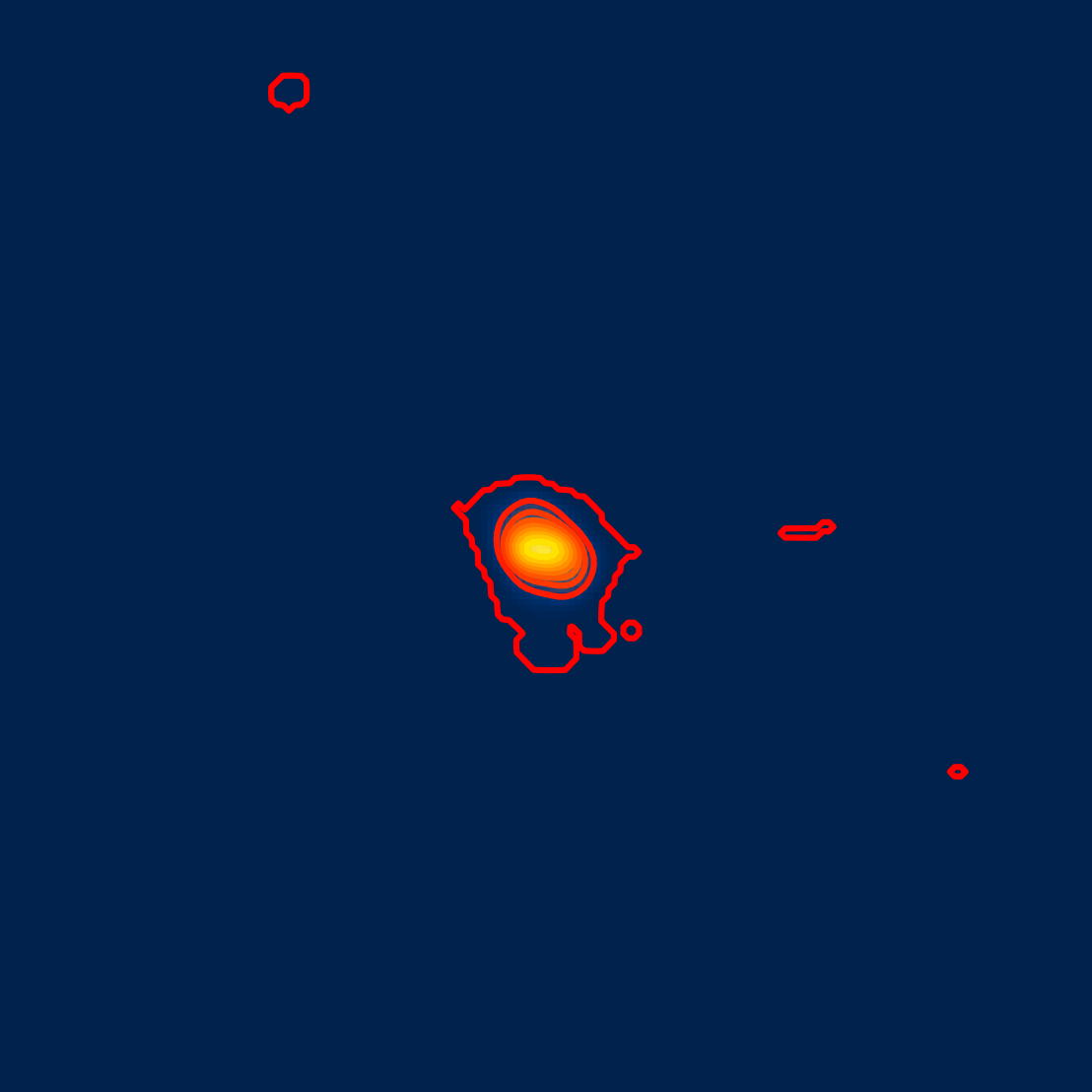} 
            \\
            \rotatebox{90}{FIRST}
            & \includegraphics[width = 0.18\textwidth]{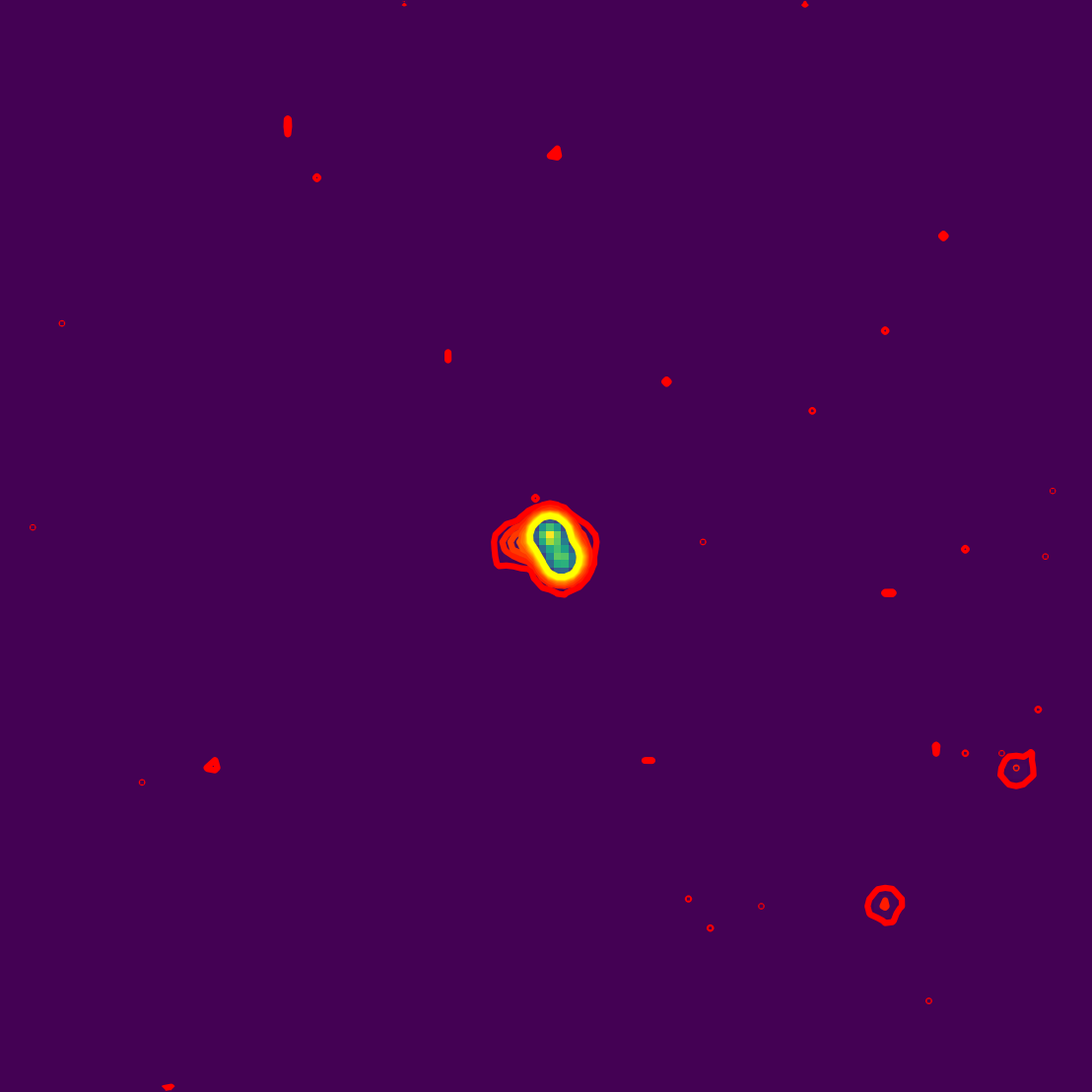}
            & \includegraphics[width = 0.18\textwidth]{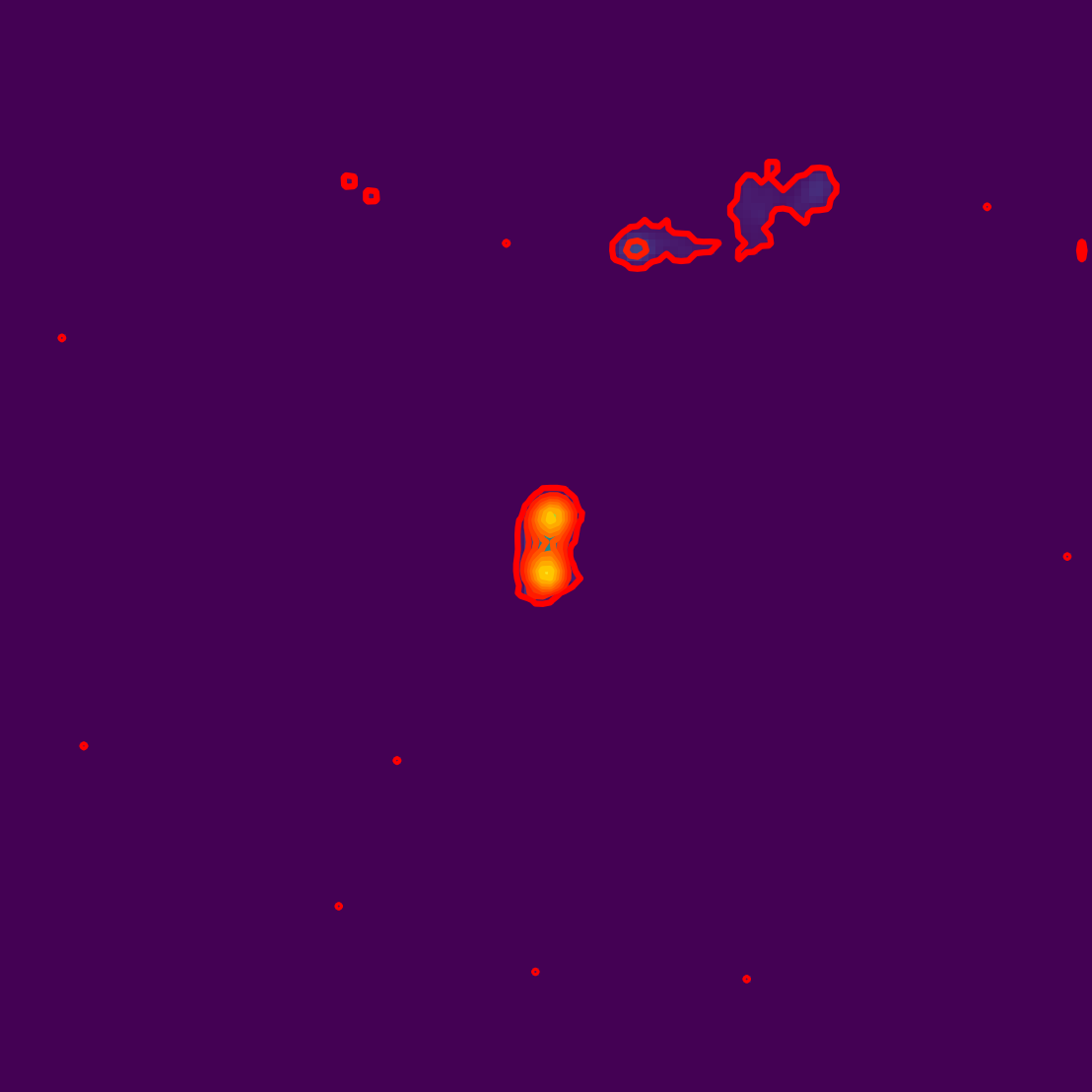}
            & \includegraphics[width = 0.18\textwidth]{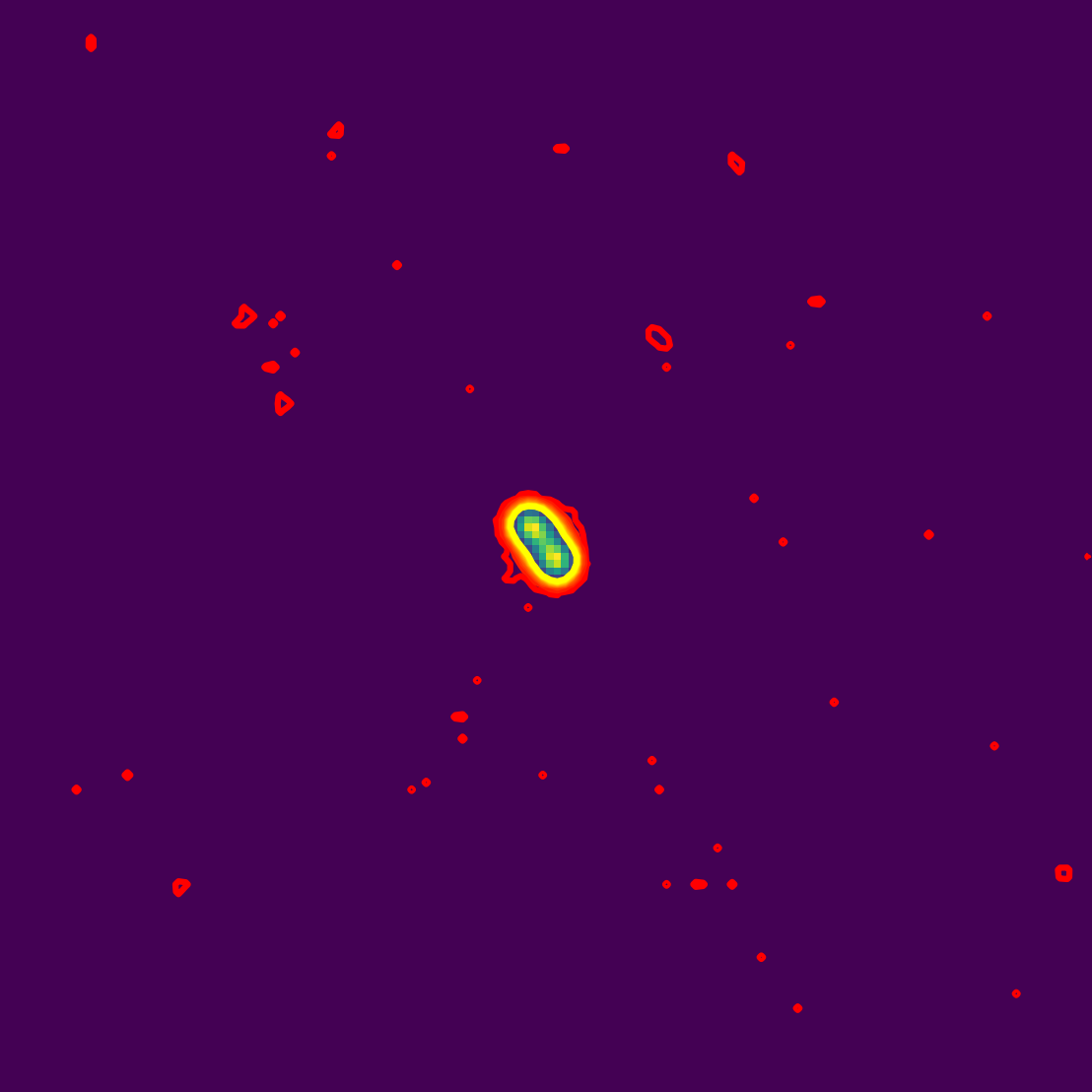}
            & \includegraphics[width = 0.18\textwidth]{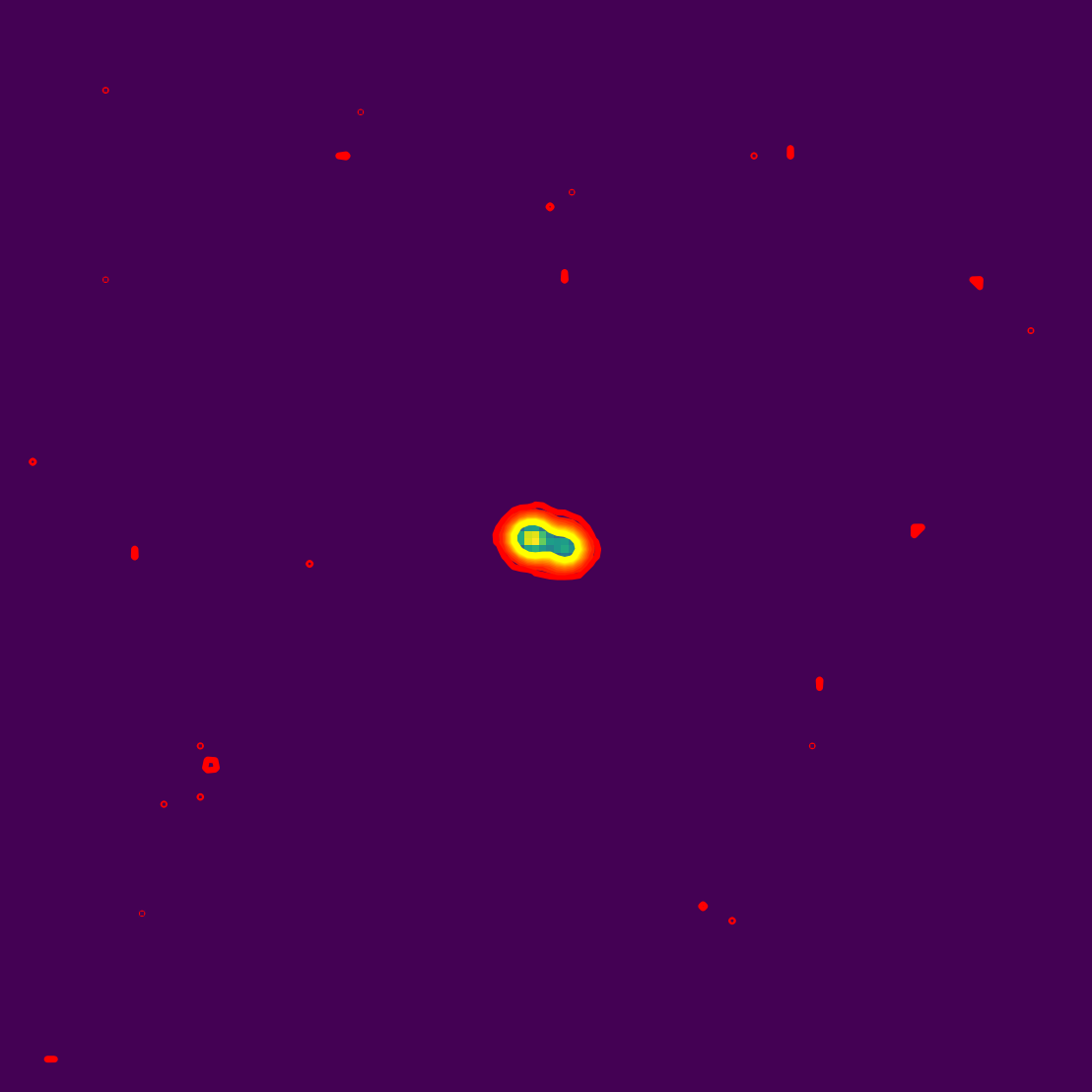}
            & \includegraphics[width = 0.18\textwidth]{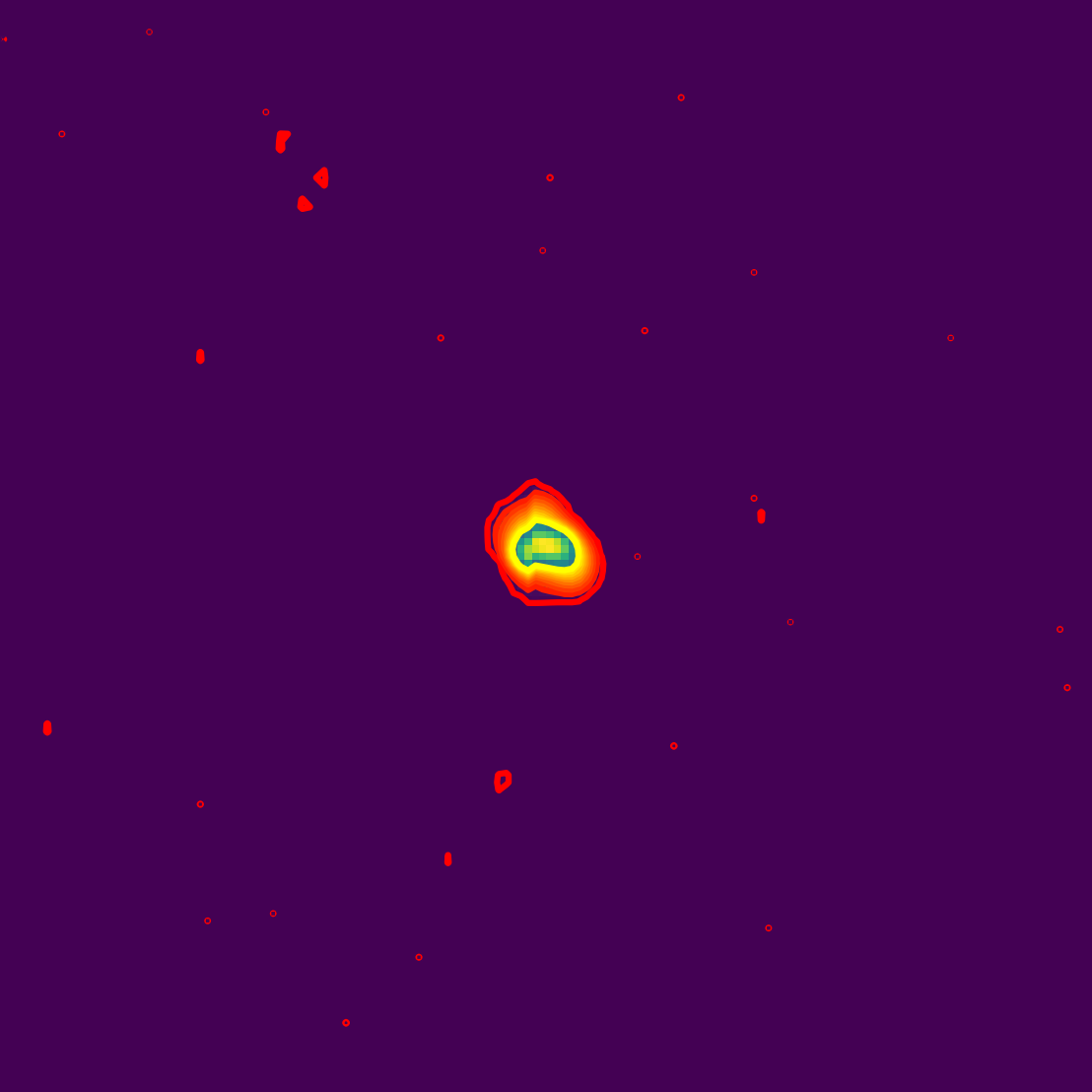}            
        \end{tabular}
        \caption{Example of RGZ\,FR FRIIs that are labelled as FRIs in LoTSS catalogues. The 4.5-arcminute radio images were obtained from LoTSS and VLA FIRST overlaid by 10 levels of contour maps spanning 3$\sigma$ to 100$\sigma$ for FIRST images and 3$\sigma$ to 3000$\sigma$ for LoTSS images.}
        \label{fig:FR2_FR1}
    \end{figure*}

Additionally, Figure~\ref{fig:lum_ps_rgz-lm} shows significant differences in the number and distribution of RGZ\,FR FRI/IIs in comparison to LoTSS FRI/IIs. We examine the different characteristics of these class-changed sources in extensive detail. There are 310 class-changed sources, accounting for 60\% of the total RGZ-LoTSS sources. The largest class-changed sources are the sources classified as FRIIs in the RGZ\,FR catalogue but identified as FRIs in the LoTSS\,FR catalogue, which are 46\% of the total class-changed sources. Both RGZ\,FR FRII-Low and FRII-High objects show a large proportion of RGZ FRII-LoTSS FRI objects, with 52\% in the FRII-Low objects and 46\% for the luminous FRII objects.

Furthermore, the RGZ\,FR FRIIs that appear as FRIs in LoTSS catalogue have a flux-angular size distribution situated on the left side of Figure~\ref{fig:lum_ps_rgz-lm} (a), indicating that they are comparatively small. From Figure~\ref{fig:FR2_FR1}, it can be seen that the RGZ FRII-LoTSS FRI objects are commonly found as extended single sources or compact double sources. The automated algorithm developed by \cite{Mingo2019} classified these small-sized sources as core-brightened (FRIs). Of the 143 RGZ FRII-LoTSS FRI objects, approximately 120 are categorised as "Small" FRIs in the LoTSS catalogue, representing 84\% of the total RGZ\,FR FRII-LoTSS FRI sources. Note that the small sources in \cite{Mingo2019} refer to sources having an angular size of less than 27\,arcsecond or an angular extent between the two brightest peaks of less than 20\,arcsecond for sources of 40\,arcsecond or smaller. On the other hand, some sources show a double lobe structure, potentially leading the model to classify these sources as FRIIs, despite the lobes not being clearly separated. 

Apart from Figure \ref{fig:lum_ps_rgz-lm}, which shows a significant number of the RGZ\,FR FRIIs classified as LoTSS FRIs, Figure \ref{fig:wise_cc} illustrates that low-luminosity FRIIs and FRIs occupy similar regions of the colour-colour plot. 
Some of these class-changed FRII sources could potentially be lobed FRIs. Additionally, \citet{YatesJones_2023} showed using simulations that FR class can be time-evolution dependent. It is therefore possible that some source sources might fall in the transitional period between two FR classes. This may confuse classification models due to the presence of lobes and changes in jet brightness.

In addition, we find that the class-matched sources have the lowest ratio of VF < 1 to VF = 1 at 0.167, while the ratio for class-changed sources is higher at 0.245. The same ratio for RGZ FRII-LoTSS FRI objects is significantly higher compared to both class-matched and class-changed sources at 0.300. It implies the increasing of the model's uncertainty in classifying the class-changed sources, particularly the sources with the RGZ FRII-LoTSS FRI label.
These differences in classification results could be affected not only by observational differences between the two surveys, but also by biases arising from the pre-training and fine-tuning of the model, which will be further discussed in Sections~\ref{sec:ptbias}~\&~\ref{sec:ftbias}.

\subsection{Colour analysis for RGZ-LoTSS sample} 

\begin{figure*}
\centerline{
\includegraphics[width=0.6\textwidth]{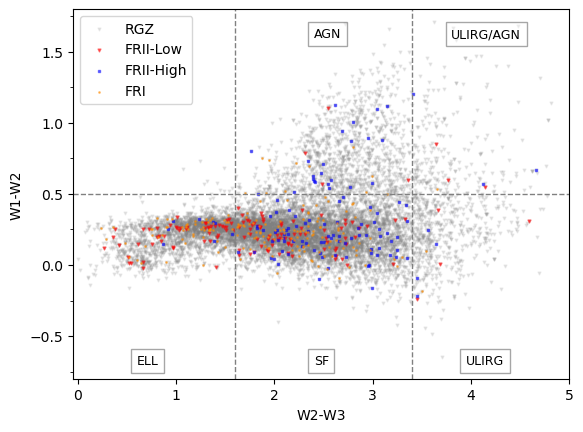}
}
\caption{WISE colour/colour plot comparing of RGZ-LoTSS sources, including FRIs (orange), FRII-Low objects (red), FRII-High objects (blue), and the RGZ\,FR sources (grey) with W1 and W2 SNR greater than 5 (no limit applied on W3).}
\label{fig:cc_LM_to_RGZ}
\end{figure*}

Figure~\ref{fig:cc_LM_to_RGZ} illustrates the distribution of all RGZ-LoTSS sources categorised into FRIs (orange) and FRIIs, using the RGZ FR label, with low (red) and high (blue) luminosity. The figure shows the distribution of sources having W1 and W2 SNR greater than 5 (no limit applied on W3).
These sources  on the WISE colour analysis plot, in comparison to the RGZ\,FR sources (grey). The proportion of sources in the SFG region rises from 49\% to 65\% for RGZ-LoTSS sources. More than 93\% of the RGZ-LoTSS FRIs lie within the area where W1$-$W2 $< 0.5$ mag and W2$-$W3 $< 3.4$ mag, indicating the dominance of FRIs in the Ell and SFG regions. FRIIs with low luminosity are commonly found in the Ell and SFG regions, similar to FRIs. On the other hand, FRII sources with high luminosity are more prominent in the SFG and AGN regions. There is evident overlap among the three classes, particularly in the SFG region. 

\subsection{Spectral Indices}

The spectral index can be used to better understand the emission properties and the evolution of different radio source populations. Radio sources with frequencies below 10\,GHz have radio spectra that are typically dominated by non-thermal synchrotron emission \citep{Duric1988}, characterised by a power-law slope of $S_\nu \propto \nu^{-\alpha}$, where $\alpha \simeq 0.7$ for optically thin emission.
In this study, we used flux densities at two frequencies to determine the spectral index: 150\,MHz from LoTSS data and 1.4\,GHz from FIRST data.

\begin{figure}
\centerline{
\includegraphics[width=0.5\textwidth]{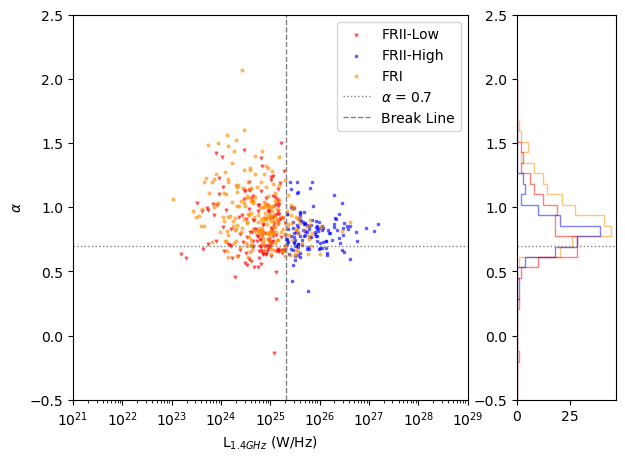}
}
\caption{Luminosity (W/Hz) at 1.4\,GHz and spectral index ($\alpha$) at 150\,MHz from LoTSS data and 1.4\,GHz from FIRST data.}
\label{fig:alpha_lum}
\end{figure}

Figure~\ref{fig:alpha_lum} depicts the relationship between the FIRST-LoTSS spectral index ($\alpha$) and the luminosity at 1.4\,GHz, for all RGZ-LoTSS sources, labelled using the RGZ FR classification. The figure suggests that there may be distinct spectral shifts between the different sub-samples. Both FRIs and FRIIs exhibit distributions skewed towards lower values of $\alpha$, and lower luminosity sources generally have higher spectral indices and show greater variance. 
The spectral index values of FRIs span approximately 0.3 to 2.1, with the highest number of FRIs being around 0.8. 
The distribution of luminous FRIIs is narrower in comparison to that of FRIs, ranging from 0.3 to 1.2, with a modal value close to $\alpha=0.8$.
On the other hand, the distribution of FRIIs with low luminosity has a prominent peak at $\alpha\sim0.65$, which is significantly lower compared to both FRIs and high-luminosity FRIIs, and their range spans -0.1 to 1.5. However, due to the relatively small sample size, it is not possible to draw strong conclusions from these data.


\subsection{Low-luminosity FRII objects}

In Section \ref{sec:ld}, we found that both classes of the RGZ\,FR galaxies are present across the entire range of luminosities, which covers approximately nine orders of magnitude. As also previously found in \cite{Mingo2019},  this challenges the conclusion from \cite{Fanaroff1974} that a sharp luminosity boundary distinctly separates these classes.
Around 58\% of the RGZ\,FR FRIIs have a luminosity below $\sim 2 \times10^{25}$ W\,/\,Hz at 1.4\,GHz, which agrees well with the percentage of FRII-Low objects in \cite{Mingo2019} at 59\%. 
Considering only a lower redshift range, $z < 0.8$, the proportion of low-luminosity FRIIs remains high as in the full $z$ range. 
However, lowering the break line to $\sim 2 \times10^{24}$ W\,/\,Hz results in a decrease in the proportion of low-luminosity FRIIs to 9.3\%, which is significantly lower than the results of \cite{Mingo2019}, who reported a proportion of 21\% for the lowered threshold.

Analysis of the RGZ\,FR radio galaxy population indicates that over 50\% of RGZ\,FR sources have spectroscopic redshifts. 53\% of low-luminosity FRIIs have spectroscopic redshift data, and FRIIs with high luminosity have a slightly smaller proportion of sources with spectroscopic redshifts, specifically 46\%. The result suggests that the distance to FRIIs with low luminosity is even more frequently determined by spectroscopic redshift than for FRIIs with high luminosity. The redshift range of RGZ\,FR sources is shown in Figure~\ref{fig:z}. The distribution of RGZ\,FR sources peaks around $z = 0.3 - 0.5$ and then decreases significantly at a redshift of around 0.8. It can be seen that more luminous FRIIs are found at higher redshifts, whereas low-luminosity FRIIs share more similarities with FRIs in terms of their redshift distribution. The difference in distribution between low-luminosity and high-luminosity FRIIs suggests that sources with lower luminosity could be less frequently found or possess characteristics that cause them to be more difficult to detect at higher redshifts.

\section{Model Biases}
\label{sec:biases}

In this work, we employed self-supervised learning pre-trained on the extensive, unlabelled dataset from RGZ\,DR1 data, followed by fine-tuning by the labelled dataset from the MiraBest catalogue. Here we consider the impact of two different generalisation biases inherent in the classification model on these results. 


\subsection{Pre-training Generalisation Bias}
\label{sec:ptbias}


The potential efficiency of SSLs model has been proposed by several studies \citep[e.g.][]{HowardRuder2018, Guo2019, Hayat2021}. On top of that, SSL is widely applied to different tasks in similar scenarios, especially in astronomy  \citep[e.g.][]{Jimenez2023, Hayat2021, Stein2022}. However, SSL can be influenced by several factors, including observational biases, biases in the fine-tuning dataset, and data shifts between the pre-training dataset and the test or fine-tuning sources, all of which are common challenges in predictive modelling. To enhance model performance, \citep{Goyal2019ScalingLearning} suggest that the most effective outcomes are achieved when the model is trained on datasets similar to the downstream test task. However, since this is often not possible, to mitigate the effect of the biases on the model's accuracy, various techniques can be employed. 

\begin{figure*}
\centerline{
\includegraphics[width=0.5\textwidth]{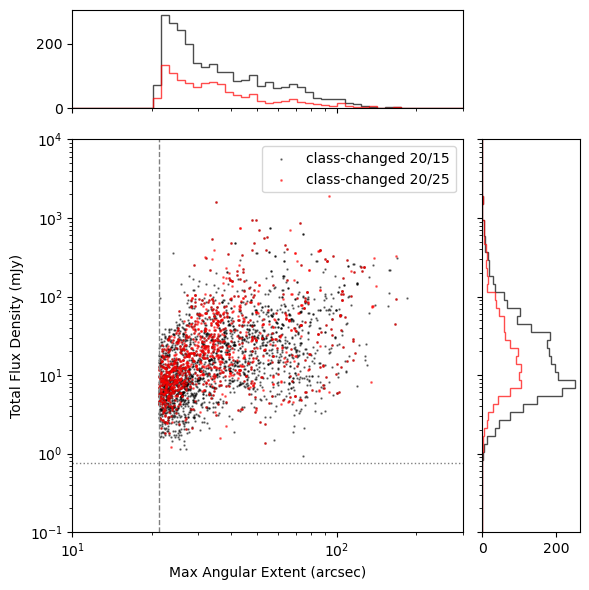}
}
\caption{Flux density and angular size distribution of class-changed sources for two different pre-training sets. Black points represent the 20/15 set (pre-trained with angular size thresholds of 20 and 15\,arcseconds), while red points represent the 20/25 set (pre-trained with thresholds of 20 and 25\,arcseconds) with the angular size and flux density thresholds of 21.2\,arcseconds (dashed line) and 0.75 mJy (dotted line). The top and right panels show histograms of the angular size and flux density distributions, respectively, for both sets.}
\label{fig:class-changed-pred}
\end{figure*}

Here we consider the effect of different choices in our pre-training data on the results of the downstream classification task using models pre-trained on data with different angular-size cuts in order to determine the impact on the predictive performance. Our baseline pre-training model uses the set of RGZ\,DR1 sources with angular sizes larger than 20\,arcseconds. In order to study the effects of different minimum angular size cuts in the pre-training data, we compare the downstream predictive performance of our baseline model to the results from alternative models pre-trained using sources with the size thresholds of 15 and 25\,arcseconds with 97,513 and 48,774 training sources, respectively, as shown in Figure \ref{fig:class-changed-pred}. The results of this analysis show that a greater proportion of sources with VF < 1 result from the 15\,arcsecond pre-training set (28\%) compared to the 20\,arcsecond (21\%) and 25\,arcsecond (22\%) pre-training sets. These results suggest that the model pre-trained using 15\,arcsecond data tends to result in predictions with a greater degree of ambiguity. However, the results for the 20-arcsecond and 25-arcsecond sets are not significantly different.

In addition, we also compare the predicted downstream FR classifications from the 15\,arcsecond and 25\,arcsecond pre-training data to those from our 20\,arcsecond pre-training baseline. We hereinafter refer to the comparison between 20\,arcsecond and 15\,arcsecond data as the 20/15 set and the comparison between the 20\,arcsecond and 25\,arcsecond data as the 20/25 set. The sources with different label predictions between two datasets are described here as ``class-changed'' sources. The 20/15 set shows a larger number of class-changed sources compared to the 20/25 set, by roughly a factor of two: there are approximately 1,100 and 2,600 sources with different labels for the 20/25 and 20/15 sets, respectively. The distribution of these class-changed sources, depicted in Figure~\ref{fig:class-changed-pred}, shows that these sources are typically small in angular size and are located in the overlapping area between FRIs and FRIIs where VF < 1 for the baseline pre-training model.
The distribution of class-changed sources from both the 20/15 and 20/25 sets shows agreement with the RGZ\,FR sources with VF < 1. 
We interpret this as indicating that although there is an effect due to the data shift between pre-training data and test sources, which influences the degree of confidence with which the model is able to classify individual sources, the lack of a systematic increase in class-changed sources in the 15 to 25 arcsecond size regime indicates that the impact of the data shift between the pre-training and fine-tuning datasets is not visibly significant.


These results imply that using a 15\,arcsecond pre-training set may lead to a decreased overall confidence in the downstream predictions, despite the availability of more data for pre-training compared to the 20-arcsecond and 25-arcsecond thresholds. This suggests that increasing the data volume may not guarantee enhanced model confidence. Despite many studies proposing that the \emph{quantity} of pre-training data is a significant factor that can improve the model's robustness \citep{ramanujan2023on} and enhance downstream task performance \citep[e.g.][]{Sun2017Revisiting, ridnik2021imagenet}, our results indicate that it the \emph{quality} of these data is also important, i.e. a larger number of unresolved sources in the training data decreases its overall quality.

This is consistent with an alternative study by \cite{Nguyen2022Quality} that suggested gathering a large amount of data is not the most effective option for constructing a pre-training dataset, but rather that filtering a noisy data source can enhance the dataset's generalisation properties. In addition, many studies observed that the models performed well when they were pre-trained on datasets that are similar to the downstream test task \citep[e.g.][]{Cole2022, Goyal2019ScalingLearning, Kotar2021}. In addition, when the distribution of data in upstream (pre-training) and downstream (fine-tuning) tasks differs, increasing \emph{diversity}, the number of unique sources encountered during training for a fixed computational budget, of the pre-training dataset may adversely affect the downstream performance, and even a significantly larger data diversity cannot compensate for the distribution shift effect \citep{Hammoud2024PretrainingDiversity}. This indicates the importance of high-quality pre-training data and suggests that it should align with the fine-tuned dataset in order to achieve better classification outcomes. These findings are also consistent with similar investigations in the radio astronomy literature that have previously found that the distribution of galaxy angular sizes in training data for semi-supervised learning can impact model performance \citep{Slijepcevic2022FixMatch}.

\subsection{Fine-tuning Generalisation Bias}
\label{sec:ftbias}

\begin{figure*}
\centerline{
\includegraphics[width=0.49\textwidth]{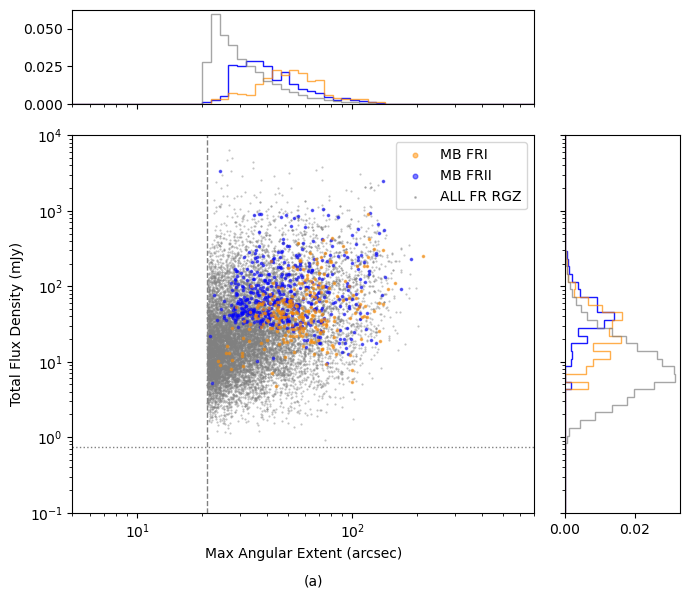}\qquad
\includegraphics[width=0.49\textwidth]{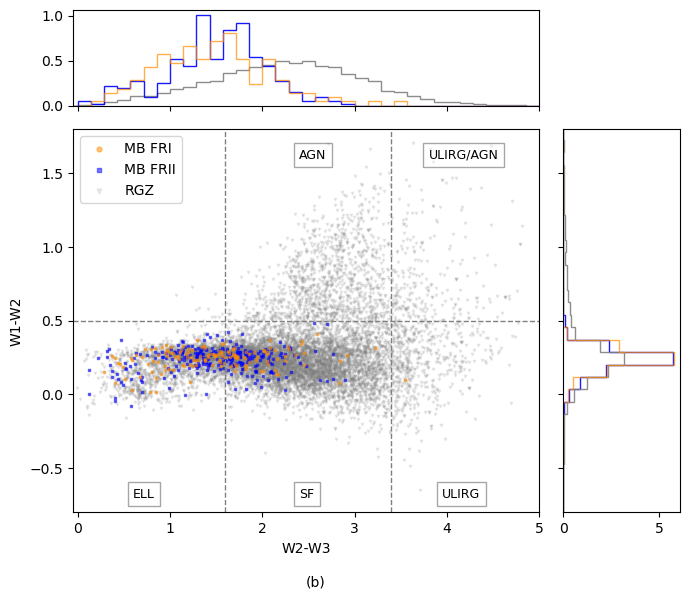}
}
\caption{MiraBest dataset, including FRI (orange) and FRII (blue) sources, compared to total RGZ\,FR sources (grey). (a) the distribution of MiraBest and RGZ\,FR sources on angular size and total flux density with normalised histograms. (b) colour analysis of MiraBest and RGZ\,FR sources with W1 and W2 SNR greater than 5 (no limit applied on W3).}
\label{fig:gen_bias}
\end{figure*}


As discussed in Sections~\ref{sec:MiraBest}~\&~\ref{sec:ptbias}, the distribution of data sources in a training set can impact on the model predictions at test time. For downstream tasks, this is also true for the training data in the fine-tuning dataset. Here we examine the distributional differences between the RGZ\,FR sources and the MiraBest dataset.
Figure~\ref{fig:gen_bias} (a) illustrates the distribution of MiraBest FRI/II sources and the RGZ\,FR sources with respect to total flux density and angular size. It can be seen that on average the MiraBest sources tend to have higher flux densities and larger angular extents compared to the RGZ\,FR sources. Moreover, FRII-type sources in the MiraBest catalogue typically exhibit higher flux density and slightly smaller apparent size in comparison to FRIs. Therefore, it is apparent that FRIIs typically appear in the upper left quadrant, while FRIs are situated in the bottom right quadrant of the MiraBest distribution, as shown in Figure~\ref{fig:gen_bias} (a). This is reflected in the classification predictions for sources with lower flux density tending to be more ambiguous, as depicted in Figure~\ref{fig:votefrac}.

We can see this behaviour echoed in Figure~\ref{fig:lum_ps_rgz-lm} (a), which illustrates the distribution of predicted FRI and FRII labels in RGZ\,FR data that are cross-matched with the LoTSS catalogue. Conversely, in the LoTSS labelling scheme, FRIIs appear preferentially at larger angular sizes, as illustrated in Figure~\ref{fig:lum_ps_rgz-lm} (b). 


We analyse the distribution of MiraBest data in comparison to RGZ\,FR VF<1 data, as well as to RGZ\,FR VF=1 data. The distribution of MiraBest data is slightly more similar to RGZ\,FR VF<1 than to RGZ\,FR VF=1. The Kernel Density Estimation (KDE) plots with Gaussian kernels of MiraBest compared to RGZ\,FR VF < 1 and VF = 1 data suggest that RGZ\,FR VF < 1 overlaps slightly less with the MiraBest data. We quantify the difference between MiraBest vs RGZ\,FR VF < 1 and MiraBest vs RGZ\,FR VF = 1 by using Kullback-Leibler (KL) divergence \citep{Kullback1951}. The results indicate the difference: KL(MiraBest || RGZ\,FR VF < 1): 0.8466, KL(MiraBest || RGZ\,FR VF = 1): 0.5378, indicating that MiraBest data are more divergent from RGZ\,FR VF < 1 than RGZ\,FR = 1.
This result implies that fine-tuning data affects model predictions, as the larger data shift leads to increased classification ambiguity.

In addition, we consider the colour-colour distribution for the RGZ\,FR sources (grey), MiraBest FRIs (yellow), and MiraBest FRIIs (blue), as shown in Figure~\ref{fig:gen_bias} (b). The figure illustrates the more localised clustering of MiraBest FRI and FRII galaxies in comparison to the RGZ\,FR sources.  The RGZ\,FR sources are more widely distributed, reflecting a broader range of characteristics. On the other hand, MiraBest data cluster tightly and mostly in the ELL and SFG regions. 
Additionally, FRII galaxies exhibit greater variability along the W2$-$W3 axis compared to FRI galaxies.
This implies that the RGZ\,FR sources may exhibit greater diversity compared to the fine-tuning dataset. It may be necessary to limit the properties of the categorised sources and require a standardised FR catalogue that includes more varieties of galaxy types in the future.


\subsection{Physical vs. Observational classification}
\label{sec:physical}

In this work, we explore the application of machine learning on the Fanaroff-Riley dichotomy, which is a rigidly discrete classification dividing radio galaxies into two categories, such as FRI or FRII. But because the radio morphologies are typically complex, there is an overlap in radio galaxy classifications. Indeed, \citet{tagsnotboxes} presented that each source can have multiple descriptors depending on the observations. \citet{Bowles2023} employed natural language processing methods to describe radio morphologies with 22 semantic tags. This technique can help identify morphologies and find rare sources with interesting morphologies. 

Furthemore, the embedding from the pre-trained SSL model and the classifications produced by fine-tuning are purely morphological, i.e. the only information currently used to distinguish FRI from FRII type galaxies is the distribution of surface brightness in their images. While this may achieve a certain level of success, it is limited in a number of ways. Firstly, the effect of resolution will affect how clearly the morphology can be recovered by the model, which will disproportionately affect important characteristics such as the degree of collimation present in the jet - especially relevant for the clear identification of FRIIs. Secondly, important information is contained in the spectral index distribution across individual sources, and this is not yet available within either the pre-training or fine-tuning data sets. This is of particular relevance for cleanly separating FRI-type objects with no backflow from their lobes. 

Even accounting for the model biases described previously, without this additional information encoded in the model, it is important to recognise that the \emph{morphological} classification is not necessarily equivalent to the \emph{physical} classification of these sources. Since it is the physical classification that is scientifically important, one might expect that future work should prioritise closing the gap between these two differing outcomes. However, given that the FR classification scheme itself is based on subjective human inspection \citep{Fanaroff1974}, one must also consider how meaningful it is to focus on closing such gaps in general. It has been proposed that using descriptive and cumulative \emph{tags} for catalogues rather than discrete classifications may be preferable for new analyses \citep{tagsnotboxes}. 

Using the structure of the pre-trained latent space more directly to probe for data associations may also achieve the same objectives, without the necessity of having to articulate an object description a priori. Uniform Manifold Approximation and Projection \citep[UMAP;][]{McInnes2018} is one methods that allows for such exploration of the latent space. UMAP effectively projects high-dimensional data into lower-dimensional spaces. This visualisation helps in clustering objects based on their features. And provides a continuum classification rather than in rigidly discrete categories. There are several studies in astronomy that used UMAP to determine the source properties or remove artefacts \citep[e.g.,][]{Mohale2024, Riggi2024}. Techniques like semantic morphology taxonomy and using UMAP can help identify complicated relationships that traditional classifications might overlook.

\section{Conclusions}
\label{sec:conclusion}


In this work, we focus on the application of self-supervised learning (SSL) in the classification of radio galaxies, particularly in radio galaxy classification by using large unlabelled datasets from the RGZ project and fine-tuning with limited labelled data from the MiraBest catalogue. We provide a catalogue of predicted FR classifications for the RGZ dataset, which includes over 14,000 objects with $\sim$5,900 FRIs and $\sim$8,100 FRIIs. In agreement with previous classification studies, we find a significant overlap in the distribution of luminosity between FRIs and FRIIs, and confirm the finding that low-luminosity FRII objects that fall under the conventional luminosity break line are more common than previously understood. 
The proportion of FRII-Low objects in our findings, 58\% of total FRIIs, aligns well with the previous study by \cite{Mingo2019}. Furthermore, the cross-matching between the RGZ\,FR and LoTSS\,FR catalogues reveals a good agreement of the LoTSS\,FR sources to the characteristics of the RGZ sources. 
A WISE colour analysis of the cross-matched sources also indicates a similarity in the host galaxy type between the sources in the two catalogues. In addition, the overlap region between FRIs and FRIIs remains large for these cross-matched sources. This overlap area between FRIs and FRIIs was the most challenging to classify by the SSL model, as indicated by the proportion of the RGZ sources with VF < 1, see Section~\ref{sec:vote_fraction}. 
Furthermore, our study identified the potential classification shifts between datasets (class-changed sources). This suggests a potential path for further investigation into the nature and evolution of these objects. For example, by identifying samples of class-changed galaxies and comparing their properties to those of class-matched gaalxies.

In order to better understand the effect of different data choices on model performance and the broader implications and limitations of SSL techniques in astronomy, we conducted additional analyses to investigate generalisation biases arising from both the pre-training and fine-tuning datasets. The differences in flux density and angular size between the RGZ and MiraBest sources show the influence of these characteristics on model predictions: pre-training using an increased number of sources with lower flux densities and smaller angular sizes tends to result in decreased confidence for model classification. In addition, our results indicate that pre-training with datasets of varying angular size thresholds can significantly impact model confidence, suggesting the need for a diverse standardised catalogue of radio galaxies to enhance classification confidence and understanding of the intrinsic characteristics of radio galaxies. 



\section*{Acknowledgements}


We thank Larry Rudnick for useful discussions that influenced this paper.

NB is supported by the Ministry of Higher Education, Science, Research and Innovation, the Royal Thai Government and the National Astronomical Research Institute of Thailand (NARIT). AMS gratefully acknowledges support from the UK Alan Turing Institute under grant reference EP/V030302/1.

This work has been made possible by the participation of more than 12,000 volunteers in the Radio Galaxy Zoo Project. The data in this paper are the result of the efforts of the Radio Galaxy Zoo volunteers, without whom none of this work would be possible. Their efforts are individually acknowledged at \href{http://rgzauthors.galaxyzoo.org}{http://rgzauthors.galaxyzoo.org}.
\section*{Data Availability}


The full catalogue of FR RGZ data, containing 14,375 radio sources, will be made publicly available at \url{https://doi.org/10.5281/zenodo.14031760}. The self-supervised learning approach used in this work was previously published at \cite{Slijepcevic2023RadioLearning}, and the code for the model is available at \url{https://github.com/inigoval/}. Standardised catalogues of labelled radio source images, including MiraBest, which were used to train the model, are accessible at   ~\url{https://doi.org/10.5281/zenodo.4288837} and \url{https://doi.org/10.5281/zenodo.8188867}, respectively.



\bibliographystyle{mnras}
\bibliography{rgz_fr_paper} 




\appendix



\bsp	
\label{lastpage}
\end{document}